# Notes from the Physics Teaching Lab: Optical Pumping

Kenneth G. Libbrecht[1]
Department of Physics, California Institute of Technology

**Abstract. We describe a series of experiments done using a commercially available optical-pumping apparatus that is currently being used in physics teaching labs at over 100 universities. Our focus here is to provide an extensive and detailed examination of the capabilities of this instrument, including numerous examples of measurements and data analysis, presented as a supplement to the manufacturer's user manual. Our hope is that instructors using this or similar optical-pumping instruments will find the experiments described here useful for designing and implementing the curricula in their own physics teaching labs.**

## Introduction

Optical pumping of rubidium vapor has become something of a standard installation in many physics teaching labs because it is so well suited as an educational tool. The phenomenon itself is a Nobel-prize-winning example of atomic state manipulation and spectroscopy, which connects very well with typical undergraduate courses in quantum mechanics. Because rubidium (Rb) is essentially a single-electron atom, the ground-state Zeeman transitions revealed by optical pumping offer a next logical step after learning about the hydrogen atom. The physics of optical pumping is complex enough to be challenging, but also quite engaging and satisfying to see first-hand in the lab.

Beyond presenting an interesting subject matter, the hardware needed for optical pumping is relatively simple and robust, making it reliable in day-to-day lab operation with undergraduate students. Moreover, the instrument avoids the eye-safety issues that come with Rb laser spectroscopy, making optical-pumping experiment well suited for use in an open laboratory setting.

A quick internet search reveals dozens of teaching-lab handouts and other materials that describe the underlying physics of optical pumping and Rb atomic spectroscopy, demonstrated using home-built instruments and especially a popular commercially available apparatus from *TeachSpin*. The *TeachSpin* user's manual additionally provides many details of the instrument construction. What is missing from all these resources, however, is a detailed description of what kinds of data can be obtained with an optical-pumping system, and what areas in parameter space provide the clearest signals and the best pedagogical experiences for students. Surprisingly, there are few journal articles (that we could find, especially modern papers) that present optical-pumping data like what one typically obtains in an undergraduate teaching lab. The aim of this paper is to fill this gap, hopefully providing a useful resource for teaching-lab instructors who may not have enough time to investigate the many facets of optical pumping physics that can be demonstrated using this apparatus.

[1] kgl@caltech.edu

In many teaching-lab handouts describing optical pumping, we have found that there is a general tendency to outline the physics theory is considerable detail and further describe the overall layout and physical concepts involved in the instrumentation. From there, however, it is often left for students to take data and compare their results with theory, using handouts that may not provide a lot of detailed experimental guidance. In our lab classes, we have found that this strategy does not work especially well. Depending on their background and motivation, many students have difficulty knowing exactly what they should do, how they should do it, and what good experimental data should look like. And they rarely have enough free time in their busy course schedules to ponder the lab experiments enough to develop a deep understanding of the underlying physics.

As a result, students often stop as soon as they have obtained some meager experimental data in the lab, with many assuming that this is the best the apparatus has to offer. This is a rather unsatisfying result for all concerned, plus it presents a substantial lost pedagogical opportunity. What works better (in our opinion) is to teach students how to set up the different investigations in detail, and then (importantly) show them what good data should look like. This gives them a concrete goal to achieve, and it drives them to better understand what signals they are seeing.

Of course, there is no single best approach to teaching, and lab instructors will doubtless have different opinions about what course materials are best suited for their students, and how that material should be presented. The investigations presented below are therefore meant to facilitate these pedagogical decisions by demonstrating what kinds of data can be generated using the *TeachSpin* optical-pumping apparatus.

## Rubidium atomic structure

This optical-pumping experiment focuses on rubidium atoms because Rb has a proven track record as a useful atomic case study, demonstrating many general physical principles by example. Rubidium is relatively easy to work with in the lab, and it has a single valence electron, giving it a relatively simple hydrogen-like atomic structure. Because Rb acts much like hydrogen in its spectroscopic behavior, choosing this atom simplifies the problem enormously right from the start.

In natural rubidium, the isotopes $^{85}$Rb and $^{87}$Rb occur in the ratio 72:28 percent, and our first task is to review the atomic level structure of these two rubidium isotopes. We will not delve into all the underlying quantum mechanics in detail here but rather focus on how it applies to the Rb atomic system.

The occupied electron orbitals of a rubidium atom are

$$1s^2 2s^2 2p^6 3s^2 3p^6 3d^{10} 4s^2 4p^6 5s$$

and the first 36 electrons are in closed shells with zero total angular momentum. The 5s electron therefore acts much like the sole electron in a hydrogen atom. The electronic energy levels of this sole electron can be described by the level diagrams in Figures 1 and 2. These show just the lowest energy S and P levels, as the higher levels are not important for our experiments.

Starting on the left in these diagrams, the ground state is an S state with an orbital angular momentum of L = 0, and the upper state is a P state with L = 1. Adding spin-orbit coupling gives the second “fine structure” column in both diagrams, with the levels shown in the usual Russell-Saunders notation $^{2S+1}L_J$, where S represents the total spin angular momentum (S = 1/2 for a single electron), L specifies the total orbital angular momentum (designated S, P, D when L = 0, 1, 2, respectively … alas, this S is different from the total-spin S), and J refers to the total angular momentum. Clearly, the optical pumping experiment is intended for students who have already taken a full-year course in quantum mechanics that covered spin-orbit coupling and related topics. One great appeal of the optical pumping in the teaching lab is how it provides a deep dive into a specific atomic case study, which is often a useful method for cementing general knowledge learned in theory courses.

After adding spin-orbit coupling, there are two optical transitions to consider, conventionally called D1 and D2 for alkali atoms. (Only D1 is labeled in the diagrams, as this is the transition we will be considering.) For rubidium, the wavelengths of these transitions are 795 nm (D1 line) and

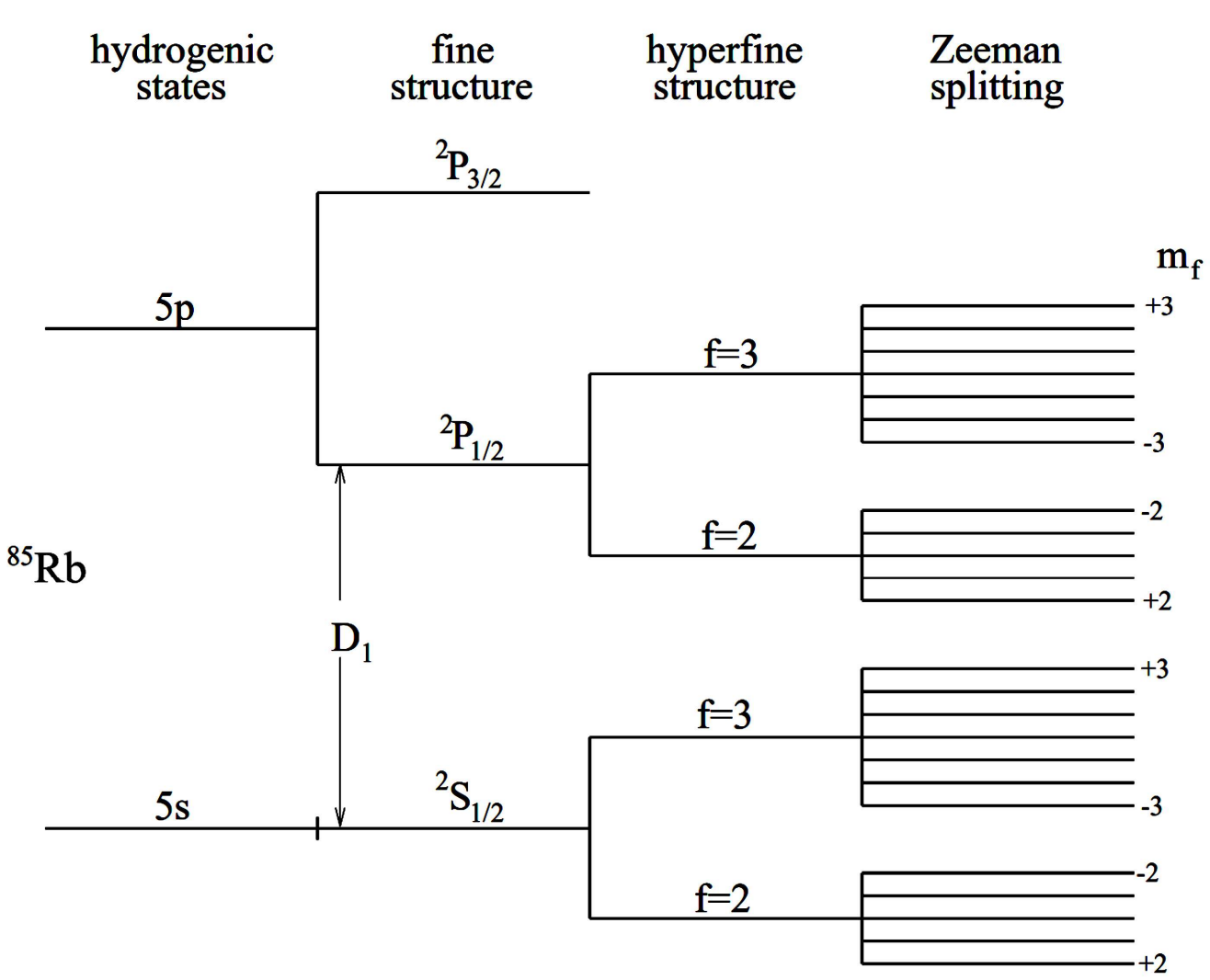

*Figure 1. The $^{85}Rb$ level diagram, focusing on the D1 line at 795 nm.*

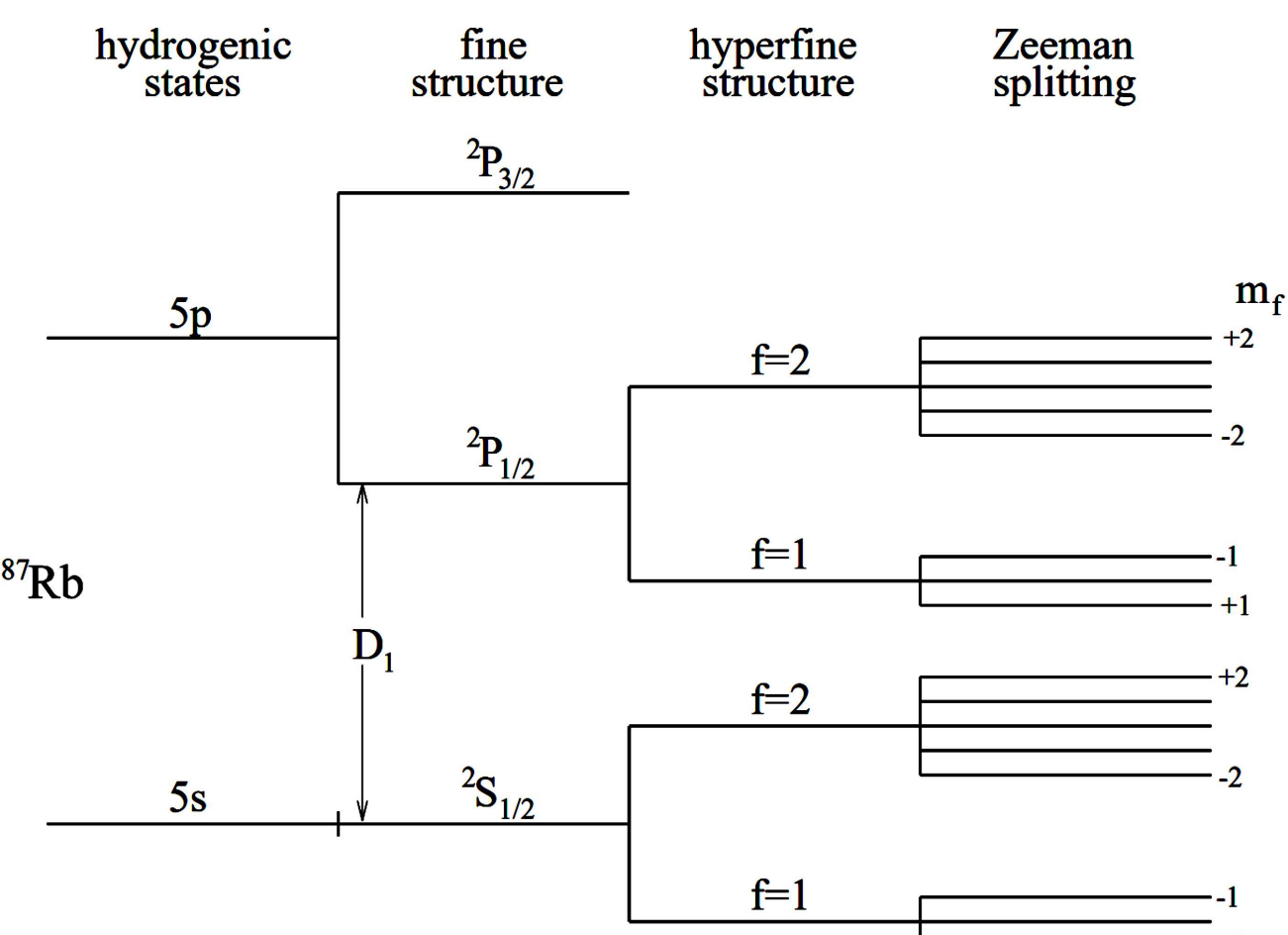

*Figure 2. The $^{87}Rb$ level diagram, focusing on the D1 line at 795 nm.*

780 nm (D2 line), and these are in the near-infrared part of the optical spectrum. In the *TeachSpin* Optical Pumping apparatus, the D2 line is suppressed using a narrow-band optical filter, so only the D1 transition is excited using optical photons.

Next we add nuclear spin, equal to I = 5/2 for $^{85}$Rb and I = 3/2 for $^{87}$Rb, giving the third "hyperfine structure" column in the diagrams. Here F shows the angular-momentum sum of J and I, and the energy splitting arises because the electron and nuclear magnetic moments interact with one another. Because L = 0 in the ground-state $^2S_{1/2}$ levels, spin-orbit coupling gives simply J = S = ½. Further adding the nuclear spin I, the only two possibilities are F = I ± ½.

Note that the different energy-level splittings are not to scale in these figures. The hydrogenic states are separated by about 1 eV, the spin-orbit splittings (between the $P_{1/2}$ and $P_{3/2}$ states) are about $10^{-3}$ eV, and the hyperfine splittings are of order $10^{-6}$ eV. Putting in some numbers, the D1 and D2 lines have transition frequencies of 377 THz (795 nm) and 384 THz (780 nm), while the ground-state hyperfine splittings are 3035.73 MHz for $^{85}$Rb and 6834.68 MHz for $^{87}$Rb. The upper-state hyperfine splittings are smaller, equal to about 361 MHz and 816 MHz for the $P_{1/2}$ states of $^{85}$Rb and $^{87}$Rb, respectively.

We note in passing that it is quite difficult to accurately calculate any of these numbers, owing to the complexities of many-body quantum-mechanical models. Spectroscopic measurements can be made with great precision, however, so all the transition frequencies are known to many significant digits. Even the hydrogen atom theory is not easy to solve with high precision, as the finite size and mass of the proton, relativistic effects, the Lamb shift (from QED), and other factors present theoretical and computational challenges. We often gloss over this inconvenient truth in our QM courses, but it cannot be ignored in the lab. The usual way to get your head around the spectroscopy problem in a finite time is to make a lot of simplifications, ignoring small perturbations whenever possible. Therefore, the discussion that follows is approximate and focuses mainly on the physics needed to understand the Optical Pumping experiment.

Proceeding to the final column in Figures 1 and 2, adding an external magnetic field produces Zeeman splittings that lift the $m_F$ degeneracy of the hyperfine states as shown. For low magnetic field strengths (Zeeman splitting $<<$ hyperfine splitting), the Zeeman levels are shifted by

$$\Delta E \approx g_F \mu_B B m_F \tag{1}$$

where $B$ is the magnetic field strength, $\mu_B = e\hbar/2m_e \approx 1.40$ MHz/Gauss (14 GHz/T) is the Bohr magneton, and the Landé g-factor is

$$g_F \approx g_J \frac{F(F+1) + J(J+1) - I(I+1)}{2F(F+1)} \tag{2}$$

with

$$g_J \approx 1 + \frac{J(J+1) + S(S+1) - L(L+1)}{2J(J+1)} \tag{3}$$

These expressions ignore the contribution of the nuclear magnetic moment, because it is about 2000x smaller than the electron magnetic moment.

## Simplifications for Rubidium

Now is a good time to pause and think about what all this means. It is helpful to have a specific case study in mind, so consider only the ground state $^2S_{1/2}$ levels of $^{85}$Rb (because this paper focuses mainly on just these levels). The hyperfine states reflect the angular-momentum addition of one electron spin (S = 1/2) with the $^{85}$Rb nuclear spin (I=5/2), giving two ground states with F = 2 and F = 3. We will often narrow our focus even more and consider just the F = 3 manifold.

Note that we refer to all the $^2S_{1/2}$ levels as "ground states" because these levels are essentially equally populated in thermal equilibrium. In contrast, the P levels are generally unpopulated, because they quickly decay (in ~ 30 nanoseconds) to the ground states. In a gas of Rb atoms, atomic collisions will soon "mix" the ground states, and this process tends to equalize the level populations. At room temperature we have $kT \approx 0.025$ eV, so collisions have more than enough energy to scramble the hyperfine levels separated by only $10^{-6}$ eV. But thermal collisions are not nearly energetic enough to excite the P levels. So, this is the normal state of affairs in a Rb gas – nothing in the P levels, while all the $^2S_{1/2}$ ground states are roughly equally populated.

Moving on to the Zeeman levels for $^{85}$Rb, the ground state $^2S_{1/2}$ levels have S = J = 1/2 and L = 0, so substituting into the above equations gives $g_J = 2$. This is just the "bare" Lande g-factor of the electron, call it $g_e$. If we set I = 0 for a moment, then F = J = S = 1/2 and the above expressions give $g_F = g_J = 2$. In this fictitious hyperfine-free atom, the ground state has only two Zeeman levels with $m_F = \pm 1/2$, and the Zeeman splitting is just the $\mu \cdot B$ energy of the electron in a magnetic field. The energy shift $\Delta E \approx g_F \mu_B B m_F$ given above means that the two $m_F = \pm 1/2$ levels are separated by $\Delta E \approx 2\mu_B B$, giving a transition frequency $\nu_{Zeeman} = \Delta E/h \approx 28$ GHz/Tesla $\approx 2.8$ MHz/Gauss. So, in the absence of nuclear spin, the Zeeman splitting arises entirely from the magnetic moment of the sole valence electron. (Note that quantum electrodynamics adds an additional correction, so real electrons have $g_e \approx 2.0023193043626$. But we ignore this small correction here.)

For the ground state $^2S_{1/2}$ levels of $^{85}$Rb, we see that Equation (2) is overly general for our needs. For these ground states we just showed $g_J = 2$, and hyperfine coupling for one-electron atoms simply gives F = I ± 1/2. Plugging this into Equation (2) then yields

$$g_F \approx \pm \frac{g_e}{2I+1} \approx \pm \frac{2}{2I+1} \tag{4}$$

where the plus sign applies to F = I+1/2 and the minus sign applies to F = I–1/2. Again, this expression applies in the limit $g_I \approx 0$, because the electron magnetic moment dominates the Zeeman splitting.

For $^{85}$Rb, we see that $g_F \approx \pm 1/3$, where the plus sign applies to the F = 3 manifold and the minus sign applies to the F = 2 manifold. As you can see in Figure 1, the F = 2 Zeeman manifold is indeed inverted from the "normal" F = 3 manifold.

Interestingly, we see that both the F = 2 and F = 3 manifolds have the same magnitude for the Zeeman splitting, because $|g_F| \approx 1/3$ for both. This makes sense because the low-field Zeeman splitting all comes from the electron spin, and there is only one electron in this problem. Also, the maximum value of $m_F = 3$ (for the $^{85}$Rb ground states) gives $\Delta E \approx \mu_B B$. Thus the maximum $m_F$ value means the electron magnetic moment is aligned along the applied $B$ field. Similarly, the minimum value $m_F = -3$ means the electron is aligned anti-parallel to the applied $B$ field. And the Zeeman energy splitting between these two maximal states is $\Delta E \approx 2\mu_B B$, the same as that for a bare electron.

In the lab, optical radiation is needed to excite the atoms from the S to P states, with frequencies of several hundred THz. But the Zeeman splittings are much smaller, so we only need electromagnetic radiation with a frequency

$$\nu_{Zeeman} = g_F \frac{\mu_B}{h} B \tag{5}$$

with $\mu_B/h = 1.3996245$ MHz/Gauss. This radiation is provided by oscillating magnetic fields that drive magnetic-dipole transitions between the different Zeeman levels. Quantum mechanical selection rules prohibit magnetic-dipole transitions except between Zeeman levels separated by $\Delta m = 0, \pm 1$.

We will mainly focus on the $^{85}$Rb ground states in this paper, and here are some parameters that often come up with this isotope:

| | |
|---|---|
| I = 5/2 | = $^{85}$Rb nuclear spin |
| $\Delta E_{hf}/h = 3035.73$ MHz | = ground state hyperfine splitting |
| $\nu_{Zeeman}/B = 467$ kHz/Gauss | = low-field Zeeman splitting |
| $g_J \approx 2$ | = electronic g-factor |
| $g_I \approx -2.936 \times 10^{-4}$ | = nuclear g-factor |

## Strong-field Zeeman splitting

If the magnetic field is strong enough, the Zeeman splitting starts to become comparable to the hyperfine splitting, and then Equation (1) is no longer an accurate representation of the energy levels. A better treatment of the Zeeman effect yields what is called the *Breit-Rabi equation* [1931Bre]. For the special case of ground-state alkali atoms (with S=1/2, L=0, J=1/2, and F = I±1/2), the Breit-Rabi equation becomes

$$\Delta E_{Zeeman} = g_I \mu_B B m_F \pm \frac{\Delta E_{hf}}{2} \sqrt{1 + \frac{4 m_F}{2I+1} x + x^2} \tag{6}$$

where $g_I$ is the small nuclear g-factor (previously neglected), $\Delta E_{hf}$ is the hyperfine splitting of the $^2S_{1/2}$ level, the ± in this expression refers to the upper (+) and lower (-) hyperfine levels, and the dimensionless factor $x$ is

$$x = (g_J - g_I) \frac{\mu_B B}{\Delta E_{hf}} \tag{7}$$

Evaluating Breit-Rabi gives the results in Figure 3, which show the Zeeman energy levels for the Rb ground states. [As a historical aside, the original Breit-Rabi paper [1931Bre] (written in English) presents a fascinating vignette into how theoretical physics has changed over the past century.]

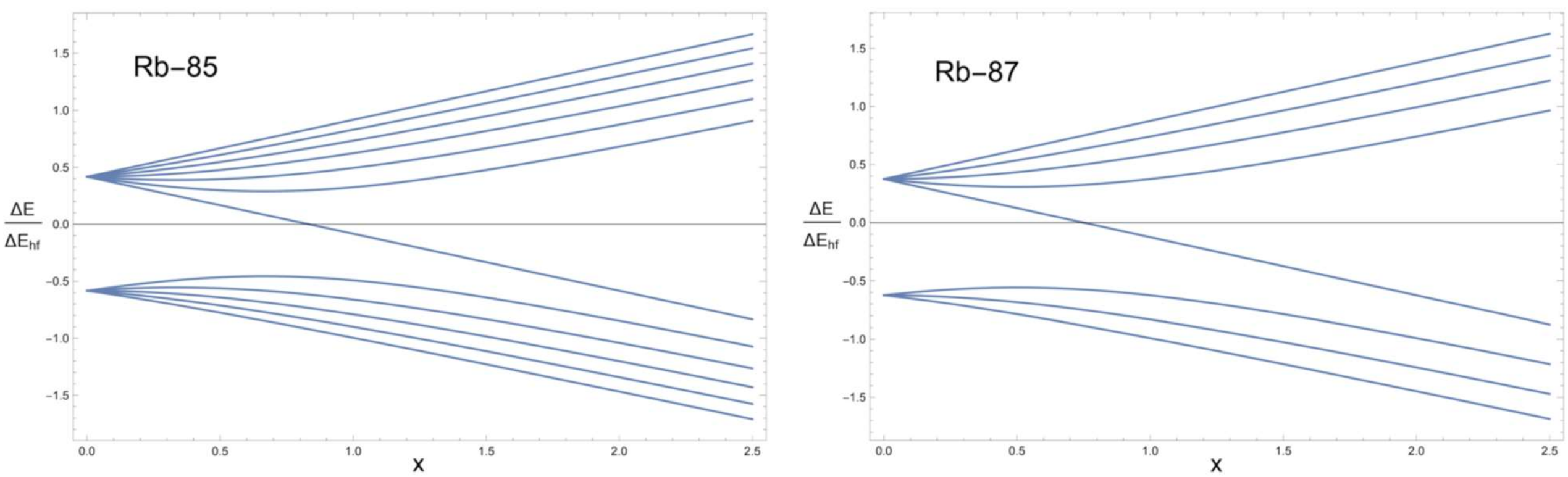

*Figure 3. The Zeeman energy levels for the $2S_{1/2}$ ground states of Rb85 (left) and Rb87 (right) according to the Breit-Rabi formula, plotting $\Delta E/\Delta E_{hf}$ as a function of $x = (g_J - g_I)\mu_B m_F B/\Delta E_{hf}$. We can observe nonlinear Zeeman splittings in the Optical Pumping lab, but our instrumental capabilities are confined to the far left in these diagrams.*

# Optical pumping

Now that we have a basic picture of the energy level diagrams for both Rubidium isotopes, the next step is to examine how light can be used to manipulate the populations of the various atomic states, a process that is sometimes called *quantum-state engineering*. The phenomenon of optical pumping is a staple of AMO physics experiments, used in laser cooling of atomic gases, trapping atoms and molecules in optical potentials, AMO searches for fundamental physics beyond the Standard Model, quantum entanglement, quantum computing, and many other applications. Once again, our goal here is to present a case study using our Optical Pumping apparatus, as this relatively inexpensive experiment demonstrates many aspects of the essential physics one often finds in modern AMO research labs.

## Level populations and transitions

In the Optical Pumping (OP) experiment, we use optical radiation near 795nm to excite the D1 transition shown in Figures 1 and 2. This is an “allowed” electric-dipole transition, restricted by the quantum selection rules $\Delta L = \pm 1$, $\Delta F = 0$ or $\pm 1$ (except $F = 0$ to $F = 0$ is not allowed), and $\Delta m = 0$ or $\pm 1$. Once an atom has been excited from $S_{1/2}$ to $P_{1/2}$, the upper state rapidly decays back to the ground state by emitting a photon, with a decay time of about 30 nsec, following the emission selection rule $\Delta m = 0$ or $\pm 1$.

The Rb lamp in this experiment uses electron collisions in a Rb plasma to excite Rb atoms to the upper electronic states, where they decay and produce D1 (and other) photons. After removing unwanted radiation using an optical bandpass filter, what is left is Doppler-broadened radiation near the 795nm emission peak. And these lamp photons pass through the Rb gas cell, where they will readily excite all the D1 transitions that are allowed by selection rules. The energy differences between the various hyperfine and Zeeman energy splittings are inconsequential when considering excitations of the D1 transitions in this experiment. Modern AMO experiments generally use narrow-band lasers that can select individual transitions, but the OP lamp spectrum includes multiple optical transitions and is further broadened by Doppler motion.

Because the hyperfine splittings are small compared to $kT$, all the ground-state hyperfine levels in the Rb cell are roughly equally populated in thermal equilibrium. In contrast, the $P_{1/2}$ state population is essentially zero in thermal equilibrium. This provides the conceptual starting point for the OP experiment – when no light is shining on the Rb atoms in the vapor cell, they are all in the $S_{1/2}$ ground state, equally distributed among the various hyperfine ground states.

This situation does not change much if *unpolarized* 795nm light is shining on the Rb atoms. These photons excite all the allowed D1 transitions, and the excited states decay back down rapidly, yielding essentially the same uniformly populated hyperfine states. The lamp intensity is low, so the excited-state $P_{1/2}$ population is always very low compared to the ground-state $S_{1/2}$ population.

## Light polarization & angular momentum

The ground-state level populations can change, however, if *circularly polarized light* is used to excite the D1 transition. Because photons are spin-1 massless particles, they carry angular momentum equal to $\pm\hbar$, and this angular momentum can be transferred to the atom depending on the polarization of the incident light wave. Any polarization state can be expressed as a linear combination of the usual horizontal $|H\rangle$ and vertical $|V\rangle$ polarization states, because these two states provide an orthogonal basis set. The electric field vector is confined to the horizonal plane for $|H\rangle$ and the vertical plane for $|V\rangle$.

Alternatively, any polarization state can be expressed as a linear combination of right-circular $|RC\rangle$ and left-circular $|LC\rangle$ polarization, because these two "helical" states also provide an orthogonal basis set. What is important for our present discussion is that light in the $|LC\rangle$ polarization state will only drive so-called $\sigma^+$ transitions that increase the $m_F$ value of the atomic state, as measured along the propagation direction of the light beam. Put another way, a single $\sigma^+$ photon will engage the $\Delta m = +1$ selection rule, and only that rule; light in the $\sigma^+$ polarization state will *not* engage the $\Delta m = 0$ or $\Delta m = -1$ selection rules. The same applies for $\sigma^-$ photons, except these photons decrease $m_F$, as you would expect.

If we apply an external magnetic field along the optical axis, then this field establishes the stable Zeeman states in the system. It is important to define a coordinate system for this discussion, and this is fixed by the on-axis magnetic field. In this coordinate basis, the $m_F$ states will be eigenstates of the system, so they will be stable in time if unperturbed. You can see why this is the case in classical physics by considering a $^{85}$Rb atom in the F=3 ground state with $m_F = +3$. In this maximal state, the electron spin is fully aligned with the magnetic field, so the magnetic dipole moment of the electron is aligned as well. Classically, a magnetic dipole in a magnetic field will precess about the B field; but a dipole along B does not precess. This is a classical-physics picture of why the $m_F = +3$ state is stable. The quantum-physics view is simply that $m_F = +3$ and all the other $m_F$ states are eigenstates of the Hamiltonian.

Note also that linearly polarized light contains equal amounts of $|RC\rangle$ and $|LC\rangle$ states because both $|H\rangle$ and $|V\rangle$ can be written as linear combinations of $|RC\rangle$ and $|LC\rangle$. The details are described by the *Jones calculus* representation of polarization states. Unpolarized light will also contain equal amounts of $|RC\rangle$ and $|LC\rangle$ states.

## The OP signal

Putting all this together, Figure 4 sketches the basic layout of the OP experiment, and Figure 5 describes the essential physics involved in optical pumping. Together these diagrams illustrate the most important concepts in OP experiments. Note the axial magnetic field $B_z$ in Figure 4 that defines the stable Zeeman sublevels (eigenstates of the Hamiltonian). The incident light propagating along the $B_z$ axis has a $\sigma^+$ polarization, which means that optical excitations are restricted to $\Delta m = +1$ in this Zeeman basis. After $\Delta m = +1$ optical excitation, however, optical decays are allowed for $\Delta m = 0$ and $\Delta m = \pm 1$.

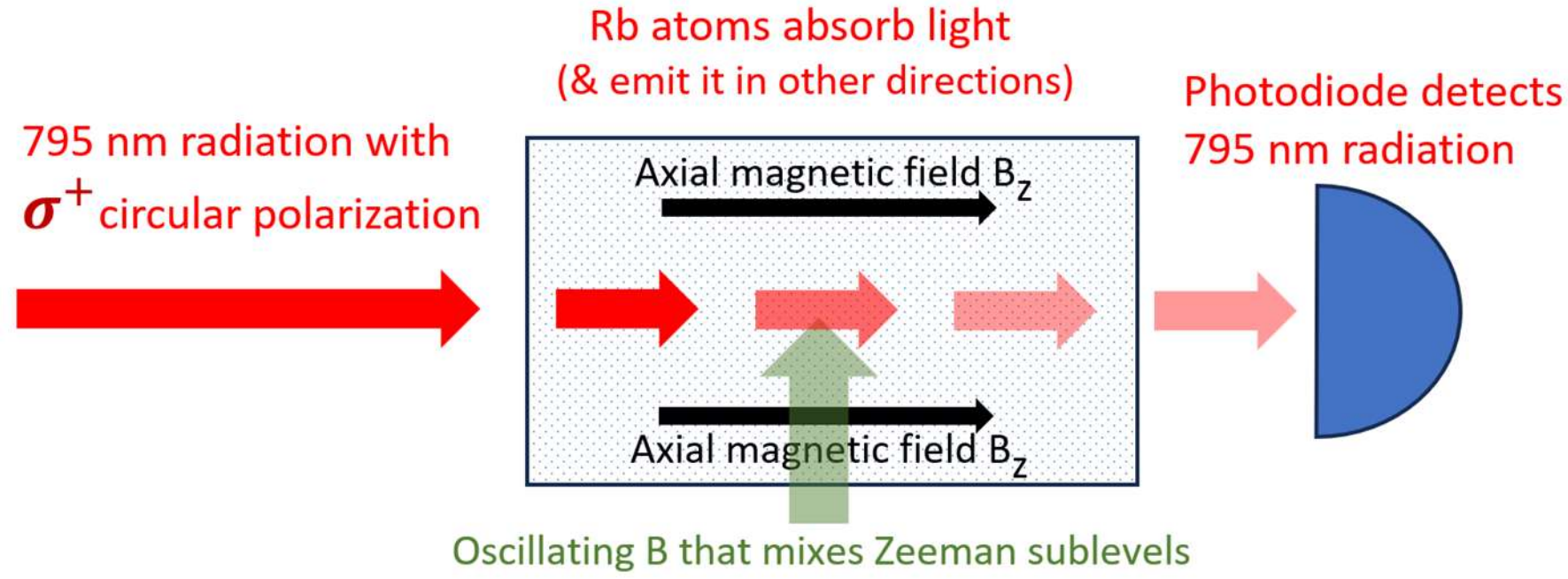

*Figure 4. The essential operation of the Optical Pumping optics. A lamp produces a beam of circularly polarized $\sigma^+$ light that passes through the Rb vapor cell and is measured by a photodiode. Some of the light is absorbed by Rb atoms in the cell and does not reach the detector. An axial magnetic field $B_z$ (which we also call the Horizontal Field) is applied along the optical axis. An additional wire coil is used to apply an oscillating magnetic field that excited Zeeman Transitions (ZT) in the Rb atoms. If the ZT frequency equals the Zeeman splitting, then transitions will mix the Rb ground-state Zeeman levels.*

As illustrated in Figure 5, each upward $\sigma^+$ transition increases $m_F$ by $\Delta m = +1$, while each subsequent decay from an upper state leaves the value of $m_F$ statistically unchanged (approximately). After many $\sigma^+$ transitions have occurred, the atoms will inevitably end up with the maximum $m_F$ value of the ground state. From this state, however, there are no longer any $\sigma^+$ transitions. This maximum $m_F$ state is often called a "dark state" in AMO jargon because atoms in this state cannot absorb (and thereby scatter) any of the incoming $\sigma^+$ light. For OP experiments, it might be better to say that the optically pumped atoms are "transparent", because their presence reduces the overall absorption of $\sigma^+$ light, thus increasing the amount of light striking the photodetector in Figure 4. The Rb case shown in Figure 6 is more complicated than the fictional atoms in in Figure 5, but the essential physics is the same.

In our OP experiment the net result of optical pumping is a manipulation of the populations of the ground-state atoms in the Rb vapor cell. With no optical pumping, the different ground-state Zeeman levels are all populated equally (in a statistical sense). In the presence of optical pumping, more atoms are in the maximum $m_F$ state, where they cannot absorb incident light. From this reasoning, we see that the amount of light hitting the photodiode in Figure 4 can be used to measure the degree of optical pumping that is present.

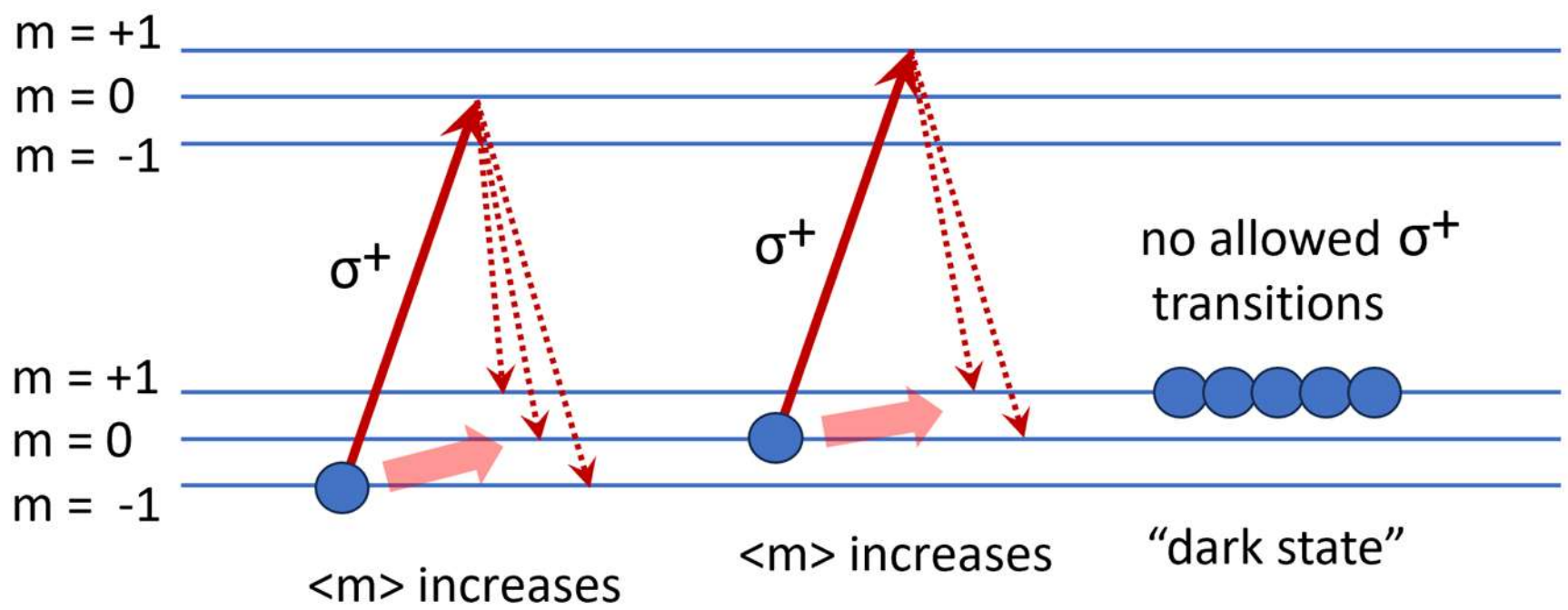

*Figure 5. The essential physics of optical pumping, illustrated using a simpler F=1 to F=1 optical transition. In the presence of a magnetic field, the Zeeman levels are split. Circularly polarized light applied along the magnetic field direction drives only $\sigma^+$ optical excitations with Δm=+1 (red arrows from the ground states to the excited states). Spontaneous decays back to the ground state can have Δm=-1, 0, or +1. After multiple excitations and decays, the atoms will be "pumped" into the m=+1 ground state. This is called a "dark" state because it can no longer be excited by $\sigma^+$ photons. If no other processes redistribute the ground state sublevels, optical pumping will soon send all the atoms into the dark state, where they can no longer absorb any of the incident $\sigma^+$ light. Rubidium atoms have a more complicated level structure, but the optical pumping process is essentially the same as shown here.*

As is often the case in laboratory physics experiments, the optical-pumping phenomenon involves numerous physical concepts all mixed together. In theory classes, one can isolate each new concept and examine it in isolation, which has many pedagogical benefits. In the experimental world, however, it becomes necessary to take a more holistic view of the apparatus and its various components. Life is more complicated in the lab, but the holistic viewpoint is often useful for solidifying one's understanding of the interplay between different physical phenomena. To this end, we next present the OP experiment and some resulting data in detail.

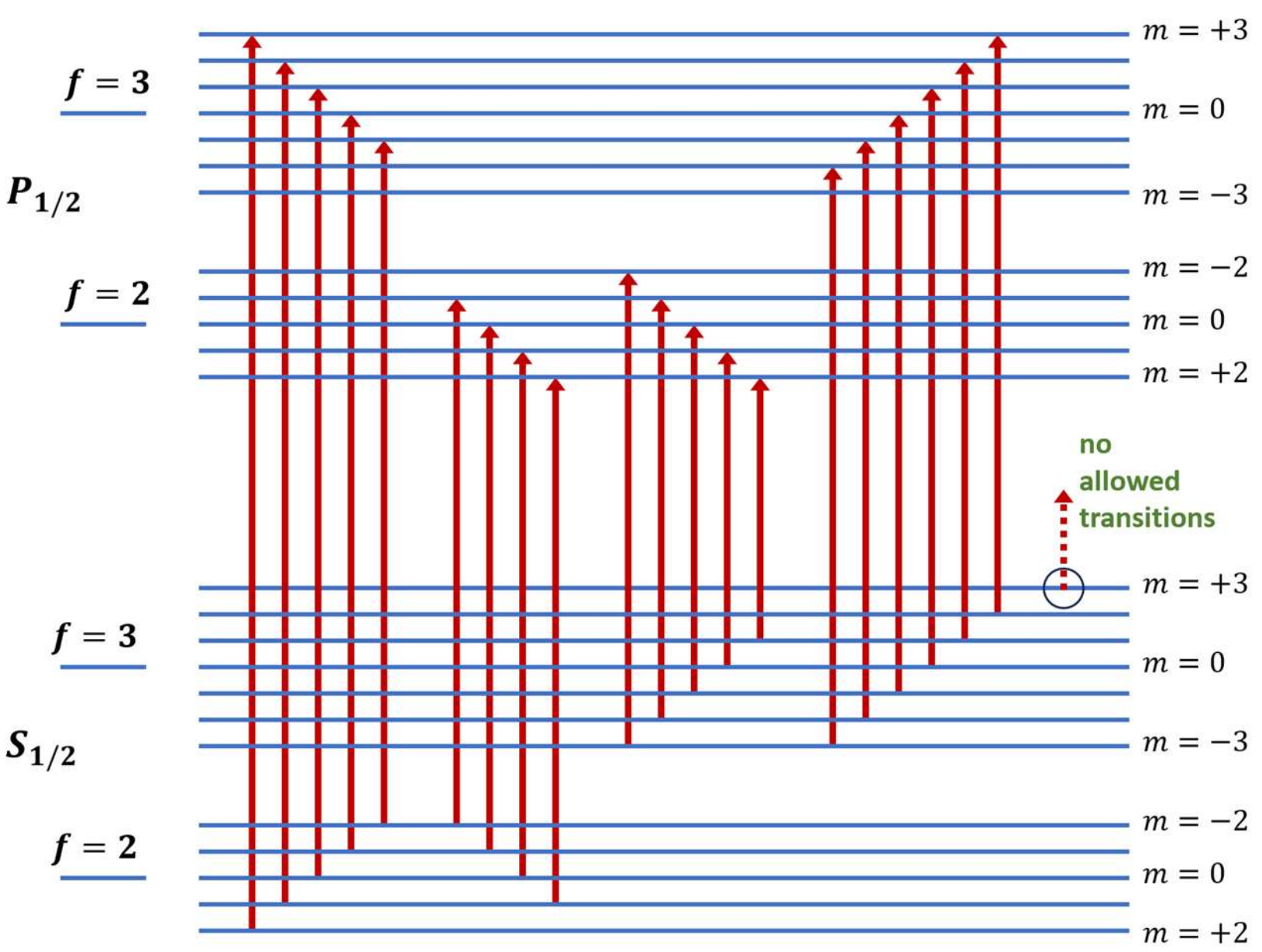

*Figure 6. The Rb case includes more Zeeman levels compared to the case shown in Figure 5, but the optical-pumping process is essentially the same. After many $\sigma^+$ excitations from the $S_{1/2}$ ground states to the $P_{1/2}$ excited states, the atoms will inevitably be pumped into the m=+3 level, where they remain trapped. If $\sigma^-$ were used instead, then atoms would be pumped into the m=-3 level, where again they would be trapped.*

## Basic optical pumping in the lab

Figure 7 shows a sketch of the optical layout of our OP apparatus, and more details about the hardware can be found in Appendix 1. Starting on the left side of the diagram, the Rb lamp consists of a heated glass cell containing Rb vapor along with an inert buffer gas. An electrodeless electrical discharge inside the lamp excites the Rb atoms into higher electronic states, from which they decay back down to the ground state. The Rb lamp emits light from many atomic transitions, but the brightest components are the D1 (795 nm) and D2 (780 nm) lines, which are a deep red color, just beyond the range of human vision.

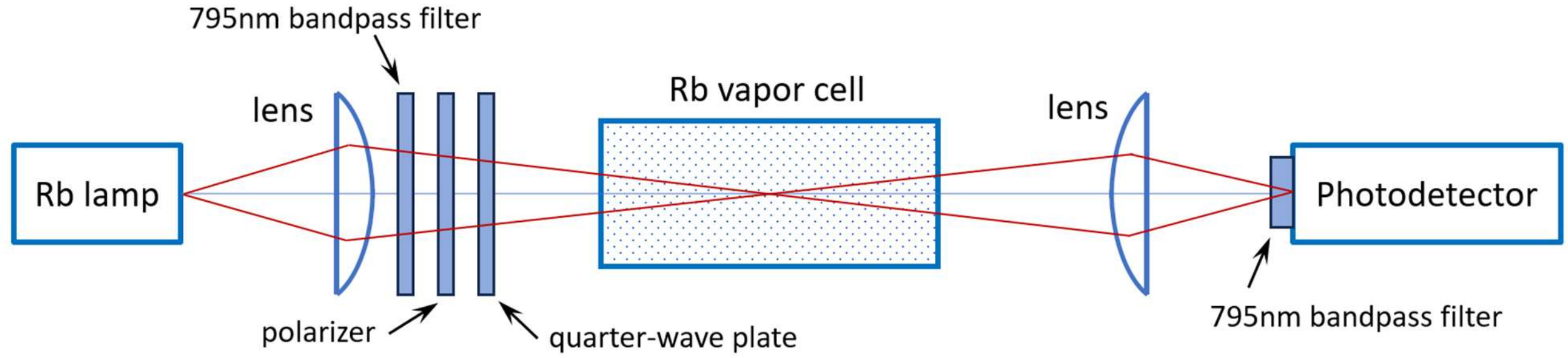

*Figure 7. The optical layout of our Optical Pumping apparatus. This is essentially the same as the original TeachSpin apparatus, although we substituted a new photodetector and an additional 795nm bandpass filter, as described in Appendix 1.*

The lamp light goes through a collimating lens, then a 795-nm bandpass filter that transmits only light near the D1 line, then a linear polarizer, and then a quarter-wave plate. The polarizer and quarter-wave plate together produce circularly polarized light that then passes through the main Rb vapor cell. The polarizer is set to zero degrees, and the quarter-wave plate is set near 45 degrees, as these are the typical operating settings for producing $\sigma^+$ light.

A quick note on eye safety. The Rb lamp produces a measured light intensity of about 100 $\mu$W/cm$^2$ at a distance of 10 cm from the lamp. Even if you put your eye quite close to the lamp, you would only get about 10 $\mu$W hitting your retina. For comparison, staring at the Sun sends about 1 mW to your retina. Thus, although the light from the Rb lamp is effectively invisible, it is about 100x too dim to present any eye-safety issues.

The walls of the Rb vapor cell have a deposit of Rb metal on the inside surfaces (but not on the windows), which is in thermal equilibrium with Rb vapor in the cell. Heating the cell produces a higher vapor density, and the buffer gas reduces the net motion of the Rb atoms. Thermal velocities are about 1km/second, so a free Rb atom would only spend about 10 $\mu$sec in the optically active central part of the cell. Collisions with the inert buffer gas increase this time, yielding sharper spectral features. After passing through the Rb cell, the light passes through a second focusing lens and a second 795-nm bandpass filter that blocks the room lights.

As described above, Figure 4 illustrates the essential components of the OP apparatus: the lamp produces a beam of 795-nm $\sigma^+$ light that passes through the Rb cell and is detected by a photodetector, and the axial magnetic field $B_z$ establishes the Zeeman axis. In addition, Zeeman-Transition (ZT) radiation – an oscillating transverse magnetic field – is sometimes supplied to mix the Zeeman sub-levels of the Rb ground states. Figure 4 gives the conceptual picture for understanding the atomic physics, while Figure 7 sketches the necessary optical elements.

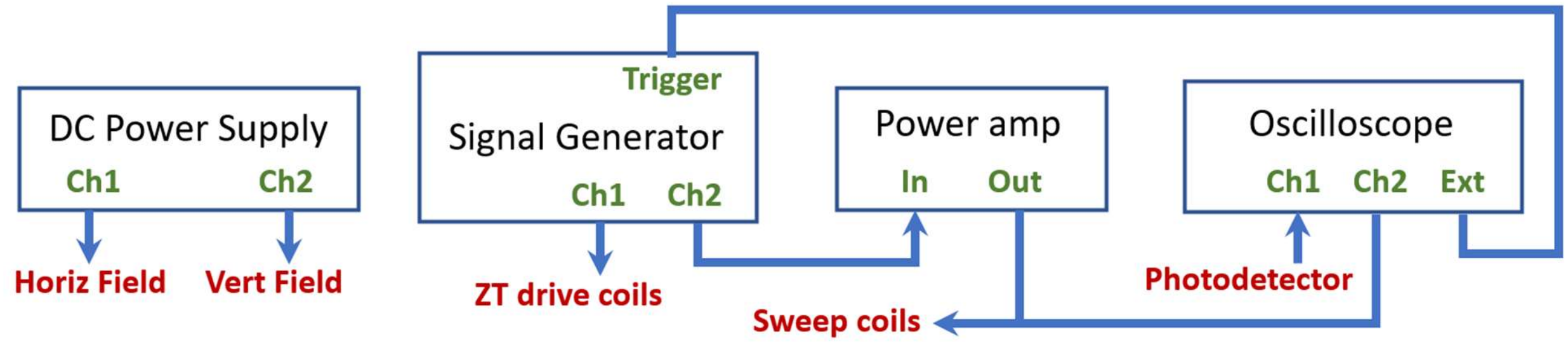

*Figure 8. A partial electronics layout used to observe the low-field optical-pumping signal. This setup is somewhat modified from the original TeachSpin hardware, and further details are described in Appendix 1.*

Figure 8 sketches the electronics that drives the OP experiment, with additional details provided in Appendix 1. Some key features include:

1) One channel of the DC Power Supply drives a set of Helmholtz coils that produce the uniform axial magnetic field $B_z$ shown in Figure 4. A small current in these coils will cancel the horizontal component of Earth's magnetic field (roughly 0.25 Gauss), giving $B_z = 0$, while a 1A current yields $B_z \approx 8$ Gauss for observing the strong-field Zeeman splittings. To produce a stable OP signal at high currents, it is best to use the power supply in its constant-current mode, as the coil resistance changes appreciably as the coils heat up.
2) The second channel of the DC Power Supply drives a second set of Helmholtz coils that cancel the vertical component $B_y$ of the earth's magnetic field.
3) The $B_x$ field (horizontal, but perpendicular to $B_z$) is reduced to near zero by orienting the apparatus so the optical axis points toward magnetic north. With a suitable orientation of the apparatus, together with the Horizontal and Vertical field coils, it is straightforward to reduce the magnetic field levels inside the Rb cell to near zero along all three axes. With this starting point, $B_z$ can then be adjusted to non-zero values, while the $B_x$ and $B_y$ field components remain at essentially zero in all our OP experiments.
4) The second channel of the Signal Generator is used to sweep $B_z$ using yet another set of Helmholtz coils oriented along the optical axis. We drive linear B-field sweeps by applying a triangle-wave signal to this set of Sweep coils, using the layout in Figure 8. The power amplifier is needed to supply coil currents that are beyond what the Signal Generator can deliver.
5) The first channel of the Signal Generator in Figure 8 drives a set of horizontal $B_x$ coils that provide an oscillating B field to excite Zeeman transitions, again as illustrated in Figure 4.

6) With the connections in Figure 8, the oscilloscope observes the OP signal from the photodetector together with the sweep waveform. The Signal Generator Trigger provides a trigger signal for the ‘scope that is synchronous with the sweep signal in ch2.

Once everything is set up (with the Rb cell at room temperature), Figure 9 shows some typical observations. Here we used a 1Hz sweep frequency, 300 mVpp to the sweep coil, and a 40 kHz, 300 mVrms sine-wave signal to the ZT drive coils. We usually give students some typical settings to get them started, as the parameter space for all the settings is quite large. [An operational detail: In normal operation, we typically keep the Signal Generator ch2 DC offset at zero, using only the DC Power Supply to adjust the DC value of $B_z$. This avoids possible confusion.]

Once students have experienced the satisfaction of setting up the hardware and finding the signal, we typically ask them to use a slower sweep speed to produce a clean OP signal like that shown in Figure 10. If they understand the signal they have just produced, it is straightforward to convert the time axis in the oscilloscope data to the applied $B_z$, as illustrated in Figure 10. If that understanding is lacking (often the case, given the substantial complexity of the OP physics and hardware), it is useful to pause at this point to contemplate the underlying physics in more detail. Having a signal on the oscilloscope with a related data-analysis exercise provides an excellent motivation for students to better understand the connection between theory and observations.

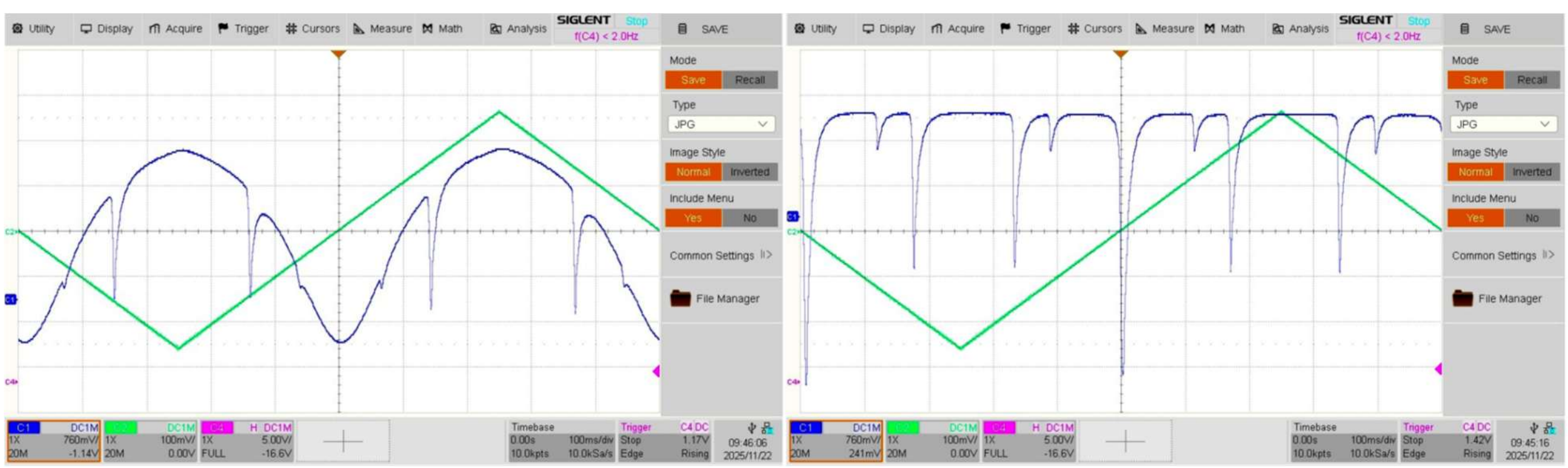

*Figure 9. These oscilloscope screenshots illustrate some typical first observations of the basic optical pumping signal, in this case sending a 40 kHz sine-wave signal to the* ***ZT drive coils*** *(see Figure 8). In both images the green trace shows the voltage going to the* ***Sweep coils*** *that modulate the horizontal magnetic field. The screenshot on the left was taken with a large vertical B field that produced a broad B=0 dip. The image on the right shows the same OP signal after adjusting the current to the* ***Vertical Field*** *coils to minimize the width of the B=0 dip. In both cases, the* ***Horizontal Field*** *was adjusted using the DC Power Supply to center the B=0 dip relative to the sweep signal.*

## Understanding the basic OP signal

Once the basic OP signal is visible on the ‘scope, it is instructive to first turn the ZT drive signal off and on, as this clearly demonstrates that the four “ZT dips” in Figure 10 are caused by the ZT radiation illustrated in Figure 4. We often refer to the large central dip (present with or without the ZT drive signal) as the “B=0” dip because it happens when the $B_z$ sweep goes through $B_z = 0$ (thus, the axial $B_z$ field changes sign when it goes through the $B_z = 0$ point).

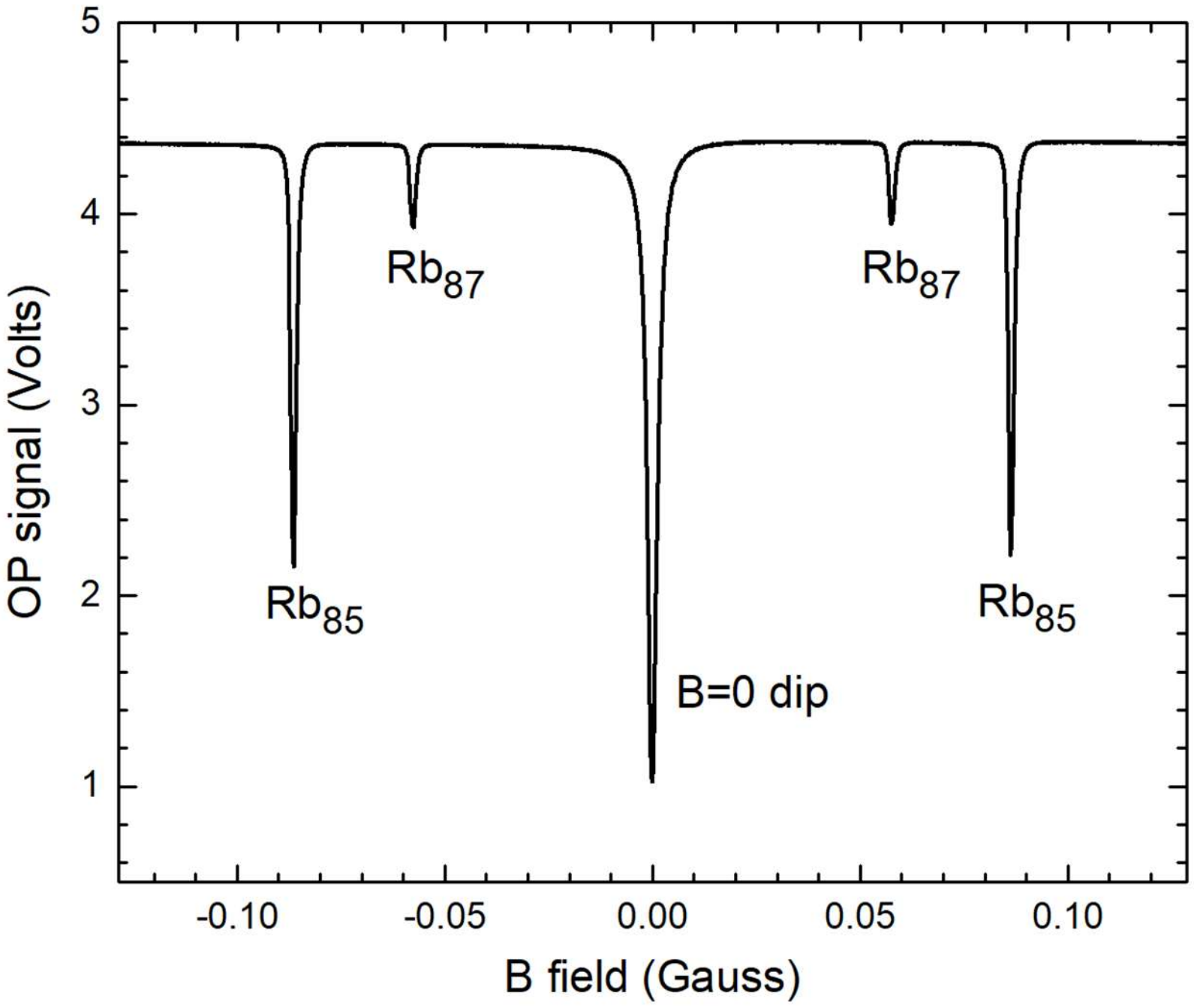

*Figure 10. This graph shows a cleaner view of the basic OP signal, obtained by using a slower sweep speed, averaging oscilloscope traces, and downloading the digital data for plotting. The position of the ZT dips was used to convert the 'scope sweep time to the applied $B_z$ in the Rb vapor cell. Here the Rb cell was at room temperature and we used a 40kHz, 300mVrms ZT sine-wave signal.*

We next consider the simplified level diagrams shown in Figures 11 and 12 to better understand the OP signal. These are basically like the diagrams in Figures 1 and 2, except stripped down to the bare essentials for the case of $^{85}$Rb. Figure 11 shows the basic optical pumping phenomenon, while Figure 12 shows why mixing the Zeeman levels using ZT radiation reduces the signal seen by the photodiode.

Note that as the $B_z$ field goes through its sweep, the photodetector (PD) signal mostly stays at a constant high level (except at the dips, obviously). This high PD level establishes the signal in the presence of strong optical pumping, as described in Figure 11. Note also that zero PD voltage does not refer to zero light level. The electronics sends the PD signal through a long-period high-pass filter that removes the DC component of the signal. In fact, the dips you see in Figure 9 are quite small compared to the large DC offset, as we investigate below.

## The ZT dips

The ZT dips appear when the ZT radiation (in the form of an oscillating transverse magnetic field in the Rb cell) drives transitions between the different Zeeman levels, as illustrated by the green arrows in Figure 12. Selection rules limit the transitions to $\Delta$m = $\pm1$, so a single ZT photon will only mix adjacent Zeeman levels. But many ZT transitions over time will soon mix all the states, roughly equilibrating the level populations.

As described in Figure 12, the transitions quickly scramble the Zeeman levels and thus disrupt the optical-pumping process. This results in fewer optically pumped "transparent" atoms (compared to the no-ZT situation shown in Figure 11), thus yielding more absorption that makes ZT dips in the OP signal. The dips only occur when Equation (5) is satisfied, and the observed dips are for the two Rb isotopes (which have different g-values). The ZT dips are symmetrically placed about $B_z =$ 0 because optical pumping works for either sign of $B_z$.

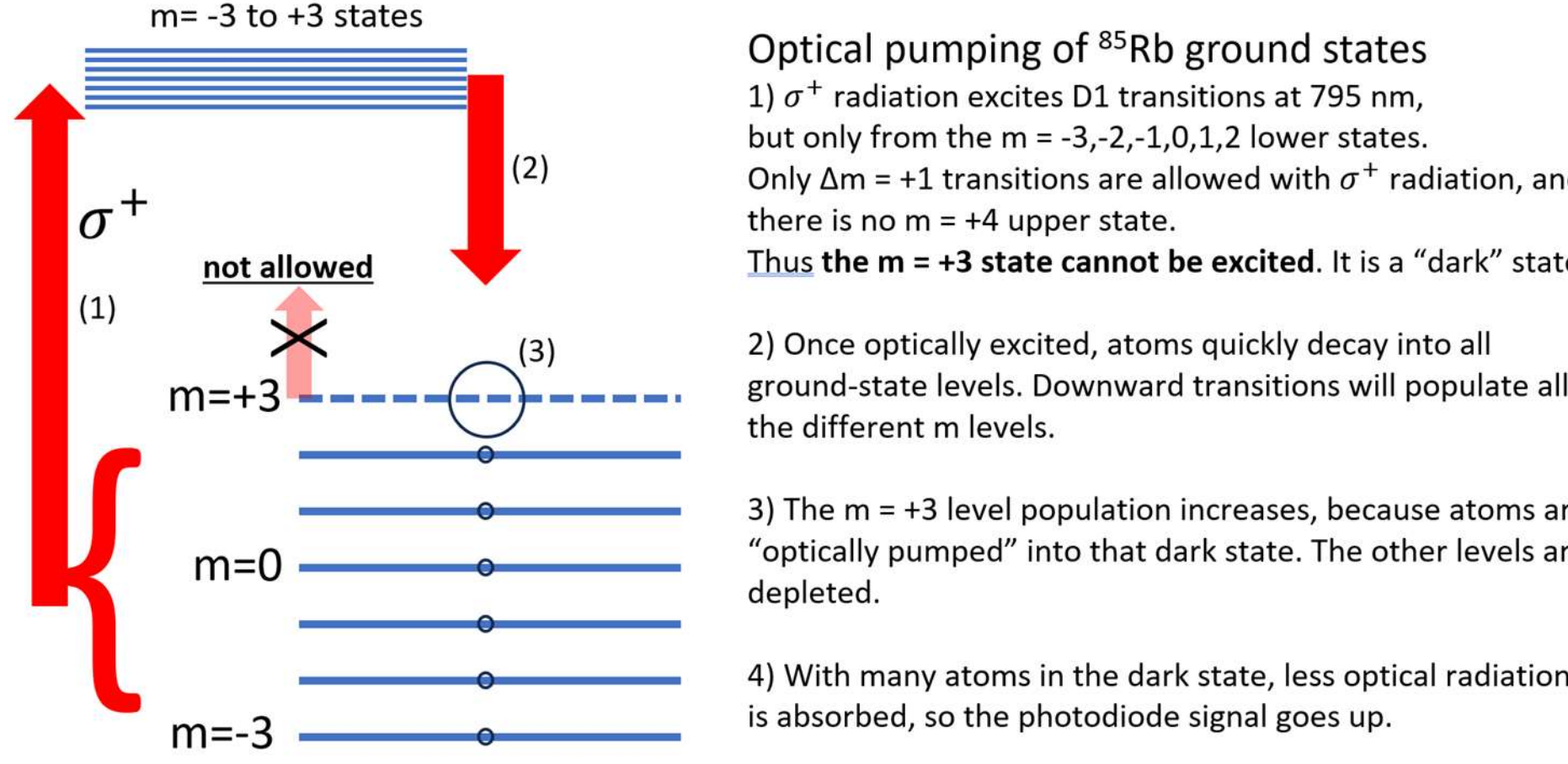

*Figure 11. The essential optical-pumping phenomenon as applied to the Rb-85 ground states when there is no mixing of the different Zeeman levels. With many atoms in m = +3 ground state (which cannot absorb $\sigma^+$ photons), the photodiode signal is high.*

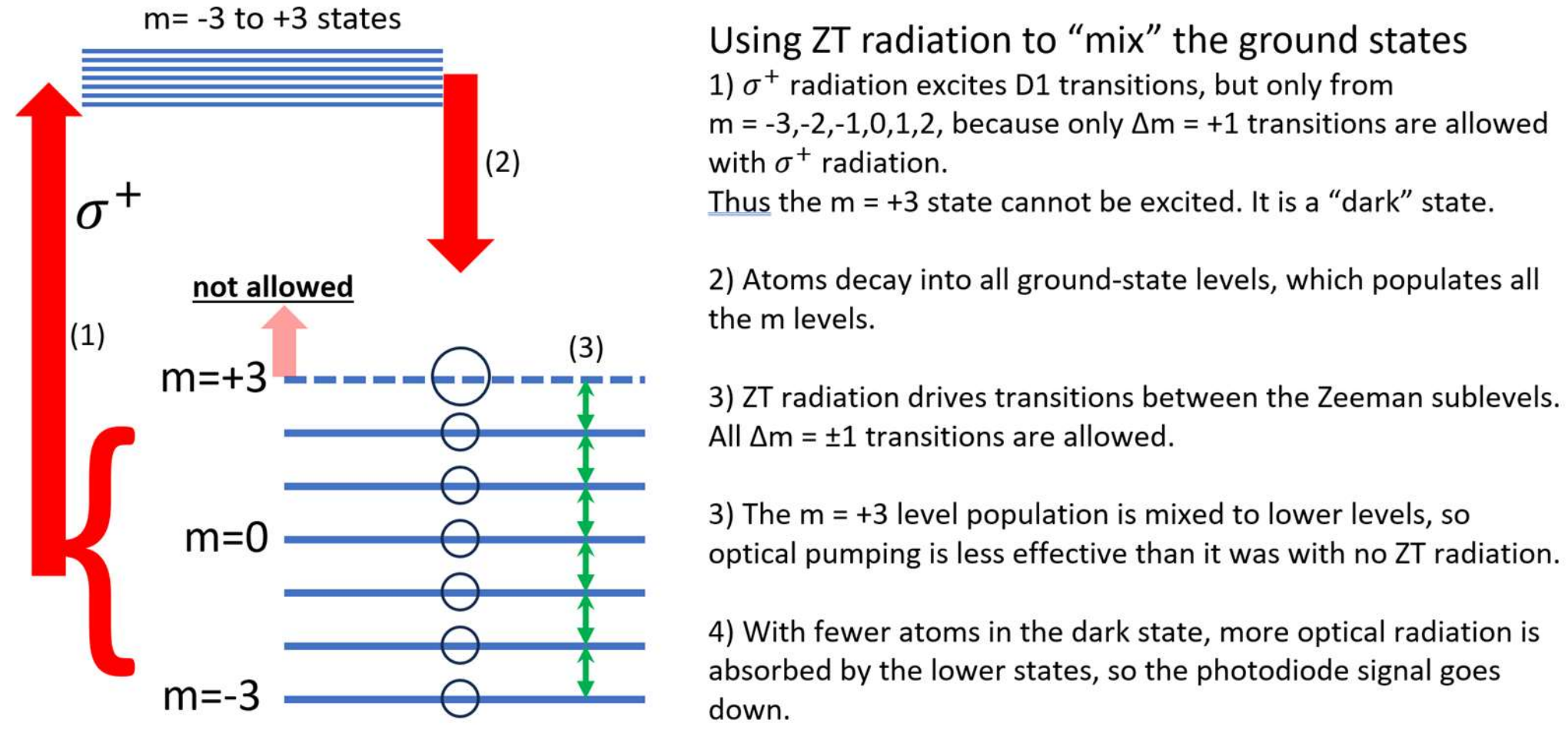

*Figure 12. The optical-pumping phenomenon as applied to the Rb-85 ground states when RF radiation is used to mix the different Zeeman levels. With fewer atoms in the m = +3 state, the photodiode signal is low. ZT not RF*

## The ZFR dip

As seen in Figure 10, a strong dip in the OP signal appears when the horizontal field passes through $B_z = 0$. This is called the "zero-field resonance" (ZFR) or the "ground-state Hanle effect" [2020Jam], and sometimes just the "B=0" dip. In theoretically perfect world, having $B_z = 0$ would not change the optical-pumping process, and no ZFR dip would be present. In the real world, however, even small stray magnetic fields will produce the dip, making it an ever-present feature in the teaching lab.

Referring to Figure 11, optical pumping sends Rb atoms into the $m_F = +3$ state, and this maximum $m_F$ value means that the electron spins are aligned along the optical axis. From the perspective of classical physics, having a large $B_z$ means that these electrons precess around $B_z$, but this does nothing because the spins are aligned along $B_z$. From the quantum perspective, the $m_F =$

+3 state is an eigenstate of the Hamiltonian, making it stable in time. In both these physical pictures, optical pumping efficiently sends Rb atoms into the dark state, aligned along the nonzero $B_z$, and there they stay, yielding a high OP signal.

But consider what happens when $B_z = 0$ and we add a small $B_y \neq 0$ field. In the classical picture, the tiny electron gyroscopes now precess about $B_y$, and this happens quickly, with precession frequencies of order 1MHz/Gauss. So, as soon as optical pumping aligns the atomic magnetic moments along the $z$ direction, precession from stray transverse magnetic fields mixes the level populations almost immediately. This fast precession means that the dark-state (transparent) population is less than it is at nonzero $B_z$, reducing the amount of transmitted light. Thus, we see a dip when the axial magnetic field passes through $B_z = 0$.

In principle, one could eliminate the $B_x$ and $B_y$ fields completely and thereby eliminate the ZFR dip. But one cannot achieve $B_x = B_y = 0$ to high accuracy in a normal laboratory environment, so the ZFR dip is always present at some level in teaching labs. In [2020Jam], the authors found that the ZFR dip was well described by a Lorentzian functional form with

$$\Gamma_{FWHM} = CB_{\perp} \tag{8}$$

where $\Gamma_{FWHM}$ is the full-width-half-maximum of the Lorentzian fit, $B_{\perp}$ is the magnitude of the transverse magnetic field, and $C = 2.072 \pm 0.003$ is a dimensionless constant (because both $\Gamma_{FWHM}$ and $B_{\perp}$ are measured in magnetic units). Confirming Equation 9 is a fine exercise for students, although note that the applied vertical field $B_y$ provided by the Vert Field coil is only one part of $B_{\perp}$. To obtain a simple linear fit as shown in [2020Jam], the two constant components of $B_{\perp}$ that do not change with the Vert Field coil current must be incorporated into the total $B_{\perp}$ [2020Jam].

To produce Figure 10, we carefully adjusted the $B_x$ and $B_y$ fields in an iterative process to find the global minimum for the width of the ZFR dip. As described below, a slow sweep is also needed to produce narrow and symmetrical dips in the OP signal. We find that the most efficient procedure for minimizing the ZFR FWHM (thus minimizing both $B_x$ and $B_y$) is to first zoom in on the dip by reducing the sweep range, as this speeds up the process. Once the desired narrow ZFR dip is obtained, one can then reproduce Figure 10 using a 10-second sweep, perhaps with some trace averaging.

In our experience, including a high-quality OP signal like that in Figure 10 in the lab handout is a beneficial teaching aid. Most students are then inspired to produce data of equal (or greater) quality, although some may be frustrated when this turns out to be somewhat challenging and time-consuming. Overall, however, we believe it is beneficial to at least show students that the OP instrument can, with care, produce some high-quality data.

One can also observe that the depth of the ZFR dip remains roughly independent of $B_y$, even as the FWHM changes substantially. Why is this? Because once you have scrambled the Zeeman levels thoroughly, scrambling them more makes little difference. Note also that the ZFR applies to both Rb isotopes, while the ZT dips are isotope specific. Thus the ZFR dip is typically deeper than either of the ZT dips, as seen in Figure 10.

A more subtle feature in the OP signal is the presence of 60-Hz noise near the ZFR dip. This is not immediately obvious on the oscilloscope, but telltale 60-Hz wiggles can often be seen on the sides of the dip if you look carefully. This noise signal happens because the OP signal is especially sensitive to small magnetic fluctuations near the ZFR dip (as one would expect), and most electronics plugged into the wall produce some level of 60-Hz magnetic noise.

Once the basic OP signal is on the oscilloscope, there are several relatively simple exercises that will help students understand the physical concepts responsible for this signal. One is simply to change the ZT drive frequency and verify that the OP signal changes as expected. Another is to convert the sweep-signal x-axis to magnetic field strength, as we have done in Figure 10. Because $g_F = 1/2$ for $^{87}$Rb and $g_F = 1/3$ for $^{85}$Rb, the positions of the ZT dips relative to B=0 should present a ratio of 1.5, and this ratio can be confirmed to high accuracy in the digital data. Time spent on such exercises is usually time well spent, because acquiring a firm physical understanding of the basic OP signal is essential before moving on to other OP investigations.

## OP Timescales

Setting up one's first OP signal is best done using a fast B-field sweep, so the setup time is not too tedious. A sweep time of ~1 second is reasonable, but the ZT and ZFR dips are clearly quite asymmetrical with such a fast sweep, as illustrated in Figure 9. Slower sweeps yield narrow, more symmetrical dips, and Figure 13 quantifies this phenomenon.

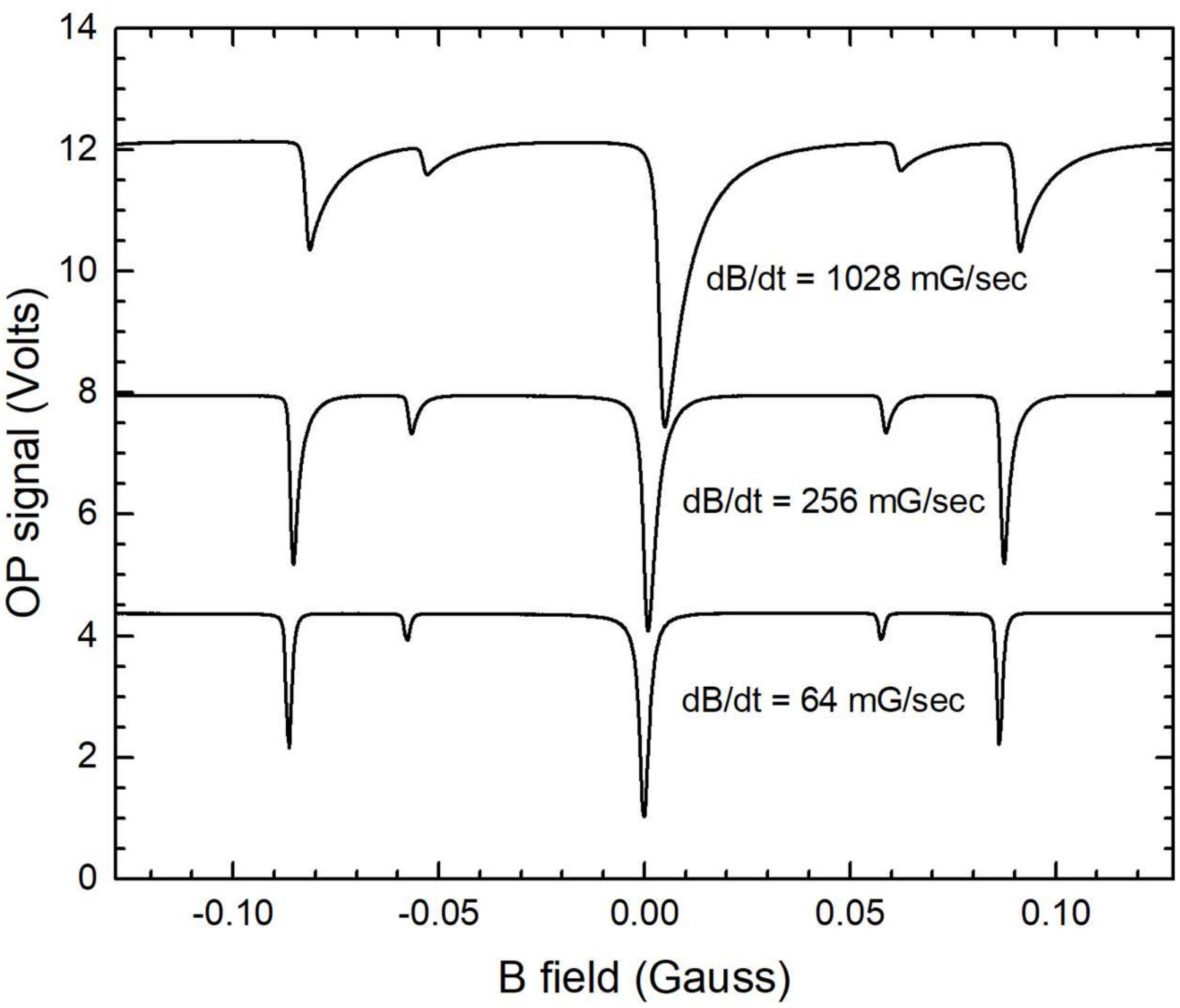

*Figure 13. These plots show how the OP signal changes with the sweep rate of the horizontal field, here using a ZT drive frequency of 40 kHz. With a fast sweep rate (top curve), the dips are broad and asymmetrical, reflecting the fact that it takes time to populate and depopulate the OP dark state. Depopulation is relatively fast, so the leading edge of each dip is abrupt. But repopulating the dark state is slower, yielding a more gradual rise of the dip signal to the steady-state OP level. As dB/dt is decreased (with no other changes), the dips become narrower and more symmetrical. Note that the different signal traces were offset vertically for clarity.*

We can understand these signals by noting that it takes time to populate and depopulate the dark state that is responsible for the optical-pumping signal. With a sufficiently slow scan, the OP signal is always in its steady-state value, reflecting a balance between optical pumping (that sends atoms into the dark state) and level redistribution effects (that take atoms out of the dark state). With a fast scan, however, the process becomes quite dynamical, producing a non-equilibrium OP signal that can be quite different from the slow-scan signal. Thus the fast-scan dips in Figure 13 are both shifted and distorted compared to the near-equilibrium slow-scan dips.

Modeling this entire process from start to finish would be a nontrivial task, involving the time-dependent interplay of many physical effects that affect the OP signal. A simpler task is to simply fit an exponential curve to the "recovery" side of the ZFR dip, as illustrated in Figure 14, revealing that it takes about 8 msec to repopulate the dark states once the scrambling has ceased. An increase in this repopulation time can be observed by using an optical attenuation filter to reduce the intensity of the incident light from the Rb lamp, and this is another possible exercise for students. In much what follows, our main focus is on the steady-state OP signal, where we typically want sharp spectral features. In those circumstances, we use relatively slow sweep times.

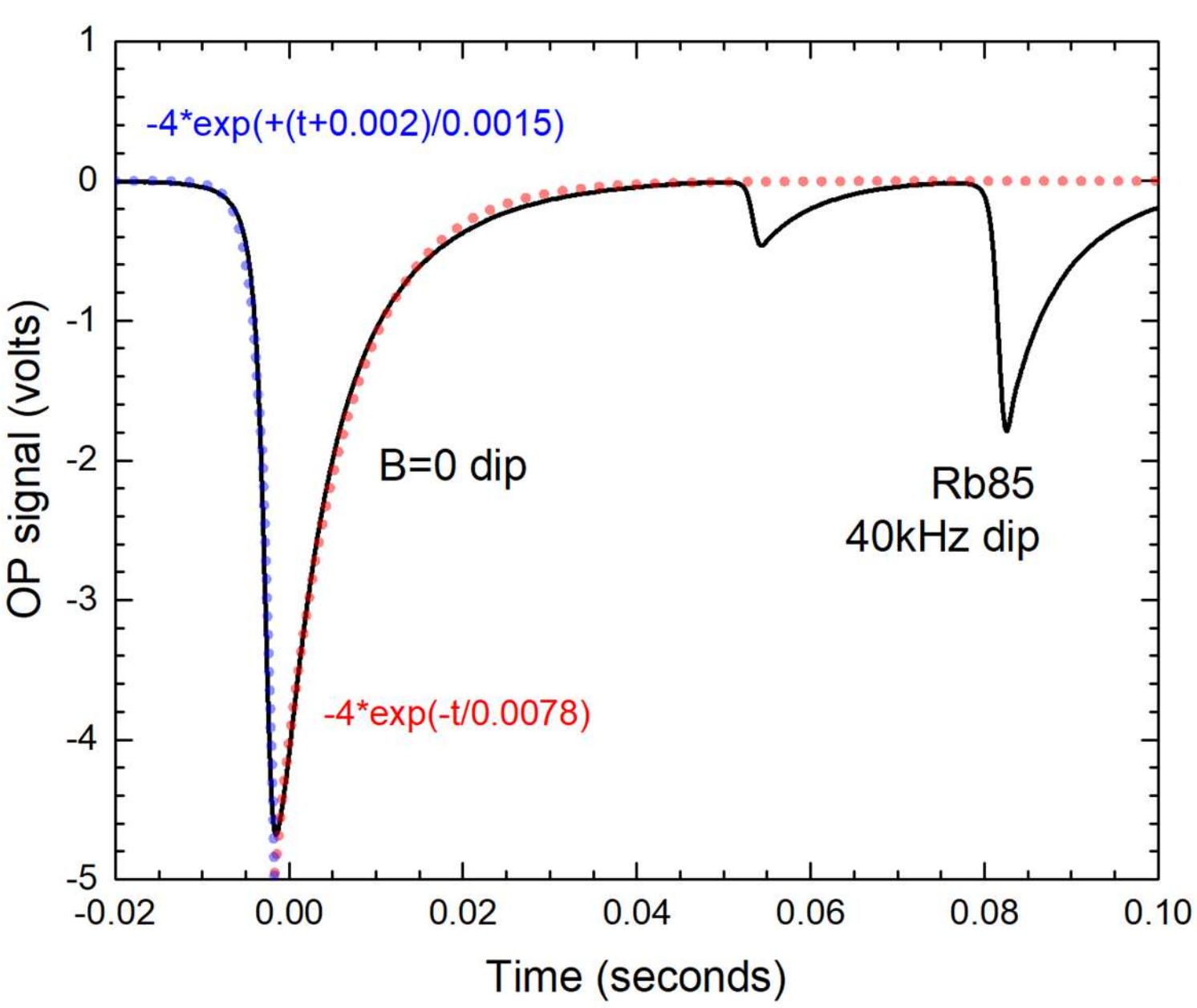

*Figure 14. This plot shows close-up of the top trace in Figure 13 as a function of time, along with fits to the leading and following edges of the ZFR dip. Here we see that depopulating the dark state took about 1.5 msec during the sweep, while repopulating the dark state took about 7.8 msec.*

## Dependence on ZT drive amplitude

Another straightforward measurement one can make using the basic OP signal is the depth of the ZT dips as a function of the ZT drive amplitude. Quantum theory informs us that the scrambling of the Zeeman levels is driven largely by *magnetic dipole transitions*, and that the rate of these transitions depends linearly on the power (a.k.a. intensity) of the incident electromagnetic radiation. Thus, as low intensities, we expect that the dip depth would be proportional to the square of the ZT drive amplitude, and this is behavior is indeed seen in Figure 15.

Figure 15 also shows us that the ZT dip depths saturate to near-constant values at high ZT drive amplitudes. This saturation occurs once the timescale for scrambling the Zeeman levels becomes much shorter than the optical pumping time that repopulates the dark states. Once the Zeeman levels are fully scrambled, scrambling them faster has little additional effect. We chose a functional form $D = D_{max}\{1 - \exp[(-A/A_0)^2]\}$ to fit the data in Figure 15, where $D$ is the dip depth and $A$ is the ZT drive amplitude. This is the simplest functional form that produces the desired asymptotic behaviors at low and high amplitudes using two adjustable parameters. We would not necessarily expect this function to fit the data precisely at intermediate drive amplitudes, and it doesn't. Creating a better model would be difficult, requiring attention to details like how the oscillating magnetic field strength varies with position inside the Rb cell and other uninteresting complications.

As a cautionary note, we have found that the stock *TeachSpin* OP apparatus can produce ZT drives signals that exhibit harmonic distortion at very high amplitudes. This distortion of the ZT drive signal can produce spurious dips at higher harmonic frequencies, and students might erroneously interpret these as multi-photon transitions. The signal generator in Figure 8 will not let you produce a distorted signal, so that is one way to avoid this potential confusion.

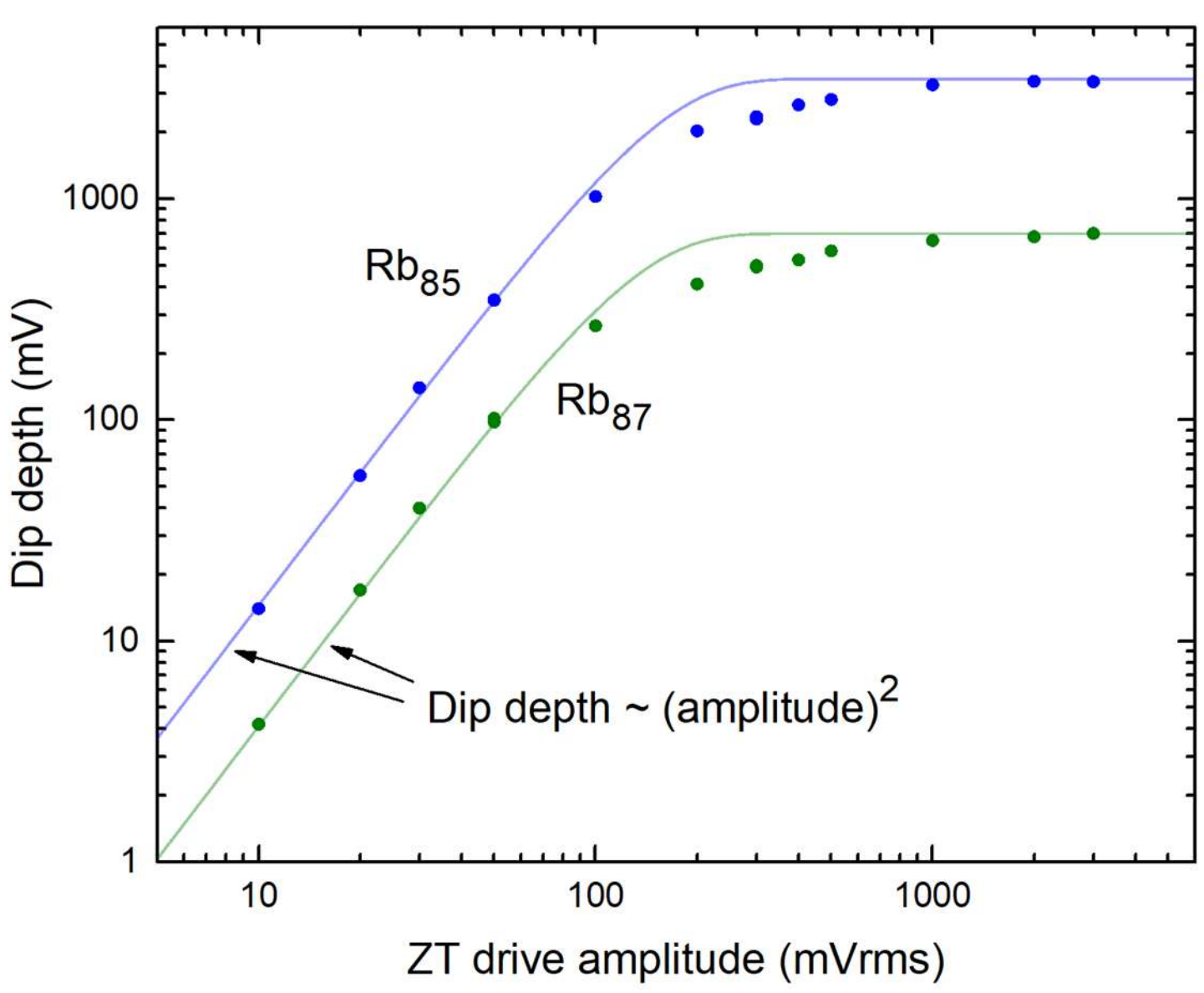

*Figure 15. The plot shows the measured depth of the Rb-85 and Rb-87 ZT dips as a function of the ZT drive amplitude. Theory tells us that the magnetic dipole transition probability increases with the square of the B-field amplitude, and this behavior is seen at low ZT drive. The dip depths saturate once the Zeeman sublevels are thoroughly mixed. The lines in this plot show simple exponential roll-offs, which fit the data reasonably well at low and high drive amplitudes. One would not expect this model to fit the data at intermediate amplitudes, where the underlying physics is more complex.*

## Dependence on Rb cell temperature

Another investigation one can perform with the basic OP signal is to measure how the various signal features vary with the temperature of the Rb vapor cell. Figure 16 some example results, extracted from a series of OP sweep data similar to that in Figure 10.

Beginning with the top plot in Figure 16, the decrease in the OP signal with increasing temperature generally reflects the increasing optical depth of the Rb cell as the vapor density increases. Modeling the measured data is a nontrivial exercise, however, as it results from both the emmision spectrum of the Rb lamp and the absorption spectrum in the Rb cell. Neither of these model components is well known, but we describe an approximate model in Appendix 2.

The fraction al dip depths as a function of temperature in Figure 16 are also difficult to model in detail, but we can understand the overall trends. At low temperatures, for example, the dips are

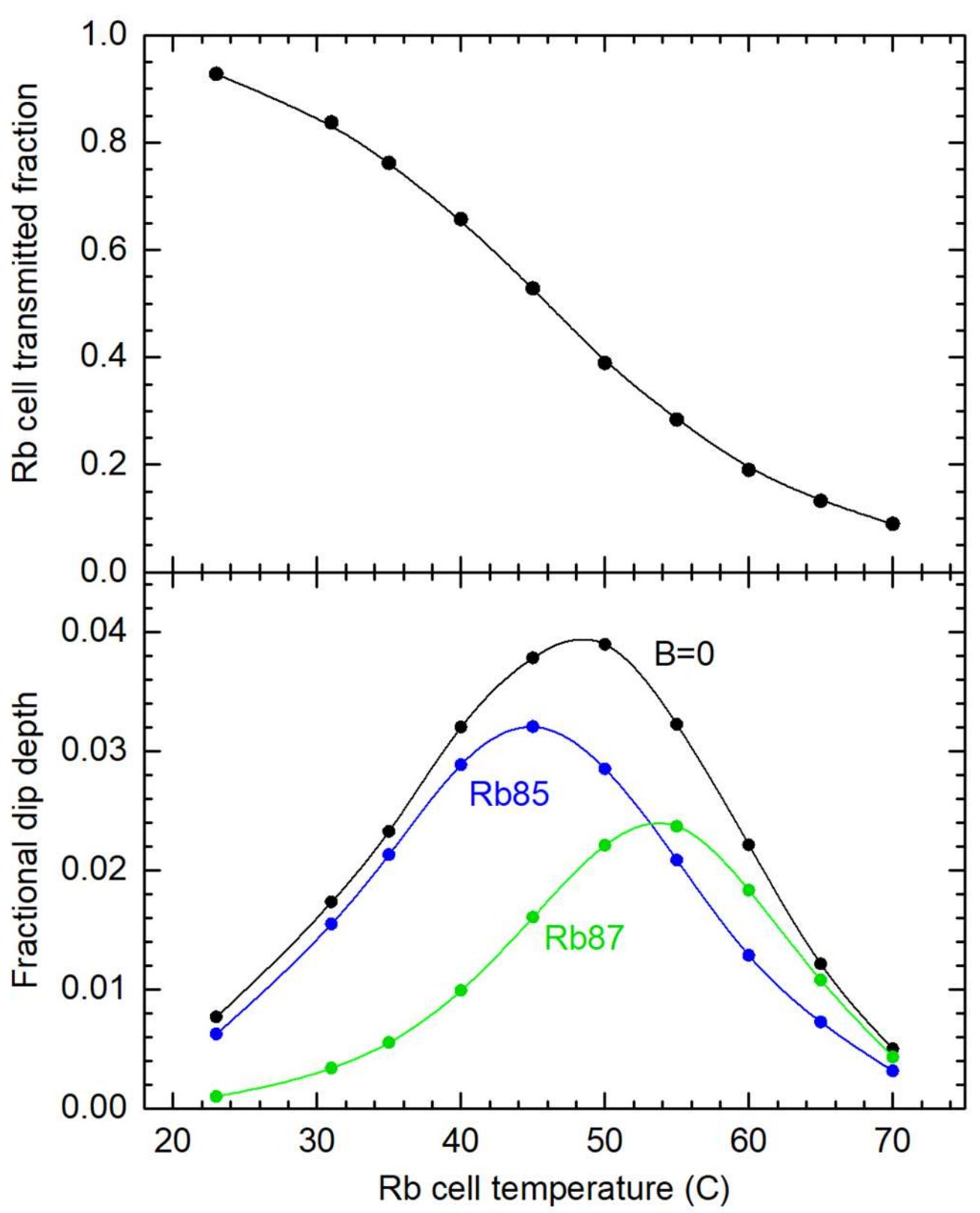

*Figure 16. The top plot on the left shows the basic OP signal away from any dips, normalized by the cell transmission with no Rb vapor. Appendix 2 describes a model of these data, which was used to extrapolate the data to the zero-vapor limit at lower temperatures. The background signal (measured by blocking the Rb lamp) produced a negligible offset in these data.*

*The lower plot shows the depth of the different OP dips (the ZT dips being near saturation, with a 40 kHz drive frequency) expressed as a fraction of the dip-free OP signal. At low cell temperatures, the dips are weak because the Rb vapor density is low. At high cell temperatures, the fractional dip depth is reduced by the transmission of off-resonant light through the cell (see Appendix 2).*

*In all plots, spline interpolation lines were added for clarity.*

generally shallow because the Rb density in the cell is low, yielding lower absorption dips in the OP signal. At high temperatures, the photons that make it through the Rb cell are mainly far off resonance (see Appendix 2). If we replaced the Rb lamp with a monochromatic, far-off-resonant laser, the laser absorption would be low in the Rb cell and the OP dips would also be relatively shallow. The model in Appendix 2 thus explains the overall trends in the data at a qualitative level, but producing an accurate of the falling dip depths at high temperatures would be challenging.

Mostly these emprical data are useful for planning OP experiments in the teaching lab. In particular, these data show the maximum OP dip signals can be obtained around 40-50 C.

## Optimizing the ZT linewidths

Throughout our various OP investigations, we found that the sharpness of the ZT dips (in the limit of slow scan times) is not especially sensitive to background magnetic fields that are perpendicular to $B_z$. This is true for both average field strengths and field gradients. However, we have observed that axial $dB_z/dz$ gradients can produce noticeable detrimental effects.

Using an external coil with its axis along the optical axis, we generate calculable $dB_z/dz$ gradients and obtained the data shown in Figure 17. In our particular lab environment, we found ZT dip widths of about 1.5 kHz with no external coil, indicating an environmental background of $dB_z/dz \approx$ 10 μGauss/m. By correcting this gradient with the external coil, we obtained linewidths as low as 0.7 kHz. From this we concluded that our lab environment contains magnetic field gradients that noticeably degrade some of the weaker OP signals. Fortunately, it is fairly easy to use an external coil to cancel the laboratory gradients to some degree, thus yielding noticeably sharper ZT features.

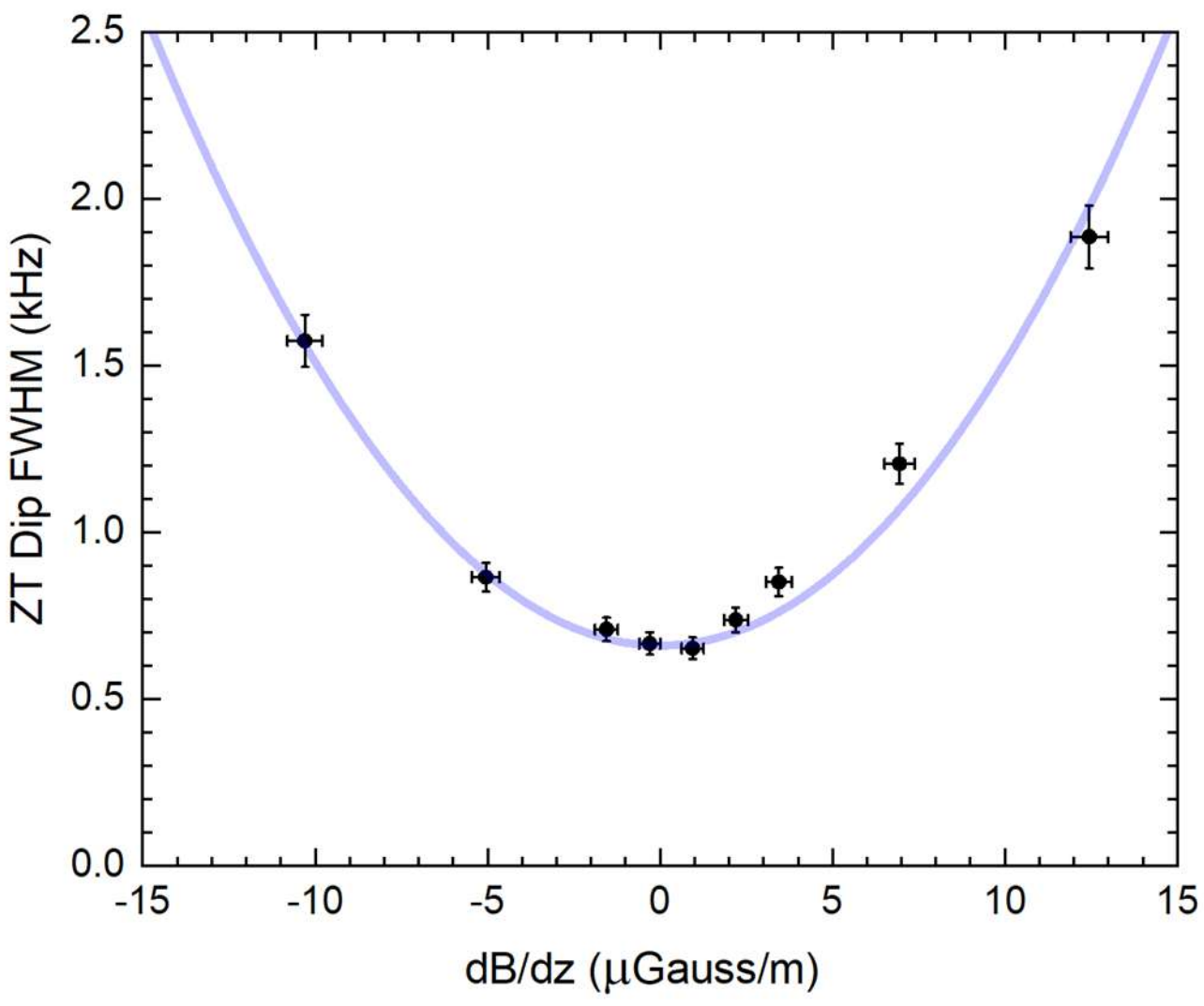

*Figure 17. The sharpness of the ZT dips can be degraded by a nonuniform magnetic field in the lab environment. This plot shows the measured FWHM of a Rb85 ZT dip when a B-field gradient along the optical axis was applied by an external coil. By setting dB/dz = 10 μGauss/m, we were able to partially correct the nonuniform B field at this location in the lab, thereby reducing the FWHM from 1.6 kHz (without correction) to less than 0.7 kHz. This trick was helpful in our observations of the nonlinear Zeeman splitting, where the ZT spectral features are somewhat weak.*

To understand these observations, note that there are several physical mechanisms that can affect the ZT dip linewidths, even with slow scan times. One is the residence time of Rb atoms in the optical beam, which can be as low at $\Delta t \approx 50$ μsec for freely traveling Rb atoms, yielding linewidths of $\Gamma \approx 1/\Delta t \approx 20$ kHz. Fortunately, the Rb residence time is much increased by an inert buffer gas in the cell, and this likely reduces this linewidth contribution to less than 0.1 kHz.

Another source of linewidth is atomic collisions that happen while Rb atoms are in the optical beam. There are more collisions at higher buffer gas densities, but the use of inert neon gas reduces the effects of these collisions (because most neon atoms have zero nuclear spin). Because researchers have optimized the buffer-gas parameters over the years, collisions also likely produce linewidths below 0.1 kHz.

Finally we have magnetic field gradients. Assuming $dB_z/dz = 10$ μGauss/m and a Rb cell size of 5 cm, this gives field variations of ~500 nG inside the cell. Further assuming $\nu_{Zeeman}/B = 467$ MHz/Gauss for the $^{85}$Rb ground states, these B-field variations produce linewidths of about 0.25kHz. So the measurements in Figure 17 are in at least rough agreement with expectations.

If you are using this (or a similar) OP apparatus, we recommend that you measure the linewidths of the ZT features with the Rb cell at room temperature. If the linewidths are below 1 kHz, then your lab environment is plenty good enough and you needn't do anything. But if the linewidths are above 2 kHz, then you may want to try moving electronics away from the Rb cell or perhaps installing an external coil to cancel the background $dB_z/dz$ as best you can. In our lab, we have found that the $dB_z/dz$ background gradient is fairly constant in time, so the external coil is something of a set-and-forget correction. Obtaining narrow ZT linewidths is not so important when investigating the basic OP signal, but it can improve measurements of the strong-field Zeeman splittings.

## Spin rotation

Observing spin rotation is a remarkably thought-provoking experiment that can be done using this optical-pumping apparatus. It involves the basic OP signal, except adding a time-dependent twist that relates to many other areas of atomic physics. We like to present students with our NMR experiment (this topic being another staple of Advanced Physics Labs [2025Lib]) and subsequently expose them to spin rotation in optical pumping. The underlying physical phenomena in these two experiments are strongly related, revealing a deep connection between Zeeman transitions and spin precession in magnetic fields. It is not immediately obvious that the precession frequency of a particle with a non-zero spin in a magnetic field should be the same as the frequency needed to drive Zeeman transitions in this same particle. But it is, and significant physical insight can be gained by understanding why this is so.

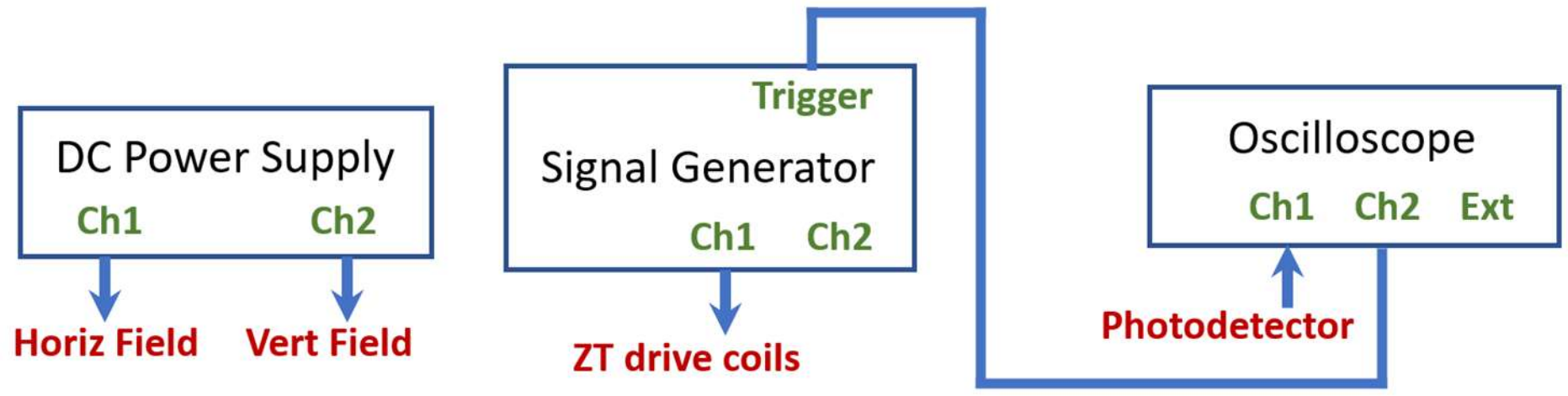

*Figure 18. A partial electronics layout for observing spin rotation of the optically-pumped magnetic dipole ensemble.*

Again, to motivate students by focusing their attention on real observations, we like to have them set up the experiment first, before describing the physics in detail. This way they have a live signal in front of them, which they can manipulate during the discussion. To this end, Figure 18 shows the electronics layout for observing spin precession, using the same optical layout shown in Figure 7.

For the experiments presented in this section, we set the ZT drive frequency to some convenient value (typically a 40 kHz sine wave, but the exact frequency value is not important), and we set the horizontal B field (using the DC Power Supply) so Zeeman transitions are excited in $^{85}$Rb. As described above, this current value should be recorded when one is observing the basic OP signal, as described above.

With this constant (not sweeping) $B_z$, we then use the Burst Mode in the Signal Generator to turn the ZT drive on and off at regular intervals, yielding the OP signal shown in Figure 19. Once the signal is seen on the 'scope, the value of $B_z$ can be tweaked to maximize the signal, guaranteeing that $B_z$ is set to maximally excite the desired Zeeman transition.

To understand the physics producing this signal, let us begin in the ZT-off state, so the OP level is high because optical pumping has over-populated the $m = +3$ dark state, so there is less absorption in the Rb cell and more light strikes the photodetector… the standard optical pumping phenomenon described above. At this point, the dark-state population is made up of $^{85}$Rb atoms with their spins aligned along $B_z$.

When the ZT signal is abruptly turned on, the oscillating magnetic field produces a spin-precession that slowly rotates the $^{85}$Rb spins from the $+B_z$ to the $-B_z$ direction. How this happens is described at great length in relation to NMR. We presented the theory in [2025Lib], but there are countless versions elsewhere in the literature, and you can find YouTube videos showing detailed animations of the spin-rotation process. The phenomenon is not obvious, but to save space we will not describe it here.

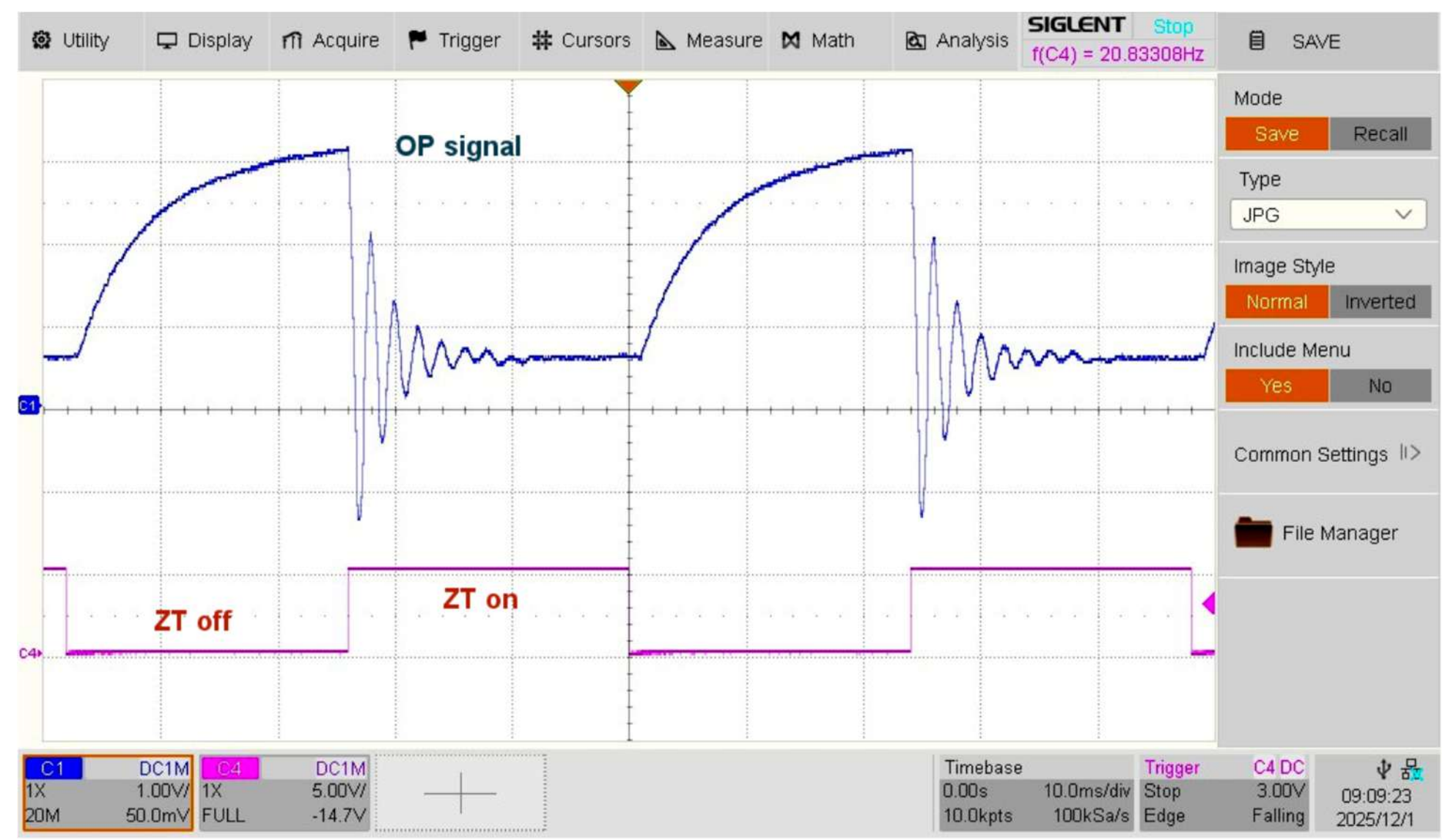

*Figure 19. A spin-rotation signal on the oscilloscope using the setup in Figure 18. Here the signal generator was in Burst mode, producing a 40 kHz sine wave for 25 msec (ZT on) followed by 25 msec with the ZT signal off. When the ZT signal is turned on, the magnetic moment of the optically pumped sample oscillates between aligned with the B-field (high OP signal) and anti-aligned with the B-field (low OP signal). This phenomenon is essentially the same as spin rotation in NMR, yielding an excellent example of spin state manipulation. After the oscillations die down, the OP signal settles at the ZT dip, yielding a low OP signal. When the ZT drive is turned off, optical pumping restores the OP signal to its non-ZT-dip (high) level after about 8 msec.*

Once you understand this fundamental NMR process, you can see that the ZT signal will continuously rotate the spin population from $+B_z$ to $-B_z$ and back again. Because there is a large sample of aligned spins in the Rb cell, this spin-rotation phenomenon is very well described by classical physics, specifically the Bloch equations. In this classical-physics picture, the spin ensemble behaves like a macroscopic gyroscope, which is easily visualized. In the corresponding quantum-physics picture, the gyroscope is replaced by a complex set of correlated Zeeman sublevels; much harder to visualize but giving the same result.

As the spin ensemble oscillates between $+B_z$ and $-B_z$, this changes the amount of absorption taking place in the Rb cell. When the spins are aligned along $+B_z$ (in the $m = +3$ dark state), the absorption is anomalously low. When the spins are aligned along $-B_z$ (in the $m = -3$ "anti-dark" state), the absorption is anomalously high.

This spin rotation describes the essential behavior of the OP signal in Figure 19. When the drive signal is initially in the ZT-off state, the spin ensemble is in the $m = +3$ dark state and the OP signal is high. Transitioning to the ZT-on state, the spins soon rotate to the $m = -3$ "anti-dark" state, and then the OP signal is low; not just low, but *lower* than the steady-state ZT-on signal. The OP signal continues to oscillate until the spins dephase and one is back at the OP level equal to the steady-state ZT dip.

In a pulsed-NMR experiment, it is not possible to observe the spin precession when the drive coil is energized. The drive coil swamps the detection coil with spurious signal. Because of this interference, the usual practice is to drive a $\pi/2$ pulse and then observe the spin precession after turning the drive coil off. In the OP apparatus, we observe the spin ensemble optically, so we can watch the spin rotation even when the ZT drive is continuously on. It's all the same physics, just in a different realm. Seeing both these phenomena, one after the other, thus helps build an overall deeper understanding of the general physical concepts responsible for both. Pedagogically powerful stuff. With some care (having a 12-bit 'scope that can record 10,000 data points helps), it is possible to obtain some beautiful spin-rotation signals, like that shown in Figure 20.

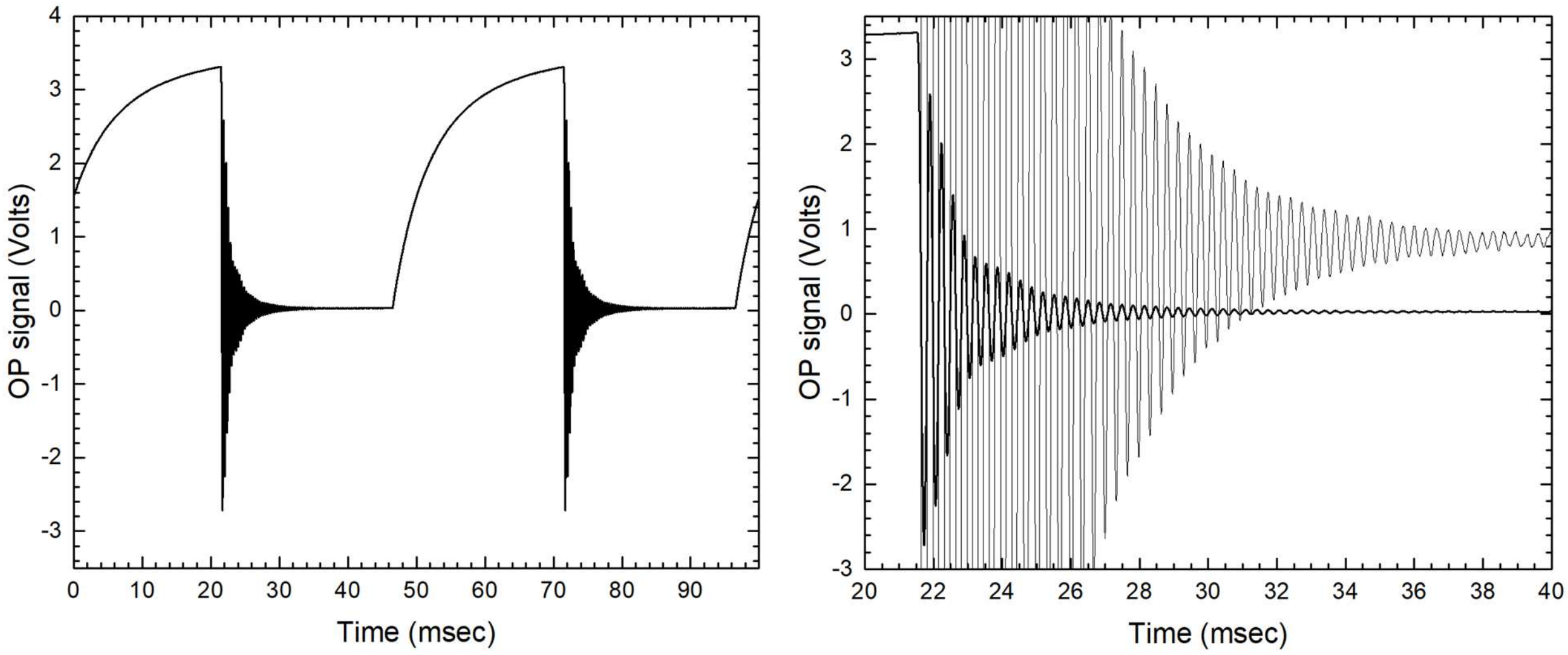

*Figure 20. With careful tuning of the system parameters, remarkably clean spin-rotation signals can be observed with this OP apparatus. The above used a ZT frequency of 40 kHz, a drive amplitude of 6 Vrms, and 500 'scope traces were averaged (taking about one minute) to produce this signal. The plot on the right is a close-up of the oscillations, with the lighter line being equal to 30 times the darker line.*

## NMR analogies

To better drive home the connection between NMR and spin rotation in OP, Figure 21 illustrates what happens when the ZT-on signal (see Figure 19) consists of just seven cycles of the 40-kHz drive. In this example, seven cycles is the minimum number needed to drive the Rb atoms from the $m = +3$ dark state the $m = -3$ "anti-dark" state, and this is analogous to a $\pi$ pulse in NMR. After being placed in the anti-dark state, optical pumping slowly returns the Rb ensemble to the steady-state dark state. As we described above, reducing the Rb lamp intensity (using an optical attenuator in the optical path) increases the optical-pumping recovery time in Figure 21, which we verified in our lab.

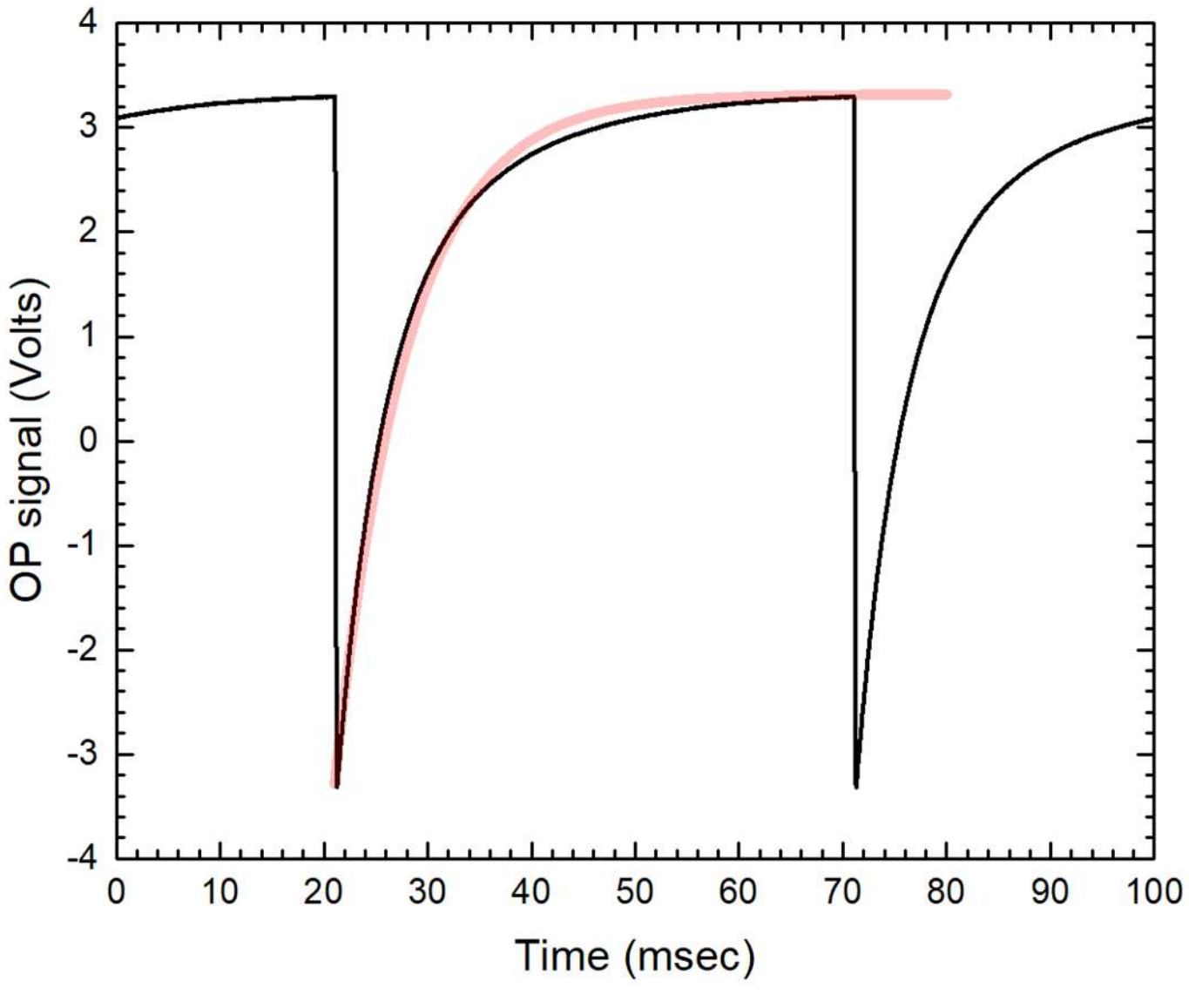

*Figure 21. This OP signal is the same as that in Figure 20, except this time we used a Burst signal with just seven cycles of the 40 kHz ZT drive signal (compared to 1000 cycles in Figure 20). The result is analogous to the application of a $\pi$ pulse in NMR, rotating the spin ensemble from the aligned state to the anti-aligned state. After the $\pi$ pulse, optical pumping slowly returned the sample to its aligned state. The red curve shows an exponential relaxation with a time constant of 7msec, similar to that found in Figure 14.*

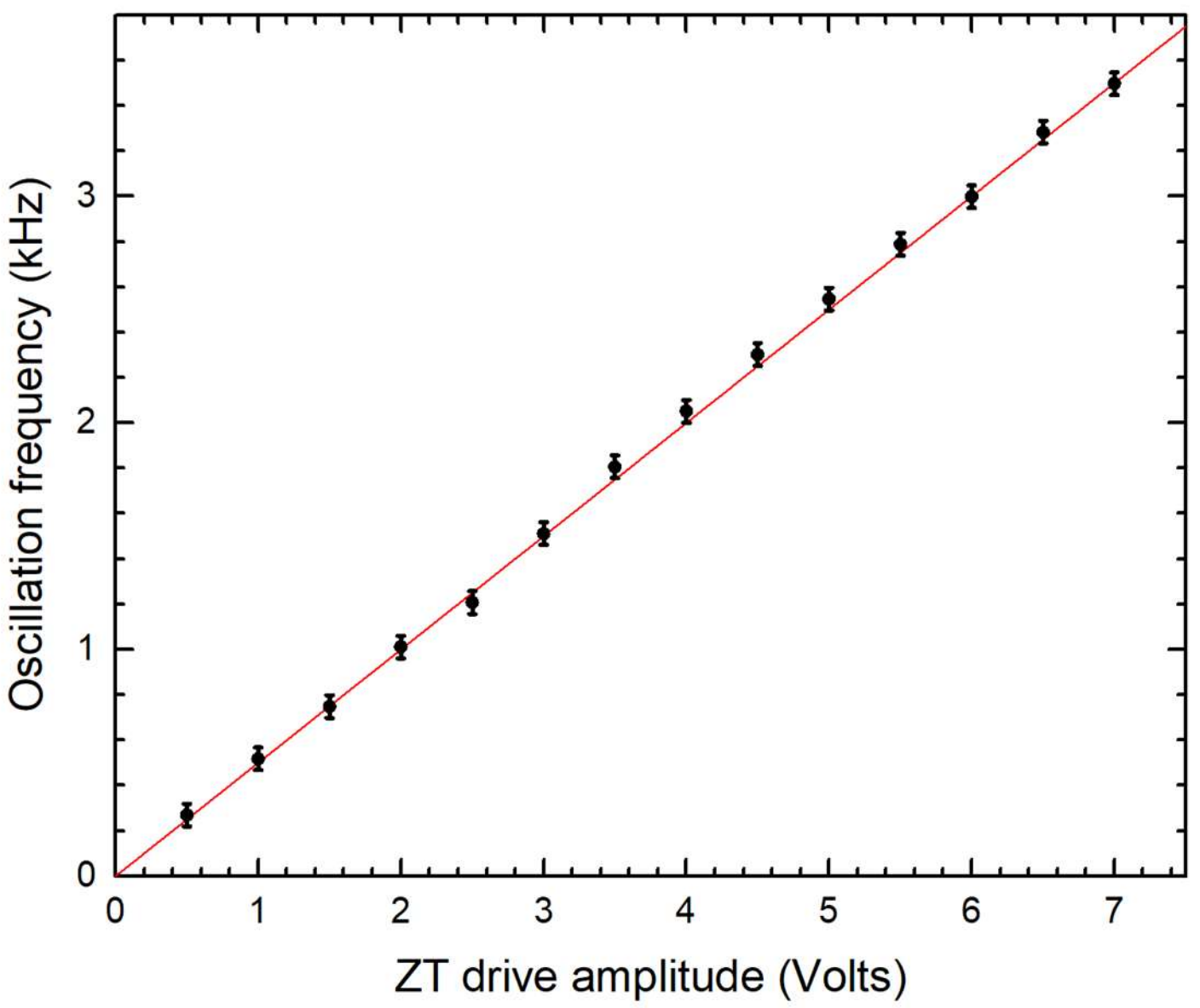

*Figure 22. This plot shows the measured frequency of the spin-rotation oscillations seen in the OP signal as a function of the ZT drive amplitude. NMR theory predicts that the frequency should be simply proportional to the drive amplitude, which nicely matches the observations. Note that the slope of this curve was fit to the data, with no zero-point offset.*

NMR theory tells us that the number of cycles needed to produce a $\pi$ pulse is inversely proportional to the amplitude of the oscillating magnetic field being applied to the sample. Applying this theory to the OP case, we expect that the frequency of the spin rotation signal should be proportional to the amplitude of the ZT drive. Confirming this expectation is a straightforward exercise for students, and Figure 22 demonstrates that some exceptionally clean data can be obtained using this OP apparatus.

## Strong-field Zeeman splitting

Optical pumping can be used to observe strong-field Zeeman splittings with remarkable clarity, allowing a significant test of Breit-Rabi theory and a good connection to atomic spectroscopy more generally. We focus on $^{85}$Rb in what follows, mainly because its ground state has more observable Zeeman transitions than $^{87}$Rb (10 compared to 6), giving $^{85}$Rb a more visually appealing Zeeman spectrum. To begin, Figure 23 shows the relevant ground-state Zeeman levels for $^{85}$Rb, and we have labeled the $\Delta m = \pm 1$ transitions (calling them ZT1 through ZT10) that we can observe using the OP apparatus. We do not include any 3 GHz transitions in this list, because we will be working only with frequencies below 10 MHz.

The OP experimental setup is the same as that shown in Figure 8, except this time we keep $B_z$ constant (no sweep), and instead sweep the ZT drive frequency using the Signal Generator's sweep mode. By creating a sine-wave signal swept from 3 MHz to 5 MHz, together with a $B_z$ appropriate for a 4-MHz ZT dip (about 8.6 Gauss, requiring a coil current of about 1A), one can readily observe an obvious (although perhaps small) signal on the oscilloscope. Once a small signal is seen, it is then straightforward to zoom in for a closer look.

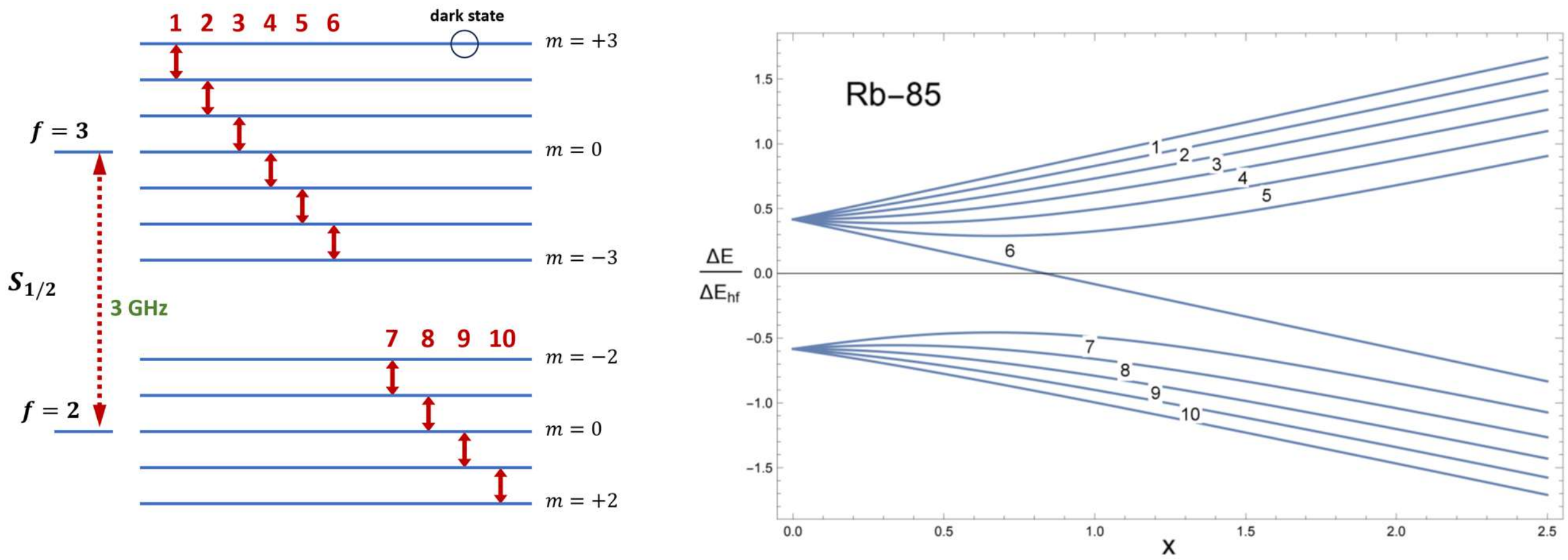

*Figure 23. These sketches label the allowed Δm=1 ground-state Zeeman transitions (ZTs) for Rb-85, ignoring all transitions between the f=2 and f=3 levels (because these occur at frequencies of ~3 GHz). The 10 transitions shown can be resolved in the OP spectra once the magnetic field exceeds a few Gauss.*

Figure 24 illustrates what a typical signal might look like using a fast sweep, after reducing the frequency sweep to 150 kHz. The physics underlying this structure is essentially the same as with ZT dip we saw above, except now the individual Zeeman transitions are resolved by the nonlinear splitting, while they were essentially degenerate (producing a single unresolved dip) at lower ZT frequencies.

To produce higher quality ZT spectra, it is necessary to use a longer sweep time, typically around 10 seconds for best results. Averaging traces also substantially improves the quality of the signal.

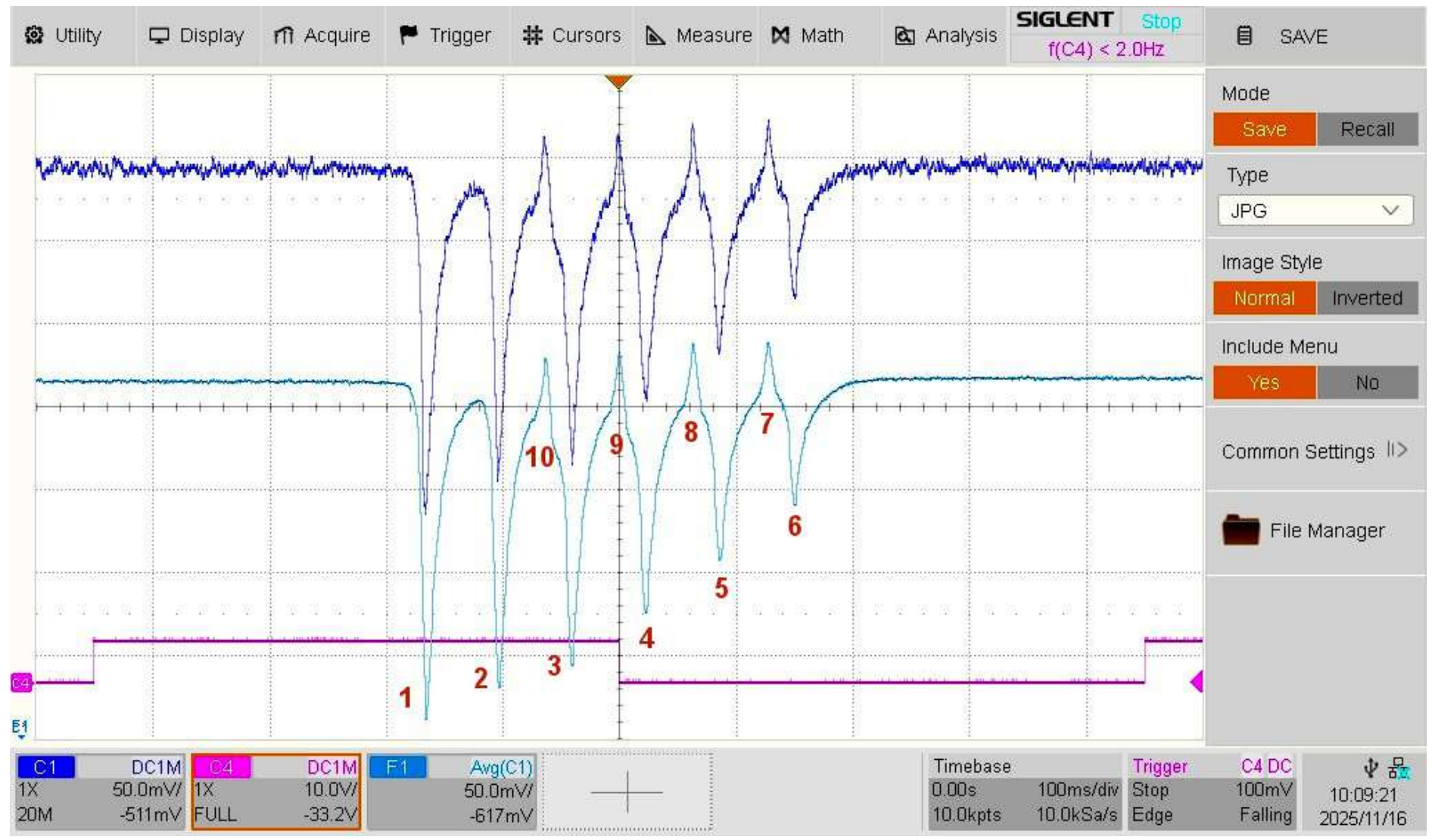

*Figure 24. This screenshot shows a typical first look at the Rb-85 strong-field Zeeman splitting, here scanning the ZT drive frequency from 3.925MHz to 4.075MHz, with a ZT drive amplitude of 300 mVrms a scan time of 0.9 seconds, with a Rb cell temperature of 50C. The top trace shows the raw photodiode signal, the middle trace averages 16 traces, and the bottom trace shows the sweep trigger signal. The Zeeman transitions are labeled following the nomenclature in Figure 23. We expect ZT1 to have the largest dip signal, as this transition couples directly to the m=+3 dark state. The ZT7-ZT10 states are rather poorly coupled to the dark state, and they appear as smaller dips or peaks depending on various experimental parameters.*

Unfortunately, taking these steps makes the data-acquisition process somewhat tedious, so there is a trade-off between fast operation and overall OP signal quality. To help navigate this trade-off in the teaching lab, we now examine how the $^{85}$Rb strong-field Zeeman splitting signal varies with different experimental parameters.

## Dependence on ZT drive amplitude

Figure 25 shows a series of ZT spectra taken using different ZT drive amplitudes. At the lowest amplitude shown, the strong ZT1-ZT6 dips are easily seen, while the much weaker ZT7-ZT10 dips are barely visible and somewhat distorted in shape. Increasing the ZT drive amplitude to 0.5Vrms, the normal ZT dips become deeper and 2-photon dips begin to appear in the spectra. Figure 26 illustrates the origin of the 2-photon dips, each involving a $\Delta m = \pm 2$ transition. We label the first of these as ZT(1/2), indicating that this dip is located midway between ZT1 and ZT2 dips.

Clearly these transitions violate the $\Delta m = \pm 1$ selection rule, which strictly only applies in the limit of weak drive signals. Multi-photon ZTs become possible at higher drive amplitudes, with

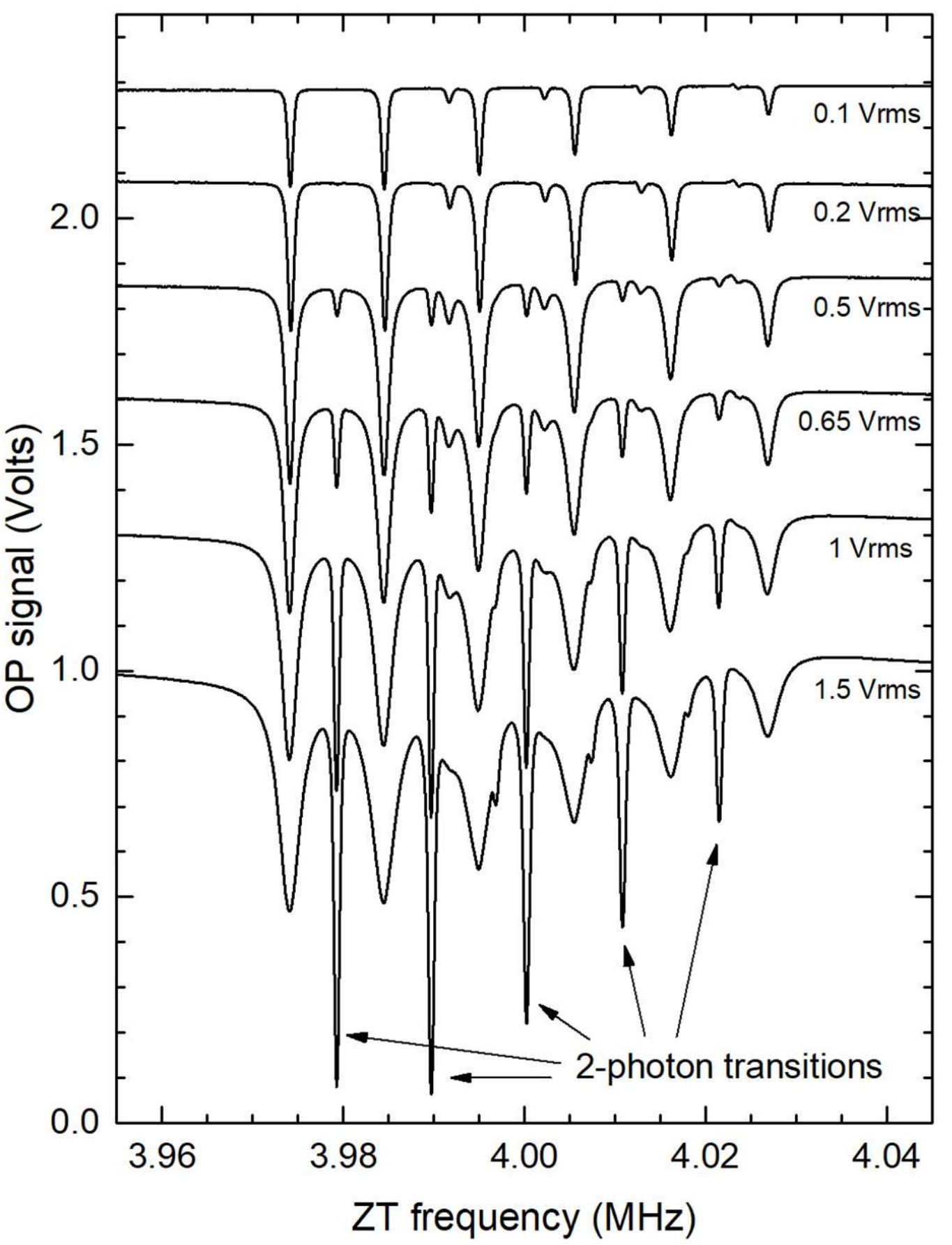

*Figure 25. These data show the Rb-85 Zeeman transitions near 4 MHz at several different ZT drive amplitudes, revealing the 10 normal ZTs along with 2-photon transitions at high drive amplitudes. Note that all eight allowed Δm=±2 transitions can be seen in the 1.5Vrms spectrum.*

*These data were taken with a Rb cell temperature of 50C, a 10-second frequency sweep over 100 kHz, and a 4-trace average. The different traces were offset in the graph for clarity.*

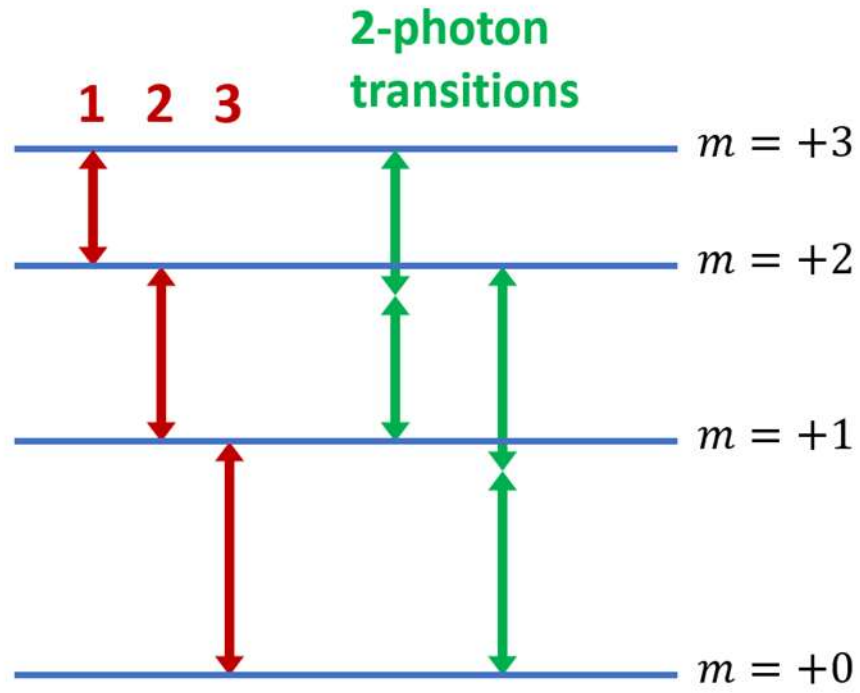

*Figure 26. The sketch on the right shows 1-photon (red) and 2-photon (green) transitions, illustrating why the latter occur at frequencies exactly midway between their corresponding normal 1-photon ZTs. Note that there is no ZT(6/7) transition in Figure 25, because this would require a 3-GHz drive frequency.*

$\Delta m = \pm 2$ being the first to appear. With the ZT drive amplitude at 1.5Vrms in Figure 25, the single-photon dips become quite broad, while the two-photon transitions are strong and sharp.

A full theoretical description of multi-photon transitions is beyond the scope of this paper, but theory does tell us that the transition probability for 1-photon transitions should be proportional to the radiation intensity (equal to the square of the drive amplitude), while the transition probability for 2-photon transitions scales with the square of the radiation intensity. Figure 27 shows that the OP data can be used to confirm these theoretical expectations. As we discussed above, the OP dip depth saturates at high ZT drive amplitudes, when the Zeeman level populations become fully mixed.

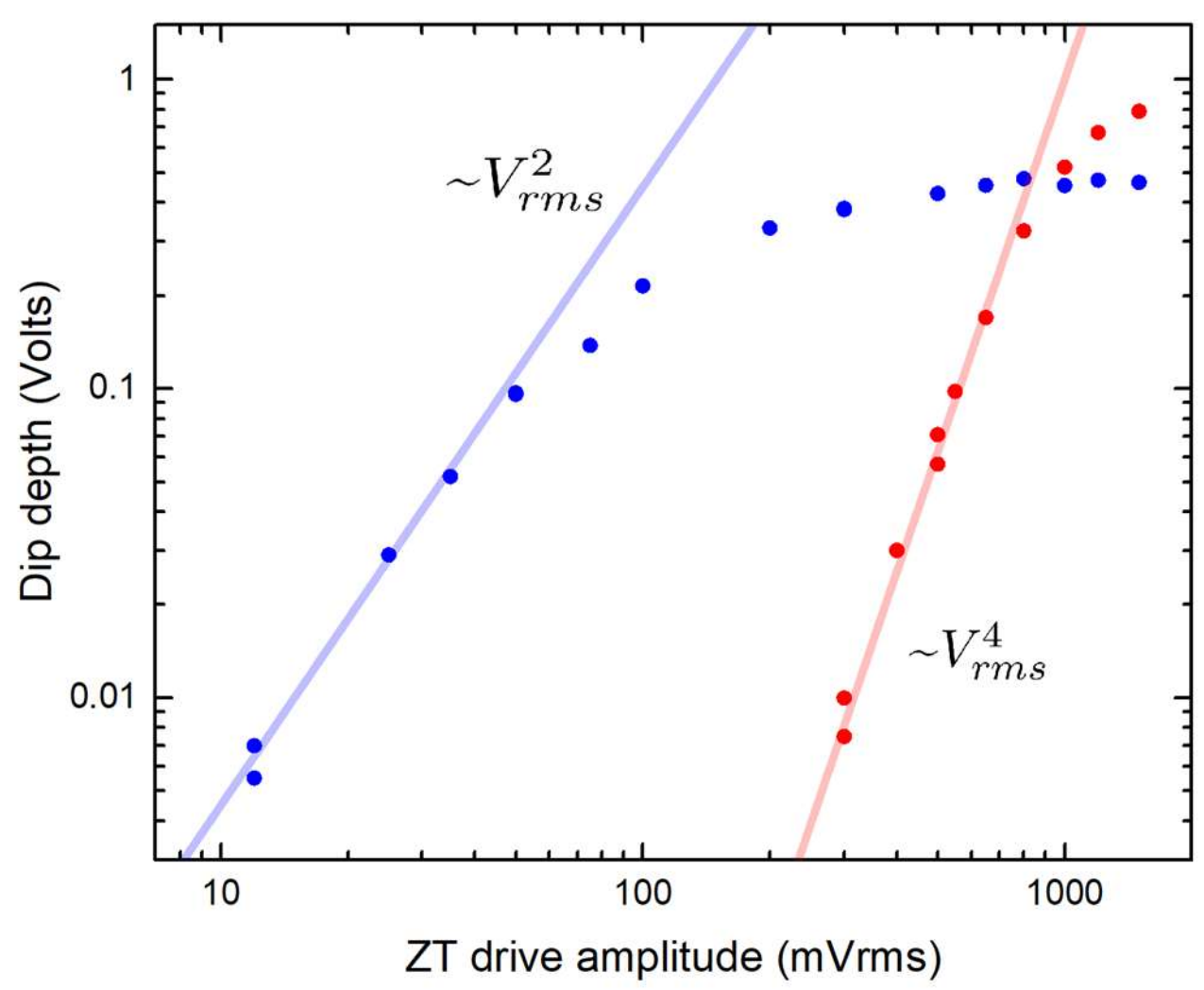

*Figure 27. These data show measured dip depths for ZT2 (blue dots) and ZT(1/2) near 4 MHz as a function of ZT drive amplitude. As found in Figure 15, ZT2 (and all the normal ZTs from ZT1 to ZT6) exhibit dip depths that are proportional to the square of the drive amplitude (far below saturation), as predicted for magnetic-dipole transitions. Theory also predicts that the 2-photon transition probabilities should be proportional to the $4^{th}$ power of the ZT drive amplitude (again far below saturation), and this trend is also observed in the measurements. These data were taken with a Rb cell temperature of 40C.*

## Dependence on scan time

Figure 28 illustrates how one can obtain higher quality strong-field Zeeman spitting spectra by using longer scan times. The physics is essentially the same as that shown in Figure 13; rapid scans can distort and shift the ZT dips because it takes time for the OP signal to recover after passing through a ZT dip.

The overall character of the strong-field Zeeman spectra can sometimes change in peculiar ways with different scan times, and we illustrate one example in Figure 29. Comparing these two spectra, we see that the ZT1 dip behaves much like we saw in Figure 13, which is qualitatively explained by the finite time needed to repopulate the $m = +3$ dark state after the Zeeman levels were scrambled passing through the ZT dip. The ZT10 transition, however, exhibits a more complex behavior that is difficult to explain. We have found that the ZT7-ZT10 features exhibit generally odd behaviors that probably arise in part because these transitions are poorly coupled to the $m = +3$ dark state. The ZT1 transition directly depopulates the dark state and thus strongly lowers the OP signal. But the ZT10 transition has little direct effect on the dark state population, so the ZT10 feature is small and exhibits some puzzling behaviors. It would probably take a sophisticated model of all aspects of the optical pumping process to explain these features fully.

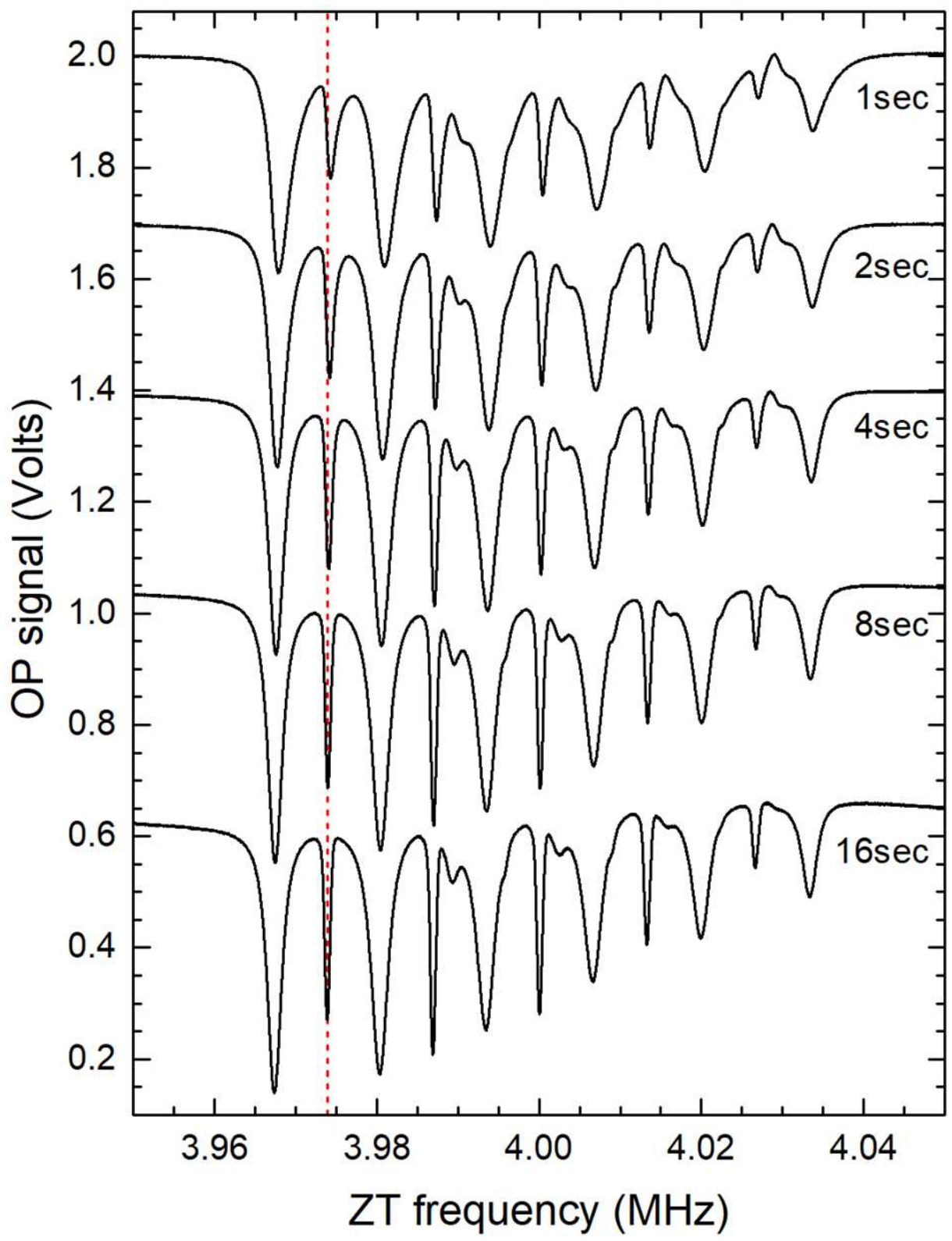

*Figure 28. These data show how the ZT spectrum near 4 MHz depends on the time used to complete the frequency scan. Like what we saw in Figure 13, these spectra demonstrate that fast scans can produce broader, asymmetrical ZT dips, shifted slightly in frequency by the fact that it takes time to restore the OP signal after passing through a dip. Slower scans generally produce higher quality ZT spectra, but we see that the differences between an 8-second scan and 16-second scan are small in this example.*

*These data were taken with a Rb cell temperature of 40C, a 32-trace average, and a ZT drive amplitude of 800mVrms.*

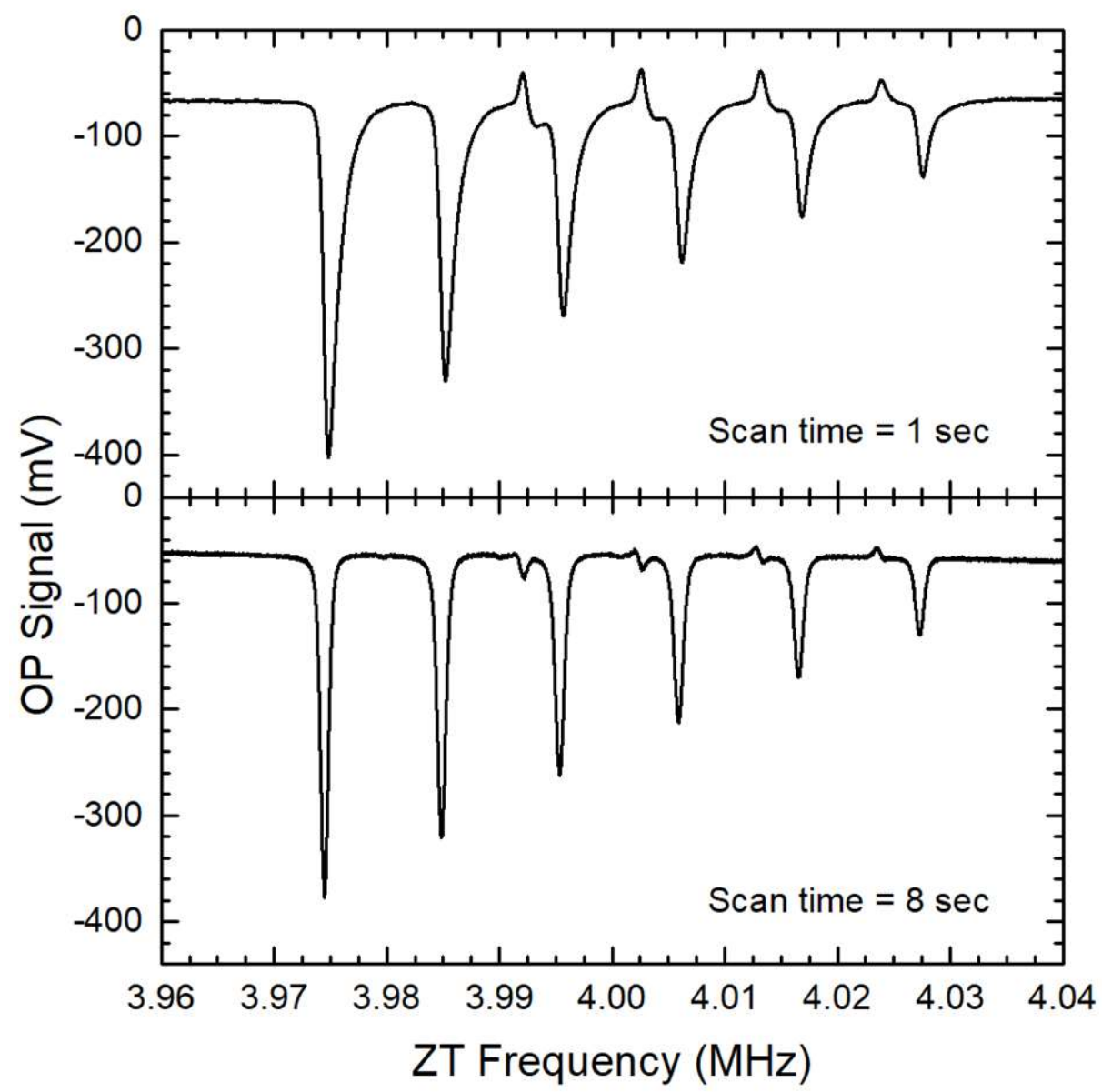

*Figure 29. The spectra on the right show how changing the scan time can turn peaks into dips in some circumstances. For example, the top panel shows a ZT10 feature that looks like somewhat distorted spectral peak using a 1-second scan time. In the lower panel, however, changing to an 8-second scan (with no other changes) produces a ZT10 feature that looks more like a distorted dip. Explaining this and other subtleties would likely require an in-depth model of the OP phenomenon, which is certainly beyond the scope of this paper. These data were taken with a Rb cell temperature of 40C and a ZT drive amplitude of 180mVrms.*

## Dependence on ZT frequency

This OP apparatus provides a great deal of interesting parameter space to explore, making the instrument well suited to the undergraduate teaching lab. At the same time, lab time is a scarce commodity, so it makes sense to focus on experiments that produce high-quality data that is easily interpreted in terms of the underlying physical concepts. Because our primary goal in this paper is to help teaching-lab instructors create the best curricula for their students, Figure 30 illustrates how the strong-field Zeeman spectra vary with the ZT drive frequency.

In the limit of low ZT drive frequency, one is in the linear Zeeman regime and all the different transitions are degenerate, yielding a single ZT dip. We call this the “basic” OP signal above, and

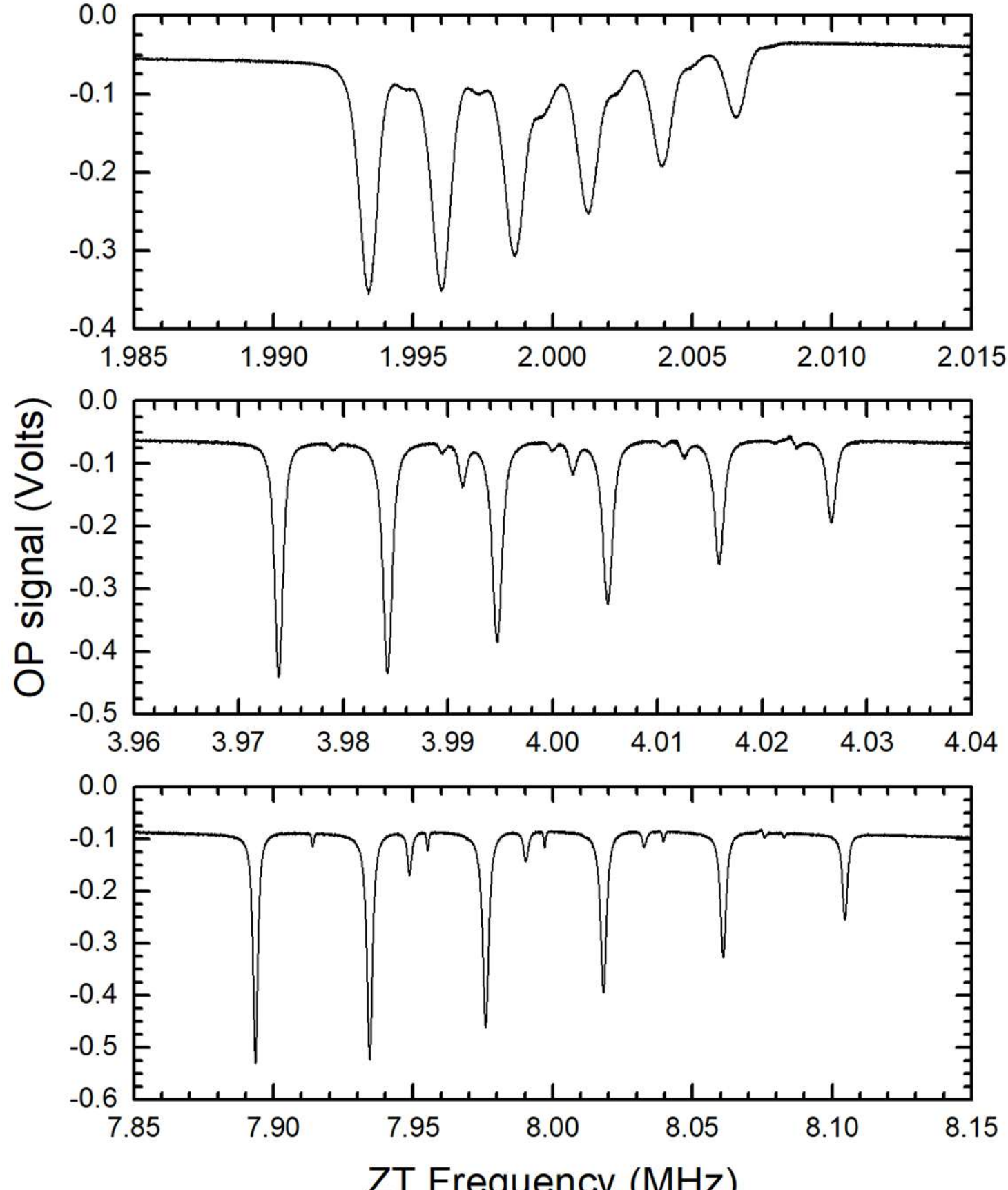

*Figure 30. These spectra show why 4 MHz is a sweet spot for observing the strong-field Zeeman splitting using this OP apparatus. With a frequency scan centered at 2 MHz (top panel), the Zeeman splitting is so small that the ZT7-ZT10 spectral features are obscured, even though the FWHM of all the main dips is ~0.8 kHz.*

*At 4 MHz (middle panel), the ZT features are nicely resolved, with the normal ZTs separated from the 2-photon transitions. Between 5-7 MHz (not shown), the normal and 2-photon transitions exhibit some overlap, making the spectra considerably more confusing to interpret. The spectral features are especially sharp at 8 MHz (lower panel), even though the FWHM of the dips is over 2 kHz. But the high B-field needed comes close to overheating the coils.*

*All these data were taken using 8-second scans with a Rb cell temperature of 50C. The ZT drive amplitudes were 150, 300, 1000 mVrms for 2MHz, 4MHz and 8MHz, yielding small 2-photon transitions in all cases.*

this is clearly the best starting point for understanding the optical-pumping process. This degeneracy is lifted at higher ZT frequencies, and the top panel in Figure 30 shows the spectrum near 2 MHz. Here the splitting is larger than the dip linewidths, but only the primary ZT1-ZT6 dips are clearly visible.

The splitting increases at higher ZT drive frequencies (using correspondingly higher $B_z$, of course), and the second panel in Figure 30 exhibits a much clearer spectrum near 4 MHz. Here the lines are sharper (relative to their separation), and the ZT7-ZT10 features are weak but visible. Importantly, the 2-photon transitions do not overlap strongly with the other ZT transitions in frequency, so one can identify every feature in the spectrum relatively easily. The overlap issue becomes clearer in the lab when you can vary the ZT drive amplitude to greatly increase and reduce the depths of the 2-photon dips in real time.

With ZT drive frequencies in the 5-7 MHz range, the 1-photon and 2-photon ZT features overlap in some cases, making it impossible to clearly identify all the transitions. We have found it best to avoid this frequency range, as the presence of overlapping spectral features results in some inevitable confusion in the lab.

At 8 MHz (the lower panel in Figure 30) the splitting is large, again yielding non-overlapping normal and 2-photon ZT features. Unfortunately, viewing an 8-MHz spectrum requires a high $B_z$, with a coil current near its rated limit of 2A (producing about 50W of heat). The coils can become quite hot operating at such high currents, and there is some risk of coil damage with prolonged use.

In our experience, we have found that 4 MHz is a good compromise for observing the strong-field Zeeman splitting, so we have mostly adopted this value in the lab.

## Dependence on Rb cell temperature

Moving on, Figure 31 illustrates how the $^{85}$Rb Zeeman spectra near 4 MHz vary with the temperature of the Rb vapor cell, with the ZT drive amplitude just low enough to avoid 2-photon transitions. In all these spectra, we see that the ZT1-ZT6 dips are clearly visible and resolved, with the dips becoming progressively broader with increasing temperature. In contrast, the ZT7-ZT10 features only appear with clear dip-like structures at 67C.

As we describe below, the ZT7-ZT10 features are not easily modeled at any temperature, while the ZT1-ZT6 dips generally fit theoretical expectations quite well. As a result, we tend to mostly ignore the ZT7-ZT10 features in the teaching lab, other than observing their somewhat puzzling existence and distorted shapes in the Zeeman spectrum. Given this reasoning, the easiest way to produce clean strong-field Zeeman spectra for further analysis is to use the Rb cell at room temperature.

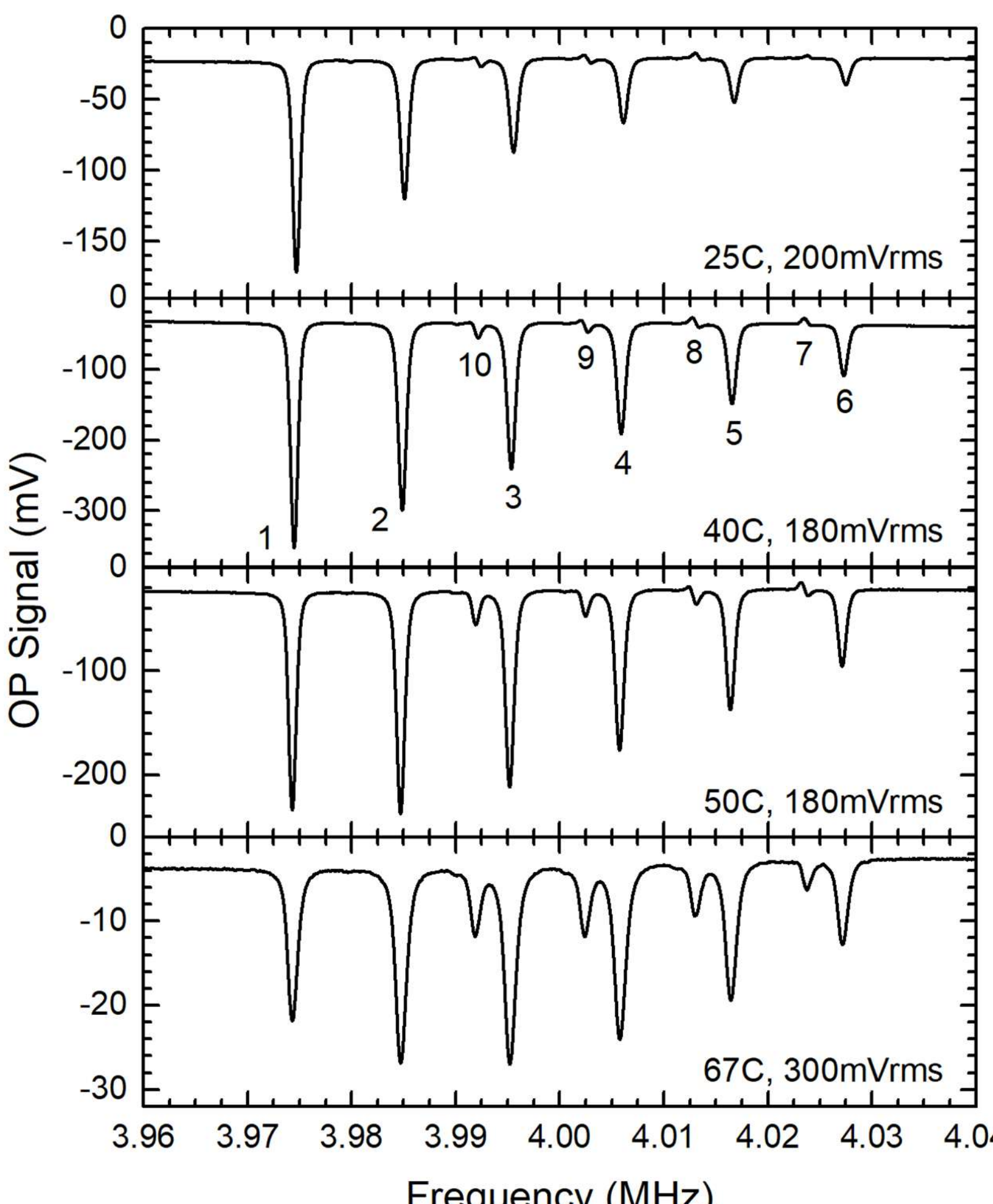

*Figure 31. These 4-MHz spectra examine how the ZT spectra change with Rb cell temperature. In each case the ZT drive amplitude was increased as much as possible while still avoiding 2-photon transitions.*

*At 25C and 40C, the ZT1-ZT6 features are especially sharp, with a monotonic progression of dip depths from ZT1 to ZT6. But the ZT7-ZT10 features are a bit odd and distorted.*

*At 50C, the ZT7-ZT10 features are clearer, with the ZT9 and ZT10 showing clear dip-like structures. At 67C, all the ZTs are nice symmetrical dips, which is ideal for modeling. However, the OP signal amplitude is much lower at this high cell temperature because the high optical depth means that not much light makes it through the cell to the photodetector.*

*These data were taken using 8-second scans and 32-trace averages.*

## Dependence on polarization

Figure 32 illustrates 4-MHz strong-field Zeeman spectra for both $\sigma^+$ and $\sigma^-$ light incident on the Rb vapor cell. As expected, in each case the deepest ZT dip is associated with the transition that directly couples to the OP dark state. We use $\sigma^+$ for teaching, as it is conceptually simpler when the dark state corresponds with the highest-energy Zeeman level.

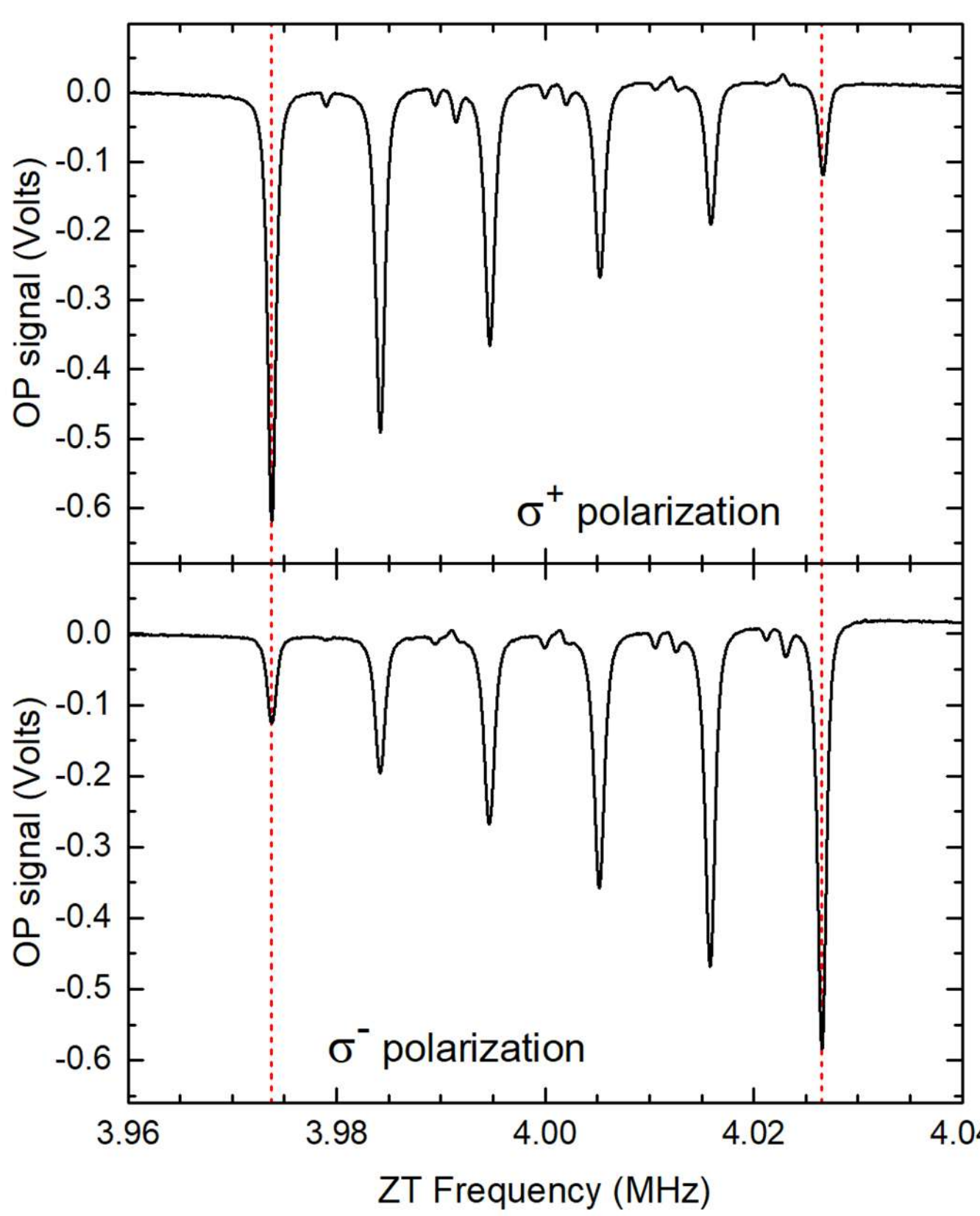

*Figure 32. These ZT spectra show what happens when you change the sign of the circular polarization entering the Rb cell (or, equivalently, the direction of the horizontal magnetic field). With $\sigma^+$polarization driving Δm=+1 transitions (top panel), ZT1 is most strongly coupled to the m=+3 dark state, while ZT6 is the least strongly coupled within the f=3 manifold. With $\sigma^-$polarization (bottom panel) the m=-3 level is now the dark state, which is most strongly coupled by ZT6. As usual, the Zeeman levels in the f=2 manifold are only weakly coupled to either dark state, so ZT7-ZT10 show only weak OP spectral features. Weak 2-photon transitions are also visible in these spectra.*

*These spectra were taken with a Rb cell temperature of 40C, a ZT amplitude of 300 mVrms, 10-second scans, and 16-trace averages. The spectra were aligned to match the ZT dips, compensating for the unknown background $B_z$ in the lab.*

## Modeling the strong-field Zeeman splitting

Having observed some strong-field Zeeman spectra, our next step is to compare the dip frequencies with expectations from Breit-Rabi theory. We begin with the exceptionally high-quality spectrum shown in Figure 33, where we have applied a number of tricks to improve the signal-to-noise ratio. Note that lower quality spectra are fine for the teaching lab, but our goal here is simply to see what the OP instrument is capable of under optimal conditions.

We next inverted this spectrum to turn the dips into peaks, and we then used a peak-finding algorithm to determine the centroid frequency of each peak along with a corresponding measurement uncertainty. The peak-finding algorithm in the Origin software package worked well for this step, but we obtained essentially identical frequency estimates by feeding the .cvs spectra data in Figure 33 into ChatGPT and simply asking the AI to find the peaks. In both cases, we did not know how the different algorithms achieved their goals, but both seemed to produce sensible results.

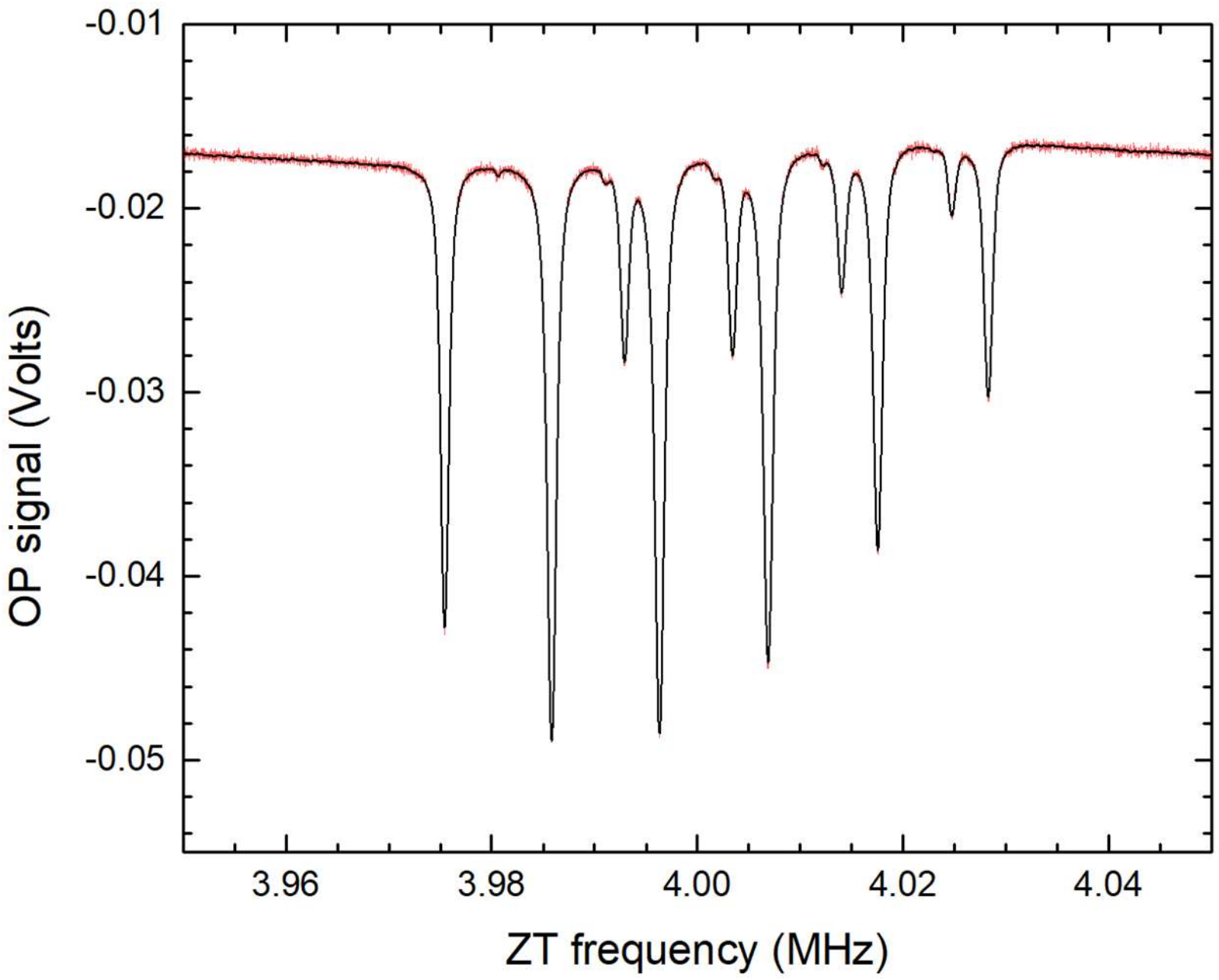

*Figure 33. This high-quality Zeeman spectrum was taken using a Rb cell temperature of 67C, yielding ten clean ZT dips. Because the signal amplitude was low, we used a 15-second scan and a 64-trace average, with data acquired using a 12-bit oscilloscope recording 10,000 spectral points. The red line shows the original trace-averaged data, and an application of 50-point Savitzky-Golay smoothing yielded the black line that was used for our analysis.*

For comparison with the Breit-Rabi equation, we did a bit of "spreadsheet modeling" shown in Figure 34. The columns in this spreadsheet are defined as:

F = the total spin of the Zeeman sublevel

mf = the m level of the Zeeman sublevel

B = the value of $B_z$, which the only independent variable in the theory

x = the dimensionless parameter defined in Equation (7)

Emf = the energy of the Zeeman sublevel given by the Equation (6)

dE = the energy difference between adjacent Zeeman sublevels (according to theory)

Nt = the ZT number defined in Figure 23

dips = the ZT frequencies measured by fitting the measured Zeeman spectrum

diff = the difference between measurement and theory

diffG = the difference expressed as a change in the magnetic field

| | F(Y) | mf(Y) | B(Y) | x(Y) | Emf(Y) | dE(Y) | Nt(Y) | dips(Y) | diff(Y) | diffG(Y) |
|---|---|---|---|---|---|---|---|---|---|---|
| Long Name | | | | | | | | | | |
| Units | | | Gauss | | MHz | MHz | | MHz | kHz | Gauss |
| Comments | | | | | | | | | | |
| F(x)= | | | | 1.3996/3035.73 | (F)-2.5)*sqrt(1 + (4/6 | (Emf)[i]-col(Emf)[i | | | 0*(col(dE)-col(dip | col(diff)/467 |
| 1 | 3 | 3 | 8.583853 | 7.9162E-03 | 1.5298701E+03 | 3.97543 | 1 | 3.975430 | 0.00138 | 2.96227µ |
| 2 | 3 | 2 | | | 1.5258947E+03 | 3.98583 | 2 | 3.985836 | -0.00175 | -3.73964µ |
| 3 | 3 | 1 | | | 1.5219089E+03 | 3.99632 | 3 | 3.996314 | 0.00489 | 10.48008µ |
| 4 | 3 | 0 | | | 1.5179126E+03 | 4.00689 | 4 | 4.006883 | 0.00326 | 6.97297µ |
| 5 | 3 | -1 | | | 1.5139057E+03 | 4.01754 | 5 | 4.017540 | -0.00126 | -2.68877µ |
| 6 | 3 | -2 | | | 1.5098881E+03 | 4.02828 | 6 | 4.028282 | -0.00608 | -13.02485µ |
| 7 | 3 | -3 | | | 1.5058599E+03 | 3015.73388 | -- | -- | -- | -- |
| 8 | 2 | -2 | | | -1.5098740E+03 | 4.02459 | 7 | 4.024755 | -0.16186 | -346.59361µ |
| 9 | 2 | -1 | | | -1.5138986E+03 | 4.01394 | 8 | 4.014041 | -0.09954 | -213.15829µ |
| 10 | 2 | 0 | | | -1.5179126E+03 | 4.00337 | 9 | 4.003431 | -0.05818 | -124.58581µ |
| 11 | 2 | 1 | | | -1.5219159E+03 | 3.99289 | 10 | 3.992935 | -0.04614 | -98.80815µ |
| 12 | 2 | 2 | | | -1.5259088E+03 | -- | | | -- | -- |

*Figure 34. With a set of ten measured ZT frequencies, a spreadsheet provides a visual and easily understood method technique for comparing with Breit-Rabi theory. The columns and analysis procedure is described in the text.*

Of course, the frequency comparison can be accomplished by a short code in any programing language, but we like the spreadsheet analysis because of its pedagogical properties – it is both visual and easy to explain.

Figure 35 shows the final result of our analysis, revealing that we obtain results consistent with Breit-Rabi theory at the part-per-million level, provided we adjust $B_z$ to fit the data and discard the ZT7-ZT10 features. Throughout our observations, we have found that the ZT1-ZT6 features exhibit clean dip-like structures, while the ZT7-ZT10 features are beset with puzzling, distorted shapes that vary widely with experimental conditions. While the ZT7-ZT10 features are mostly dip-like at a Rb cell temperature of 67 C, Figure 35 suggests that precise centroid fits of the dip positions are still affected by the distortions that are obviously apparent at lower temperatures. In addition, OP theory explains these distortions because the ZT7-ZT10 transitions are so poorly coupled to the dark state. All these considerations justify an exclusion of the ZT7-ZT10 transitions when making precise comparisons with Breit-Rabi theory.

Beyond comparing spectral measurements with an already-well-tested theory, this exercise shows how the physics of optical pumping can be applied to make a precision magnetometer, yielding ppm sensitivity with absolute accuracy. There are many engineering details to deal with, of course, but this is one example of how atomic spectroscopy can be used in a broad variety of precision metrology measurements. If you imagine OP measurements using lab-on-a-chip devices than incorporate semiconductor lasers, optics, and sensors assembled on opto-electronic integrated circuits, it is easy to see why precision atomic spectroscopy is still an area of active research.

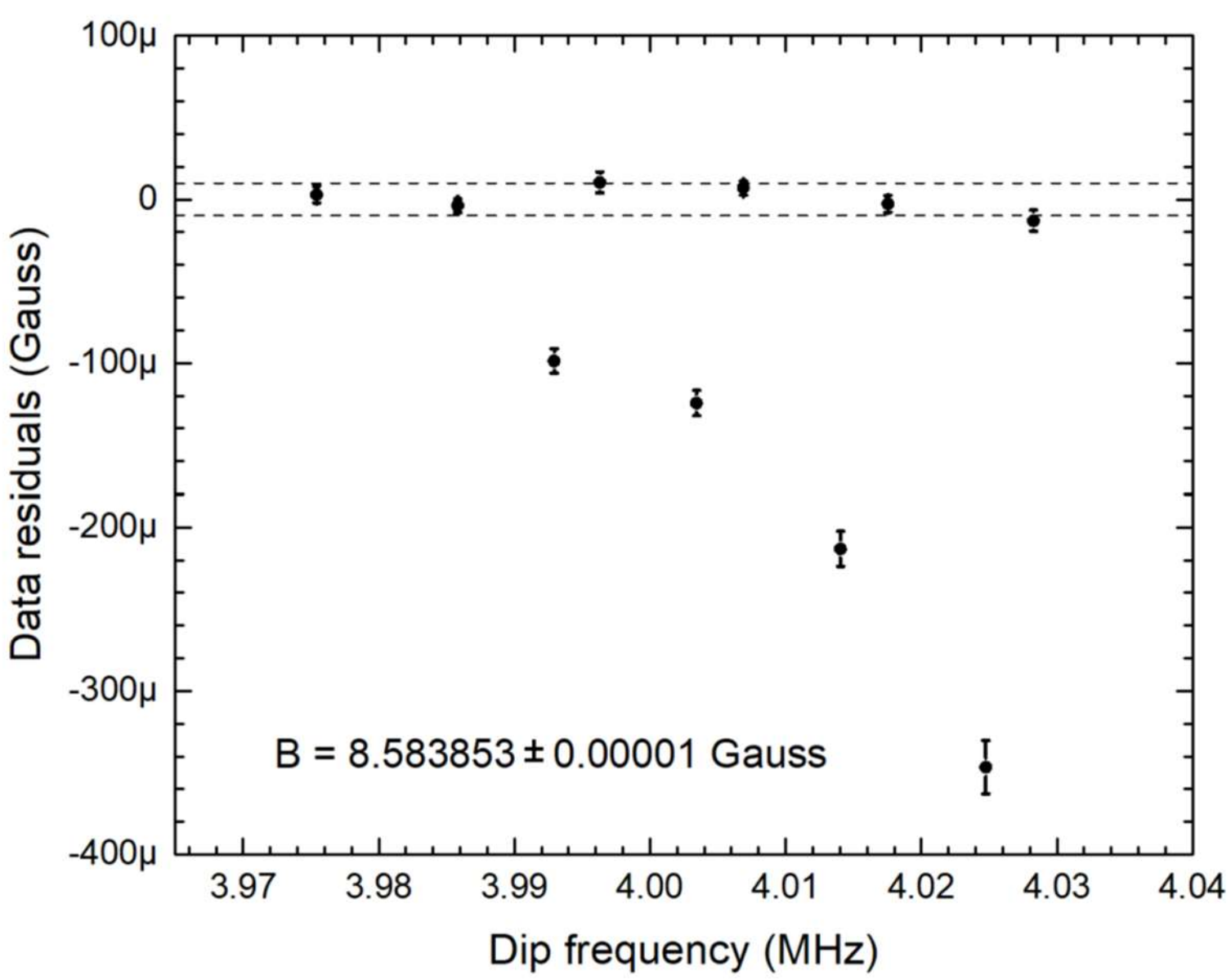

*Figure 35. Here we see the data residuals (theory – experiment) after adjusting the input $B_z$ level to produce the smallest differences for the ZT1-ZT6 frequencies. If we use these six transitions to make a measurement of $B_z$, the residuals show that this measurement can be made with an accuracy of about 1 part per million. The ZT7-ZT10 measurements do not fit the theory, which we attribute to the distortion of the ZT dips seen throughout our measured spectra. These distortions are smaller at 67C than at other temperatures, but they are still present, with ZT7 having the largest distortion in all cases.*

## Summary

There is no shortage of reasons why the Optical Pumping of Rubidium vapor atoms has enjoyed so much popularity in physics teaching labs around the world. With just this simple apparatus, a great many fascinating atomic physics phenomena can be observed and compared with detailed predictions from quantum mechanics. The material is well connected to beginning quantum mechanics courses, and modern digital oscilloscopes can collect remarkably accurate data. Going through the process of understanding the physical concepts, setting up the experiments, collecting data, and analyzing the results provides a fine experience for potential experimental physicists. It has been sixty years since Alfred Kastler was awarded the Nobel Prize in Physics "for the discovery and development of optical methods for studying Hertzian resonances in atoms." And Kastler's ingenious experiments are still being widely used in undergraduate physics teaching labs.

# Acknowledgements

This work was supported in part by a generous donation from Beatrice and Sai-Wai Fu to the Physics Teaching Labs at Caltech, together with Caltech's long-standing support of outstanding laboratory instruction across many STEM fields.

*Contact:* For corrections, comments, or just to compare notes, please contact Kenneth G. Libbrecht, *kgl@caltech.edu*, or mail to: Mail-stop 264-33 Caltech, Pasadena, CA 91125.

# Appendix 1 – Hardware details

The experiments described in this paper we done using the *TeachSpin* Optical Pumping apparatus, with some additional modifications described in this Appendix. The stock OP optical hardware from *TeachSpin* is shown in Figure 36, and Figure 7 sketches the various components. This hardware is described in detail in the *TeachSpin* user manual.

One substantial change we made to this setup was to install a different photodetector (described below) with a second 795nm narrow-band filter (*Thorlabs* FBH800-40) right in front of the photodetector, as shown in Figure 7. Our replacement photodetector included an active electronic high-pass filter with a 30-second time constant, and the additional optical filter blocks much of the room light striking the photodetector. Together, these changes yielded an OP signal with greater overall stability and automatic background subtraction, making data collection substantially simpler for students. Once the optical elements are set up and aligned, we also typically cover the device with a black cloth to further reduce stray light contributions.

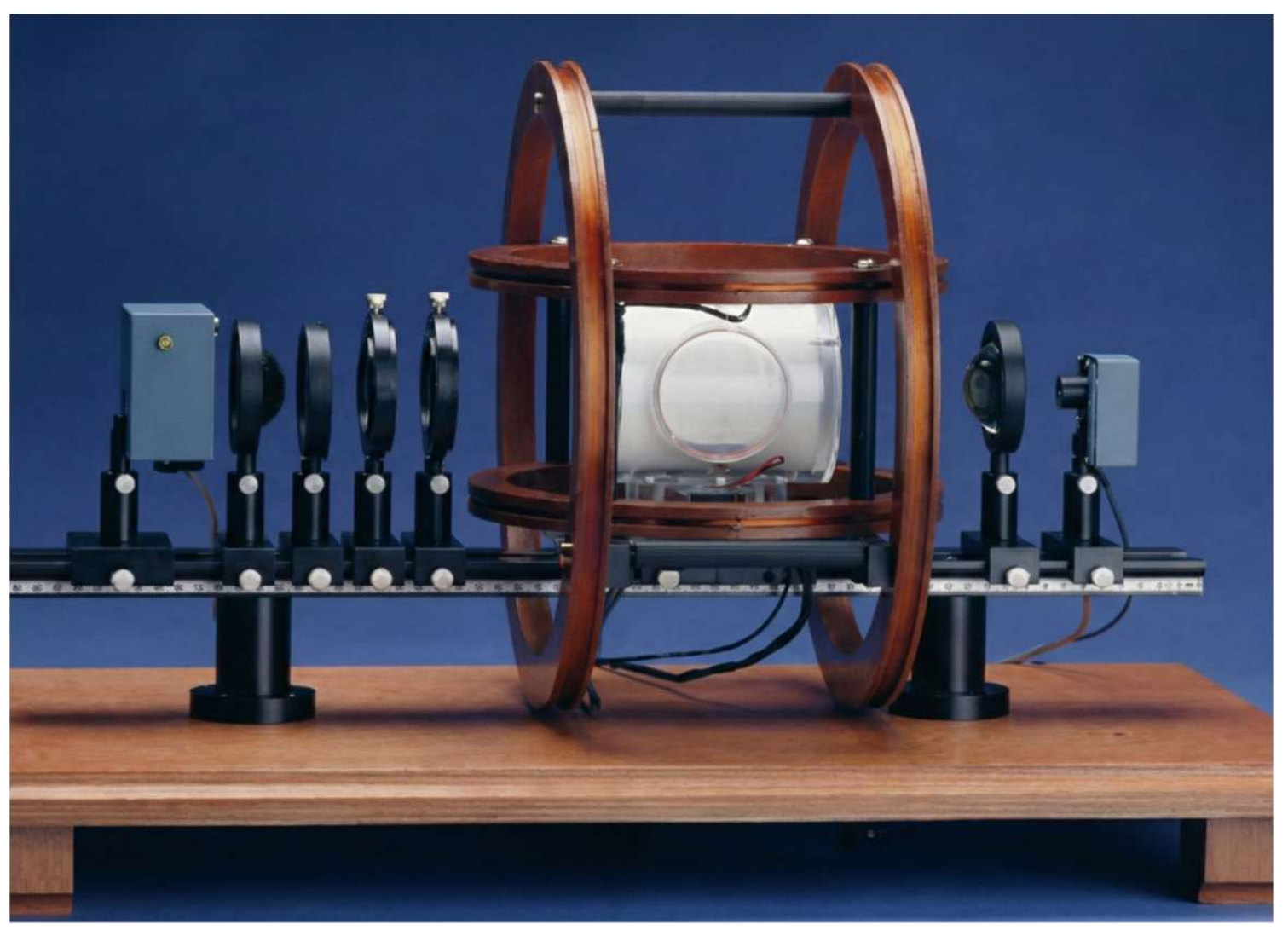

*Figure 36. This photo shows the optical hardware included in the TeachSpin Optical Pumping apparatus (sketched in Figure 7). We used this instrument for all the results in this paper after replacing the stock photodetector with the homemade circuit described below.*

We also ended up replacing the entire *TeachSpin* electronics chassis, in part because our unit was quite old (one of the first made) and not functioning well anymore. Plus we wanted to switch to all digital electronics for easier setup and operation. More generally, we prefer using commercial test equipment to drive experiments in our teaching labs for pedagogical reasons, as this is what our students will likely encounter when they move on to graduate research labs.

In place of the *TeachSpin* electronics chassis, we used a small desktop power supply (a *Wanptek* Variable 0-32V 0-10A DC supply) to drive the Rb lamp. The lamp requires 28VDC, 0.53A at start-up, with the current dropping to 0.25A after warm-up. Because the Rb lamp is independent of the rest of the electronics, electronic noise from this inexpensive switching power supply produced no noticeable noise problems.

To control the Rb cell temperature, we used a *Thorlabs* TC300B general-purpose temperature controller. While this gave no better performance than original-issue *Omega* controller, it was somewhat easier to use. The *TeachSpin* Rb cell uses a T-type thermocouple, which the *Thorlabs* controller did not support, but selecting the K-type thermocouple supported by the controller was close enough for our purposes. We obtained reliable performance overall with these set parameters: Sensor = Thermocouple Type K, Heater mode, (P, I, D) = (0.5, 200, 100), Period = 0.1 sec.

To drive the constant Horizontal and Vertical B fields, we used a *Siglent* SPD3303X DC power supply, as sketched in Figure 8. This delivers digital voltage and current set-point resolutions of 1mV and 1mA, respectively, and we inferred current drifts of about 50μA over several hours (much better than guaranteed in the instrument specifications) by observing drifts in the ZT frequencies. In terms of the OP signal, we measured that the 4 MHz ZT dip positions were stable to about 0.2 kHz over several hours, making it possible to average many traces when collecting data. We typically kept this power supply at least one meter from the Rb cell to avoid problems associated with stray 60-Hz magnetic fields.

For sweeping the Horizontal B field, we used a *Siglent* SDG 2042X signal generator to produce a triangle-wave signal, sending this signal through a *Juntex* DPA-1698 Power Amplifier to drive the Sweep coils. This inexpensive *Juntex* amplifier (readily available on eBay) can drive up to 1A, 20V at 100kHz, and the noise performance is quite good for this application. One quirk we found was that the amplifier did not power up properly with a low input voltage and a low impedance load. A workaround for this is to set the input DC offset to 2V (using the signal generator) to "kick" the amplifier into starting up, then immediately turn the DC offset back to 0V for all our experiments.

We typically use a symmetrical triangle-wave signal for sweeping $B_z$, as this produces a clean OP signal at the turn-around points (see Figure 9). In contrast, an asymmetrical triangle wave (a.k.a. a ramp signal) produces spurious dips at the turn-around points, which can be confusing for new students. Even when averaging traces with slow sweep times, the ramp signal has no advantage over the triangle-wave signal, given how triggering works for each of these options. Producing a DC offset on the sweep can be done using either the Horizontal Field power supply or the signal generator offset setting, but we typically use only the former to keep things simple.

We drove Zeeman transitions using the second channel of the *Siglent* SDG 2042X signal generator, as shown in Figure 8. This output signal is strong enough to drive the ZT drive coils directly, and the Burst and Sweep modes are well suited for performing the spin rotation and strong-field Zeeman splitting measurements.

The OP signal can be observed with any oscilloscope, of course, but we mostly used the *Siglent* SDS1104X HD to collect the data in this paper, because it has 12-bit resolution and will store 10,000 points per trace (or higher). These features are useful for making low-noise measurements, but certainly not essential for observing the essential OP physics. This 'scope also displays both a Live Trace plus the corresponding Trace Average signal simultaneously, which is useful for averaging traces. No other data acquisition system was needed to produce the results in the paper. In the teaching lab, we use the Keysight EDU 1052A because it is a more economical choice that also works very well for collecting OP data.

We should point out that none of these hardware changes were necessary, as the stock *TeachSpin* apparatus should be capable of making all the observations described in this paper. Using commercial test equipment as we describe provides additional control of the OP signals, and having digital accuracy is useful for setup and reproducibility. The only change that made the experiments noticeably easier was our new photodetector, which we describe next.

## An improved photodetector

Because we abandoned our old TeachSpin electronics chassis, we also needed to replace the stock OP photodetector. This presented an opportunity to develop an automatic background subtractor (a.k.a. a high-pass filter) circuit that would produce a high-gain OP signal without having to frequently monitor and manually adjust the DC signal subtraction. While the manual adjustment worked fine and was instructive for students, we also found it was somewhat distracting in practice, taking attention away from understanding the OP physics.

The circuit we developed is shown in Figure 37, which ended up making the OP operation noticeably easier and more straightforward for students. The AC output includes an active high-pass filter with a lengthy 30-second time constant, removing the DC signal by sending a cancellation current into the front-end trans-impedance amplifier (TIA). This input subtraction allowed a higher TIA gain, which reduced the overall noise contribution from the amplifiers. In large part, the output noise derives almost entirely from resistor thermal noise (specifically in R9 and R5).

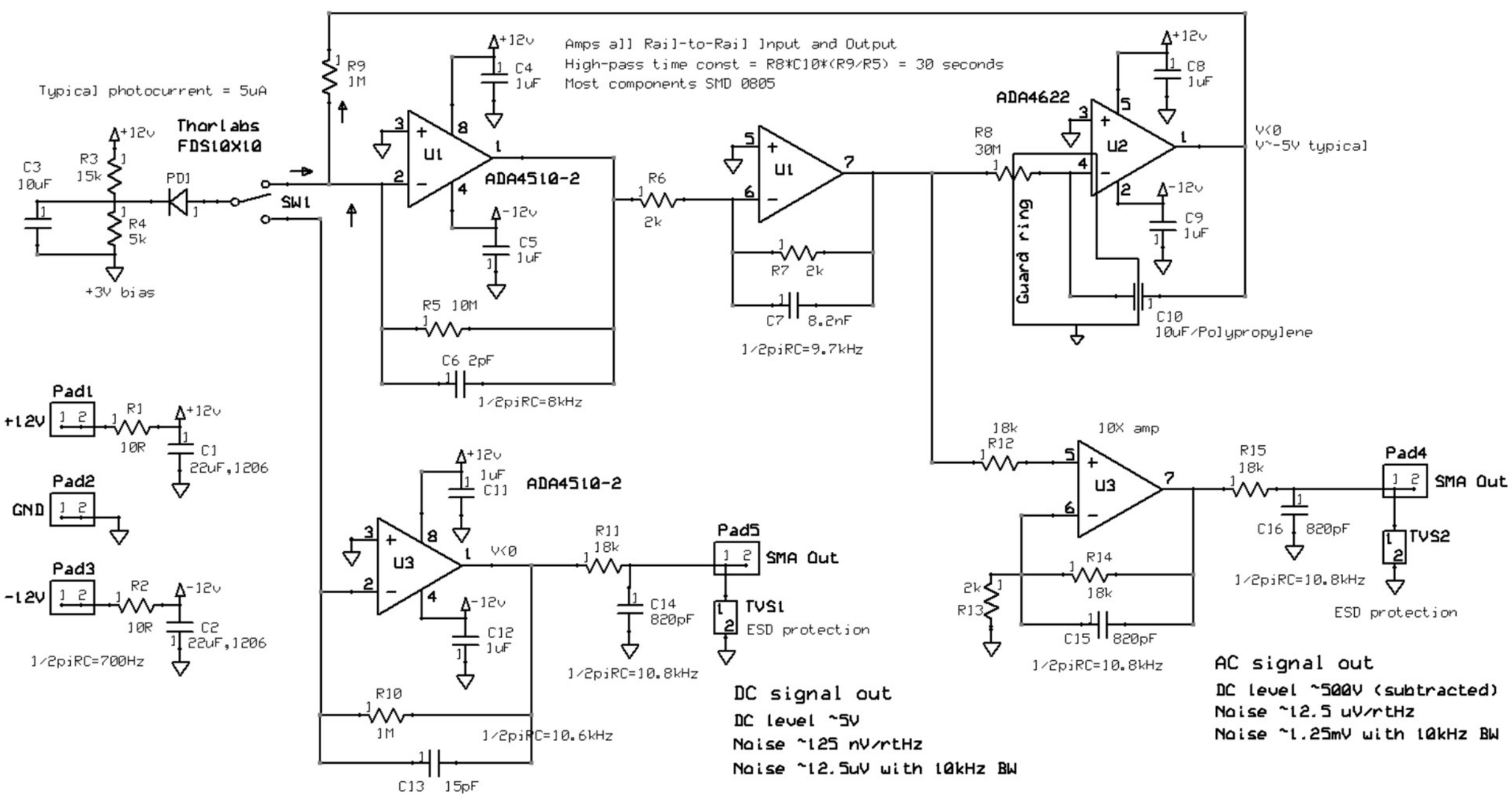

*Figure 37. A circuit schematic for the photodetector we used in the OP experiments described in this paper. In addition to good noise performance, a key feature of this circuit is an active high-pass filter with a 30-second time constant, which automatically subtracts the large DC background present in OP signals.*

The exceptionally long high-pass time constant means that small leakage currents inside C10 can produce rather large offset voltages in the AC output. Using a normal ceramic capacitor for C10, for example, can result in offsets of 1V or higher. To avoid this problem, we used a polypropylene capacitor, as these are known for having exceptionally low leakage currents. For good measure, we also added a guard ring on the printed-circuit board that reduces leakage currents coming from dirt on the surface of the printed-circuit board. With these design features, we measured DC offsets in the AC output of no more than a few mV.

We chose a 10 kHz bandwidth for the AC output based on the spin-rotation observations described above. We found it difficult to observe spin-rotation signals above a few kHz, so we reduced high-frequency noise by cutting off the signal at 10 kHz. Most of the OP signals appear at much lower frequencies, and in those cases we often applied additional low-pass filtering during data analysis. In particular, we found that Savitzky-Golay smoothing substantially reduced high-frequency noise with little distortion of the underlying signal.

The AC output in this circuit is sufficient to observe all the main OP signals described in this paper. However, it is also useful to have a direct DC output, in part because one needs the DC signal to optimize the optical alignment of the system. The DC signal in Figure 37 also has a good noise performance, but it would need additional amplification and a DC subtraction circuit to view small OP signals. The DC output is quite useful for alignment and diagnostics, but students generally do not use this feature unless they want to study the DC signal versus temperature.

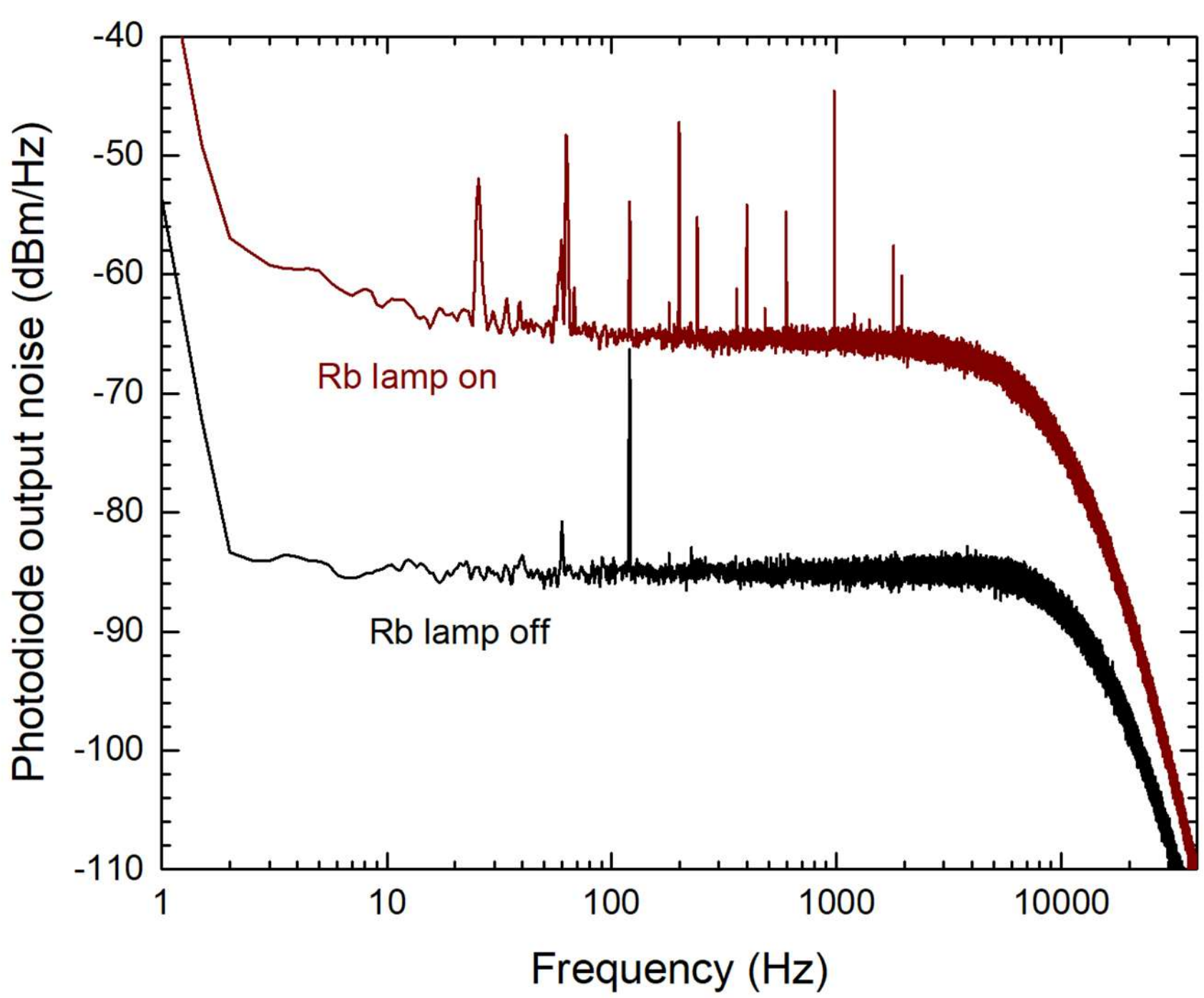

*Figure 38. Measured noise spectral densities (trace-averaged) for the photodetector circuit in Figure 37, both with no incident light on the photodiode (Rb lamp off) and with the Rb lamp on (with the Rb vapor cell at room temperature). The lamp-off spectrum agrees with calculations, the observed noise arising mostly from thermal noise in the TIA resistors. The lamp-on noise is likely dominated by plasma instabilities in the Rb lamp, with the noise level being about 10x higher than photon shot noise. When the Rb cell is heated above room temperature, the lamp signal is reduced by absorption in the Rb cell, and the resulting noise curve is between the two curves shown above.*

Figure 38 shows noise spectra for the AC signal output when the Rb lamp is off (so essentially no light strikes the photodiode) and with the lamp on. For both measurements we used a Rohde & Schwarz MXO4 oscilloscope to average FFT traces, using a resolution bandwidth of RBW = 1 Hz. With the lamp off, the expected noise spectral density is about 12.5 µV/rtHz, equivalent to -85 dBm/Hz in Figure 38. This noise floor is well modeled by thermal noise in the TIA resistors, with the op-amps contributing quite little to the observed noise.

With the lamp on and the Rb cell at room temperature, the measured photocurrent was typically slightly above 5 µA. This produces a shot noise of about -85 dBm/Hz, which is about 10x lower than the observed noise floor. It appears that this noise results mainly from plasma instabilities in the Rb lamp, but we have not investigated this further.

Once the Rb lamp had warmed up for ~15 minutes, its output (measured by the photodetector) exhibited a remarkably good short-term stability. Over one-minute timescales, the AC signal typically fluctuated by less than 5 mV, which is about $10^{-5}$ times smaller than the 500 V DC background (equal to 5 µA x 10MΩ x 10) that would be present in the AC channel if not for the DC subtraction. We have no ready explanation for why this lamp reliably exhibits this remarkable 10 ppm short-term stability. It is difficult to make an LED lamp with comparable performance.

Figure 39 shows an example of single-trace OP signal taken using this photodiode circuit, demonstrating its overall performance. High-frequency noise can be reduced in post-processing as shown, or trace-averaging can achieve this goal in real time (see Figure 33). Signal drifts of a few millivolts are typical, as seen in Figure 39, but larger drifts are greatly reduced by the high-pass filter. Moreover, because the high-pass filter has such a long time constant, it produces little distortion of the OP signal. In practice, this all means that the OP signal remains stable while the oscilloscope is left at the set-and-forget settings of DC coupling and zero voltage offset.

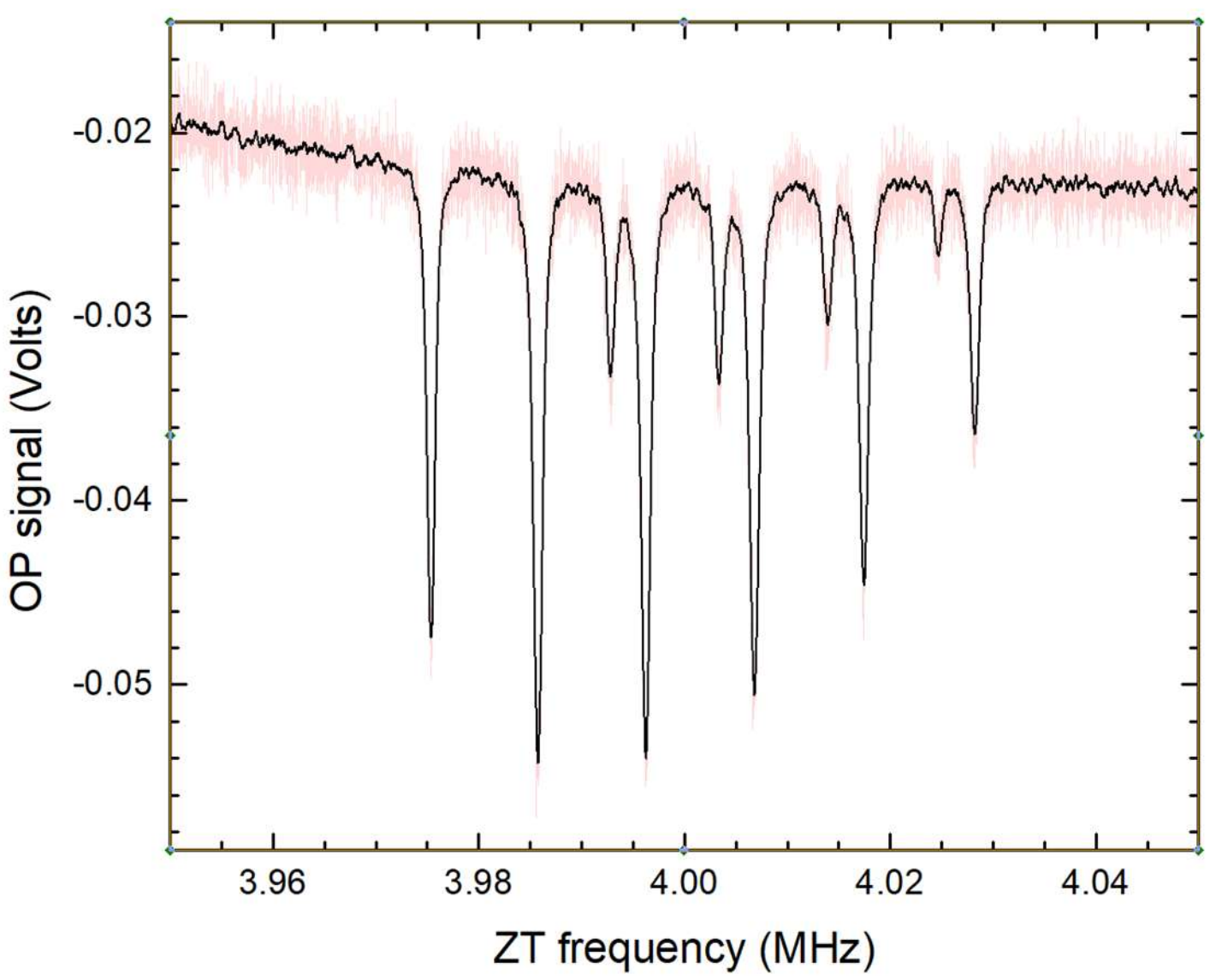

*Figure 39. The red line in this graph is like that in Figure 33, except showing a single oscilloscope trace. The voltage noise is about 2mVrms with a 10kHz bandwidth, reflecting a combination of electronic noise in the photodetector circuit and Rb lamp noise after substantial absorption in the Rb cell. The black line shows the same data with 50-point Savitzky-Golay smoothing to reduce the bandwidth. This example shows that even low signal amplitudes can be observed with good signal-to-noise using the OP hardware.*

# Appendix 2 – Modeling the OP signal brightness level

To model the OP overall signal level as a function of the Rb cell temperature (Figure 16, top graph), we start with the Rb vapor density $\rho_{Rb}(T)$ (number of atoms per unit volume) in equilibrium with solid Rb metal (which is always present on the walls of the vapor cell). Stat-mech theory suggests a roughly Arrhenius form

$$\rho_{Rb}(T) \sim \exp(-T_A/T_K) \tag{10}$$

where $T_K$ is the cell temperature in Kelvin and $T_A$ (also in Kelvin) is an empirical Arrhenius constant. This relationship is not a perfect representation of $\rho_{Rb}(T)$ from experiments, but it is sufficient for our purposes, especially given that the Rb cell temperature is not controlled with high absolute accuracy in the OP apparatus. We assume a value of $T_A \approx 10000$ K for both Rb isotopes, which should be accurate to a few percent [1973Gal].

We next treat the Rb optical spectrum as if it consisted of a single two-level transition, as this is sufficient to produce an approximate toy model. Including the usual Doppler broadening from thermal motion, we can write the transmitted light absorption coefficient $Abs(\Delta f) = I_{out}(\Delta f)/I_{in}(\Delta f)$ as

$$Abs(\Delta f) = \exp\left(-\tau_{0,cell} e^{-x^2}\right) \tag{11}$$

where $\Delta f$ is the light frequency relative to the 2-level transition frequency, $\tau_{0,cell}$ is the on-resonance optical depth of the Rb vapor cell at temperature $T$, $x = \Delta f/\sigma$, and $\sigma$ is the Doppler broadening parameter. To simplify matters a bit more, we assume that $\sigma$ is essentially independent of temperature.

Because $\tau_{0,cell}$ is proportional to the Rb gas density, we can write the absorption coefficient

$$Abs(\Delta f) = \exp\left[-A \cdot \exp(-T_A/T_K) \cdot e^{-x^2}\right] \tag{12}$$

where $A$ is a constant that depends on the overall length of the cell and the properties of the atomic transition.

To calculate the OP photodiode signal, we next need a suitable model of the Rb lamp spectrum. We again assume a simple 2-level approximation of the Rb atoms in the lamp and write

$$I_{lamp}(\Delta f) = C \cdot \left[1 - \exp\left\{-\tau_{0,lamp} e^{-x^2}\right\}\right] \tag{13}$$

where $C$ is an overall normalization constant and $\tau_{0,lamp}$ is the optical depth of the lamp when $\Delta f = 0$.

The model parameters are not well constrained at this point, so we choose some reasonable estimates to facilitate the discussion: $T_A \approx 10000$ K, $A \approx 3 \times 10^{13}$, $\sigma \approx 320$ MHz, and $\tau_{0,lam} \approx 5$. Using these values yields the plots of $I_{lamp}(\Delta f)$, $Abs(\Delta f)$, and $I_{out}(\Delta f)$ shown in Figure 40. At the highest temperatures, we see that most of the light transmitted through the cell is from the edges of the Doppler profile (green curve in Figure 40).

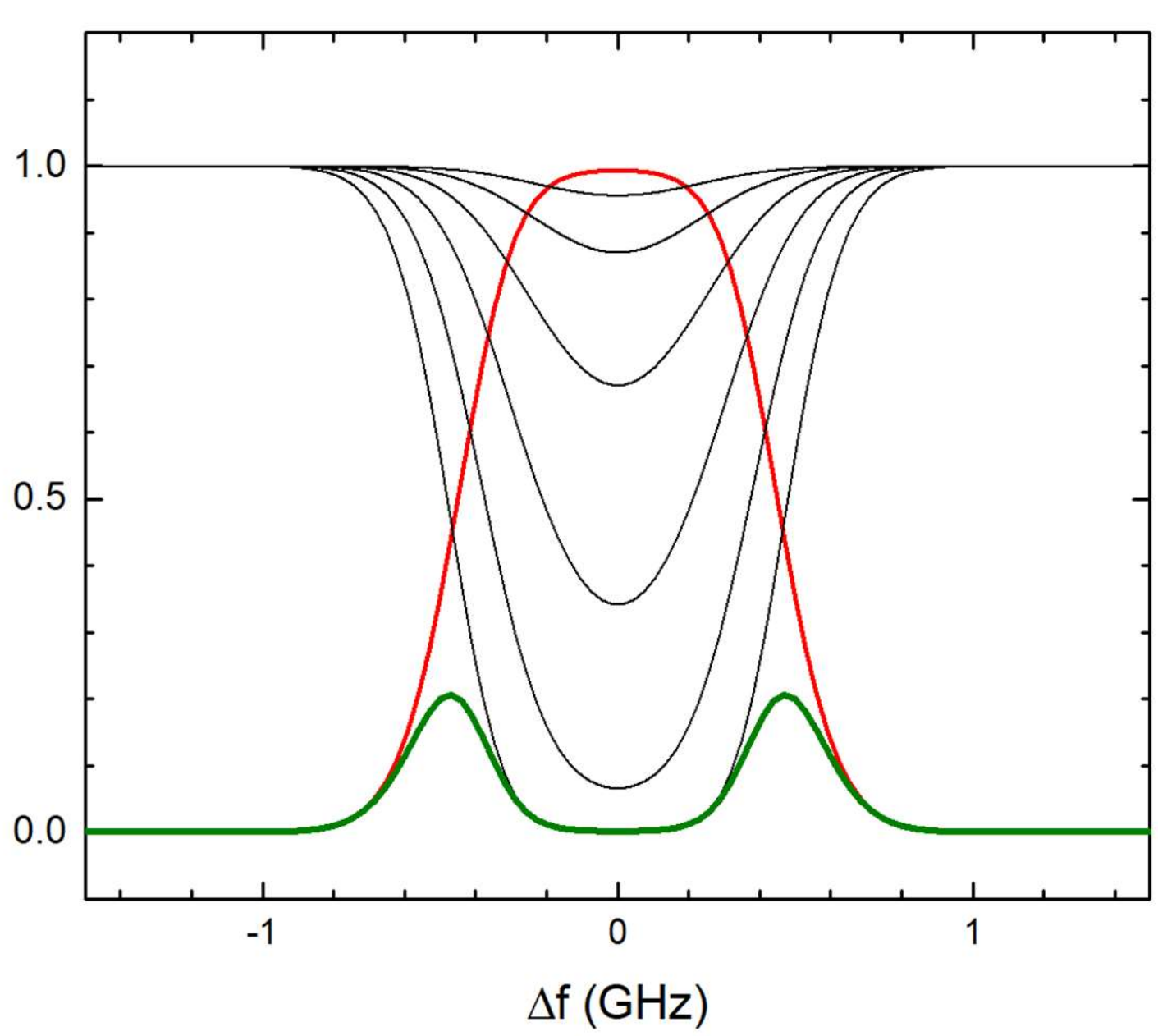

*Figure 40. Model plots of the lamp emission spectrum (red), the Rb cell absorption coefficients at 20, 30, 40, 50, 60, 70 C (black), and the product (lamp spectrum)*(absorption coefficient) at 70C (green). These plots are mainly for illustration purposes, as the actual model parameters are not well known.*

Finally, we integrate over the spectrum to obtain the total optical power measured by the photodiode

$$P_{PD}(T) = C\int_{-\infty}^{\infty}[1-\exp\{-\tau_{0,lamp}\exp(-x^2)\}] * \exp\{-A\cdot\exp(-T_A/T_K)\cdot\exp(-x^2)\}dx \tag{14}$$

It is instructive to also consider a "laser" model in which we replace $I_{lamp}(\Delta f)$ with a monochromatic laser at the Rb resonance frequency. This produces a simpler function that can easily be adjusted to match experimental data:

$$P_{PD,laser\ model} = C\ \exp\{-A\exp(-T_A/T_K)\} \tag{15}$$

Figure 41 shows both these models along with the measured OP signal. Clearly the "laser" model, while conceptually simpler, greatly underestimates the OP signal at high temperatures. The laser model calculates only the signal at line center, while Figure 40 illustrates that much of the final signal comes from the edges of the Doppler profile.

The full model in Figure 41 better represents the data at all temperatures, but the model is still too simple (using a single Rb transition, for example), plus the model parameters are not well constrained. One conclusion we reached is that this modeling exercise is not well suited for the teaching lab. The details are somewhat complicated, the model parameters are poorly constrained, and the model does not provide a simple functional form that can be adjusted to fit the data. On the other hand, the full model does qualitatively explain the overall trends in the observations, and the exercise is useful for outlining the most relevant physical processes involved.

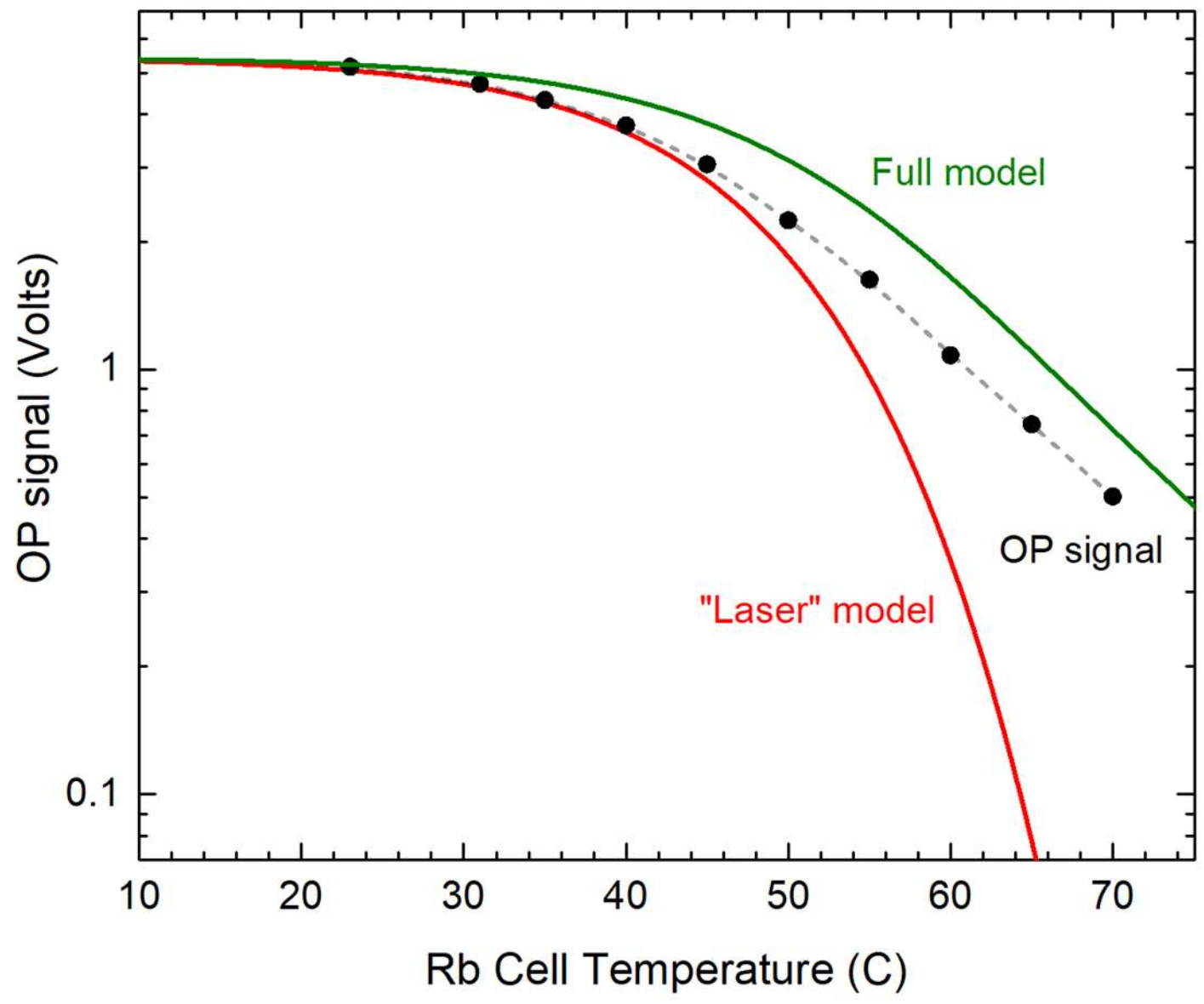

*Figure 41. This data points in this graph show the measured OP signal (same as in Figure 16, but unnormalized here). The green line shows the full model described by Equation 16, and the red line shows the "laser" model in Equation 17. The laser model greatly underestimates the measurements at elevated temperature, while the full model does a better job. But even the full model is too simplistic to accurately reproduce the data. Note that the measured background signal was below 10 mV in these data.*